\documentclass[acmtog,balance=false]{acmart}
\usepackage{mathtools}
\usepackage{amsmath}
\usepackage[]{lineno}
\usepackage[T1]{fontenc}
\usepackage{upquote}
\usepackage{colortbl}
\usepackage{multirow}

\setcitestyle{authoryear,open={[},close={]}} 

\setcopyright{none}
\begin{document}

\newcommand{\loss}{\mathcal{L}}
\newcommand{\usrc}{u_\text{src}}
\newcommand{\phase}{\phi}
\newcommand{\phasequant}{\psi}
\newcommand{\quant}{q}
\newcommand{\quantset}{\mathcal{Q}}
\newcommand{\prop}{f}
\newcommand{\amp}{a_\text{in}}
\newcommand{\red}[1]{\textcolor{red}{#1}}
\newcommand{\blue}[1]{\textcolor{blue}{#1}}
\newcommand{\propASM}{\mathcal{P}_{\textrm{\tiny ASM}}}
\newcommand{\Fourier}{\mathcal{F}}
\newcommand{\transfer}{\mathcal{H}}
\newcommand{\target}{a_{\text{target}}}
\newcommand{\sy}[1]{\textcolor{blue}{#1}}

\title{Holographic Parallax Improves 3D Perceptual Realism}


\acmSubmissionID{410}

\author{Dongyeon Kim}
\email{dongyeon93@snu.ac.kr}
\authornote{Authors contributed equally to this research.}
\orcid{0000-0003-4141-304X}
\author{Seung-Woo Nam}
\email{711asd@snu.ac.kr}
\orcid{0000-0001-9031-6049}
\authornotemark[1]
\affiliation{%
  \institution{Seoul National University}
  \country{Republic of Korea}
}

\author{Suyeon Choi}
\email{suyeon@stanford.edu}
\orcid{0000-0001-9030-0960}
\authornotemark[1]
\affiliation{%
  \institution{Stanford University}
  \country{USA}
}

\author{Jong-Mo Seo}
\email{callme@snu.ac.kr}
\orcid{0000-0002-1889-7405}
\affiliation{%
  \institution{Seoul National University}
  \country{Republic of Korea}
}

\author{Gordon Wetzstein}
\email{gordon.wetzstein@stanford.edu}
\orcid{0000-0002-9243-6885}
\affiliation{%
  \institution{Stanford University}
  \country{USA}
}

\author{Yoonchan Jeong}
\email{yoonchan@snu.ac.kr}
\orcid{0000-0001-9554-4438}
\affiliation{%
  \institution{Seoul National University}
  \country{Republic of Korea}
}

\renewcommand{\shortauthors}{Kim, D., Nam, S.-W., and Choi, S. et al.}

\begin{abstract}
Holographic near-eye displays are a promising technology to solve long-standing challenges in virtual and augmented reality display systems. Over the last few years, many different computer-generated holography (CGH) algorithms have been proposed that are supervised by different types of target content, such as 2.5D RGB-depth maps, 3D focal stacks, and 4D light fields. It is unclear, however, what the perceptual implications are of the choice of algorithm and target content type. In this work, we build a perceptual testbed of a full-color, high-quality holographic near-eye display. Under natural viewing conditions, we examine the effects of various CGH supervision formats and conduct user studies to assess their perceptual impacts on 3D realism. Our results indicate that CGH algorithms designed for specific viewpoints exhibit noticeable deficiencies in achieving 3D realism. In contrast, holograms incorporating parallax cues consistently outperform other formats across different viewing conditions, including the center of the eyebox. This finding is particularly interesting and suggests that the inclusion of parallax cues in CGH rendering plays a crucial role in enhancing the overall quality of the holographic experience. This work represents an initial stride towards delivering a perceptually realistic 3D experience with holographic near-eye displays.
\end{abstract}





\begin{teaserfigure}
  \includegraphics[width=\textwidth]{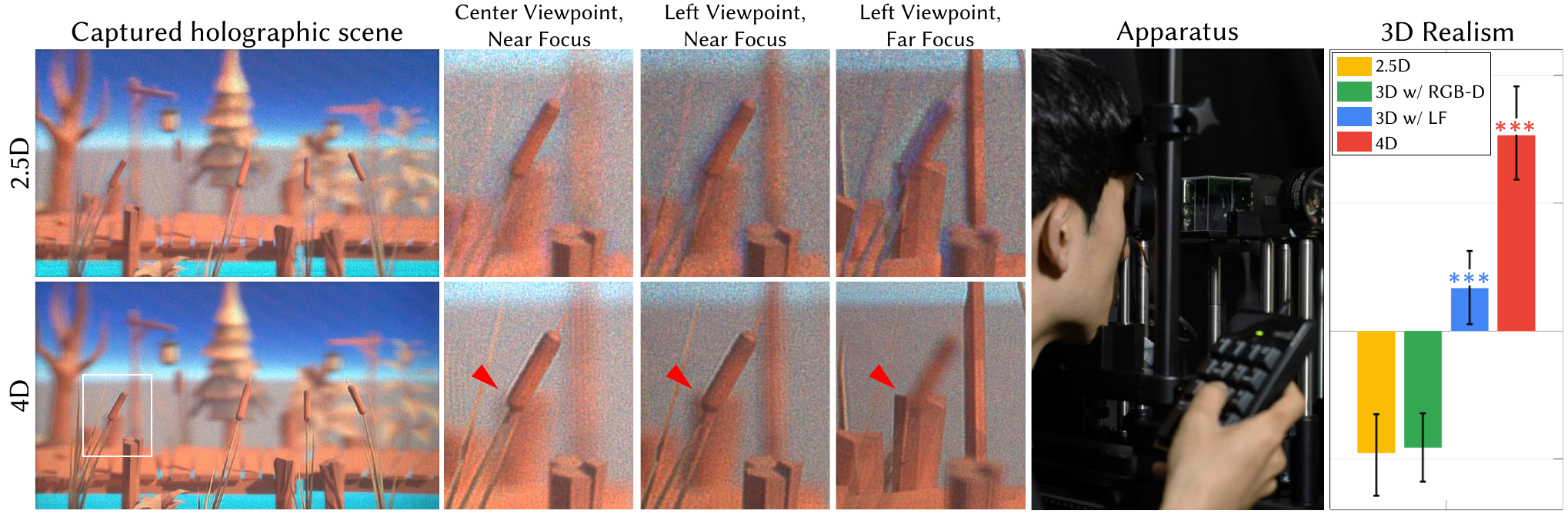}
  \caption{\textbf{Experimental results from our perceptual holographic testbed} (\textit{left}). With recent advancements in computer-generated hologram (CGH) algorithms, the image quality of holographic displays has surpassed the threshold required for conducting robust user studies, enabling us to investigate the perceptual impact brought by modern holographic displays. While a significant amount of effort has been made for specific target formats in ideal viewpoints, they lack to deliver correct parallax cues under natural viewing conditions (\textit{top row}). In contrast, light field holograms successfully convey these cues (\textit{bottom row}) highlighted with red arrows in the enlarged images provided with different capture settings (position, focus). We build a full-color, high-quality holographic testbed, where we conduct a user study examining 3D perceptual realism across various 3D CGH algorithms (\textit{right}). Our results reveal that CGH algorithms designed for specific types of targets significantly lag in perceptual realism, whereas the light field hologram notably outperforms other formats. Asterisks (red: 4D over the other formats, blue: 3D w/ LF over 2.5D and 3D w/ RGB-D) indicate the statistical significance of the difference (***: $p$<0.001), and the errorbars denote 95$\%$ confidence interval.}
  \label{fig:teaser}
\end{teaserfigure}

\received{23 Jan 2024}
\received[revised]{23 Jan 2024}
\received[accepted]{17 April 2024}

\maketitle

\section{Introduction}
Holographic displays offer great potential as the next-generation platform for augmented and virtual reality displays~\cite{Maimone:2017,jang2024waveguide} due to their versatile functionalities providing high-resolution volumetric images with aberration and vision correction~\cite{kim2021vision} capabilities arising from complex amplitude modulation of light. Nevertheless, achieving superior holographic visualization through the utilization of spatial light modulators (SLMs) that operate solely on either phase or amplitude has been a long-standing challenge in the field. Recently, several significant breakthroughs in holographic image quality incorporating machine learning-based approaches have shown a promising path toward a renaissance of computational holography~\cite{peng2020neural,chakravarthula2020learned,shi2021towards,yang2022diffraction, nam2023depol}. However, most evaluations of these holographic displays have been conducted using camera-based experiments, which only consider specific viewing conditions.

Unlike stationary cameras, the human eye is constantly in motion, involuntarily experiencing pupil contractions and relaxation~\cite{bahill1975main}. In contrast to incoherent displays, which have developed gaze-contingent approaches to address eye movement~\cite{mercier2017fast,guan2022perceptual}, holographic displays possess unique capabilities in controlling the plenoptic function of light~\cite{choi2022time, park2020efficient}. However, the limited size of the eyebox~\cite{Jang2018Holographic} and computational load of holographic displays often leads to approximating the 3D scene based on the center view and overlooking the impact on other views~\cite{shi2021towards}.

These two facts pose a number of fundamental questions to the field of holographic displays: Does an approximated holographic 3D scene, optimized for a camera or an ideal viewpoint, truly provide a satisfying 3D experience for users? How robust are these approximations, and are the perceived differences discernible even within the current optical settings of holographic near-eye displays with limited \'etendue? Secondly, if we aim to determine the optimal format for high-quality 3D holographic scenes by addressing the aforementioned question, what criteria must be met in order to surpass perceptual thresholds?

In this study, we investigate the perceptual realism of 3D scenes presented through holographic near-eye displays, while considering natural viewing conditions. Our approach includes simulating the perceptual quality of the 3D holographic scenes with varying computer-generated holography (CGH) target content types such as 2.5D RGB-depth maps, 3D focal stacks and 4D light fields, and pupil conditions and accounting for the impact of eye movements on sampled signals. To find the best CGH supervision format for realistic 3D holographic scenes, we conduct user studies. Our results show that incorporating parallax cues significantly enhances the 3D user experience, even with limited head movement. This study represents a first step in the field of 3D visual experience with holographic near-eye displays and provides guidelines for creating perceptually realistic 3D holographic scenes. (See Fig.~\ref{fig:teaser})


The contributions of this study are as follows.
\begin{itemize}
\item We simulate the impacts of eye movement, pupil size fluctuation, and directional sensitivity of the retina on the perceived 3D holographic scenes, which implies discrepancies between camera-based experiments and evaluations involving humans.
\item We design and conduct user studies under various viewing conditions to determine the optimal formats that holographic displays need to reproduce in order to achieve 3D perceptual realism. To this end, we build a perceptual testbed of a holographic near-eye display with high-quality, full-color 3D holographic scenes.
\item The user studies reveal the findings indicating that the 3D CGH supporting parallax cues significantly improves 3D perceptual realism in various viewing conditions, even with limited head movements. 
\end{itemize}



\section{Related Work}
\subsection{Computer-generated holography}

Computer-generated holography (CGH) encompasses algorithms that generate holograms for spatial light modulators (SLMs), manipulating the complex-valued incident wave field to achieve desired light field distributions for viewers. These algorithms have been developed to accommodate various 3D data formats, including image layers~\cite{Zhang:2017, shi2022end}, RGB-D~\cite{Chen:21, shi2021towards, choi2021neural3d}, focal stacks~\cite{choi2022time, kavakli2023realistic, yang2022diffraction}, or polygons~\cite{Matsushima:2009:polygon, wang2023high}, with essential occlusion handling~\cite{symeonidou2015computer}.

It is noteworthy that intensity-based data formats do not inherently impose constraints on the phase distribution of holograms, introducing uncertainty to their plenoptic function. One can assume a random phase to simulate diffused light~\cite{lohmann1967binary}, or a smooth phase which might result in better contrast~\cite{Maimone:2017, shi2021towards} but at the cost of a reduced eyebox size~\cite{choi2022time}.

Recent studies have emphasized the tradeoff between image quality and the eyebox. Yoo et al.~\shortcite{yoo2021optimization} investigated controlling randomness to strike a balance, and stochastic pupil sampling can ensure consistent 2D image appearance across the eyebox~\cite{chakravarthula2022pupil}. However, these approaches have limitations in accurately expressing spatial-angular information across the \'etendue~\cite{kuo2020high,park2019ultrathin}. Light field holograms, also known as holographic stereograms~\cite{Shi:2017, choi2022time, Padmanaban:2019, park2020efficient, Kang:2016, zhang2015fully}, present a promising solution to address this challenge.

\subsection{Visual perception}
Virtual reality (VR) systems aim to offer immersive experiences by understanding the human visual system and perceptual studies build guidelines for the relevant communities. Recent advancements in visual difference predictors (VDP) ~\cite{mantiuk2021fovvideovdp} are used to estimate perceived image quality, considering different display configurations and observance models.


\begin{figure}[t!]
  \centering
\includegraphics[width=\columnwidth]{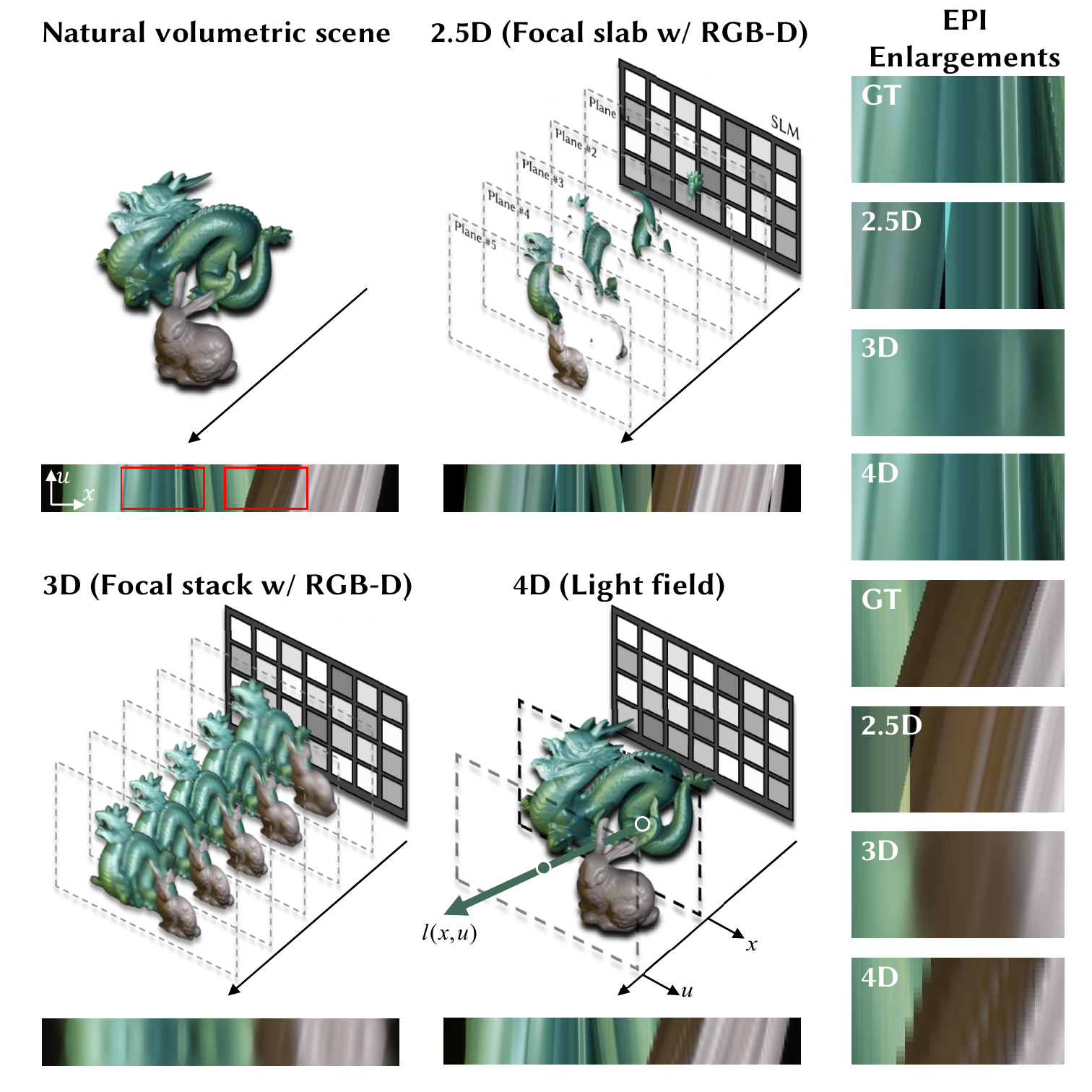}
  \caption{\textbf{Various CGH supervision targets} (2.5D, 3D, 4D) \textbf{for holographic displays to realize the natural volumetric scene} (ground truth, GT). The reconstructed epipolar plane images (EPIs) of the individual data formats are provided to demonstrate the angle-dependent spatial information and the red-boxed regions are enlarged to demonstrate the differences. The EPIs are reconstructed with 25 horizontal views for the GT case, 5 planar images for 2.5D and 3D cases, and 5 view images for 4D case. Dragon, Bunny: credit to Stanford Computer Graphics Laboratory.}
  \label{fig:concept}
\end{figure}

Depth perception mechanisms are vital for evaluating the perceptual realism of 3D scenes. Various cues, including binocular disparity, accommodation, convergence, and motion parallax, contribute to depth perception~\cite{cutting1995perceiving}. Aligning these cues is crucial to reduce visual fatigue~\cite{hoffman2008vergence} during prolonged VR device use. Images with binocular disparity are most sensitive to human perception, providing the primary cue for depth perception within arm's reach. Motion parallax, perceived through retinal motion, is the strongest depth cue supported for objects approximately one meter or more away. Retinal motion-driven depth perception is aided by the smooth pursuit eye movement~\cite{nawrot2003eye,naji2004perceiving}. Rendering VR scenes, taking into account ocular parallax~\cite{konrad2020gaze}—the change in viewpoint due to eye rotation—has improved perceptual realism. However, the evaluation was done with stereo 3D head-mounted displays.

Assessing 3D realism can be subjective, incorporating various cues for image and depth perception, relying on individual visual behavior. However, the primary aim of VR displays persists in achieving \textit{3D perceptual realism} and successfully passing the visual Turing test~\cite{wetzstein2016factored}, particularly when assessing volumetric scenes under natural viewing conditions. Recent studies~\cite{zhong2021reproducing, march2022impact} have conducted visual Turing tests using dual-plane stereo displays, representing progress in advancing next-generation display technologies.

\subsection{Perceptual 3D holographic testbed}

Conducting meaningful user studies with holographic displays has historically been challenging due to low image quality, characterized by speckles and imperfect representation of the complex-valued field, resulting in low contrast compared to other displays. However, recent advancements in CGH and SLMs have significantly improved image quality through techniques like time-multiplexing and calibration~\cite{lee2020wide, choi2022time, lee2022high, curtis2021dcgh, peng2020neural, chakravarthula2020learned}. These advancements enable more accurate and robust user studies with holographic displays. Kim et al.~\shortcite{kim2022accommodative} conducted a user study with holographic near-eye displays, enhancing accommodation response using CGH supervision with a regularizer on the contrast ratio of two-dimensional (2D) images. However, the study focused solely on 2D content and did not consider parallax cues. For a comparison of 3D perceptual testbeds, please refer to Table S1.

\definecolor{mg}{rgb}{0.639,0.984,0.722}
\definecolor{my}{rgb}{0.996,0.875,0.643}
\definecolor{mr}{rgb}{0.941,0.561,0.620}
\begin{table}[ht!]
    \centering
    \begin{tabular}{c|c|c|c|c}
        \multirow{2}{*}{} & multiple points  & \multirow{2}{*}{retinal blur} & view \\
         & in a single ray &  & dependency \\
         \hline
       2.5D  & \cellcolor{mr} no & \cellcolor{my} approx. & \cellcolor{my} approx. \\        \hline
        3D w/ RGB-D & \cellcolor{mg} yes & \cellcolor{my} approx. & \cellcolor{my} approx. \\ \hline
        3D w/ LF & \cellcolor{mg} yes & \cellcolor{mg} correct & \cellcolor{my} approx. \\ \hline
        4D & \cellcolor{mg} yes & \cellcolor{mg} correct & \cellcolor{mg} correct \\ \hline
    \end{tabular}
    \caption{\textbf{Assessment of data formats in terms of supported visual cues}. In 2.5D and 3D formats, the blur behavior is constrained by the phase profile or the pupil size used in rendering the focal stack. Conversely, the defocus behavior in light fields would be accurate, though blur size could be limited by the \'etendue supported by the display system. The light field format and supervision uniquely enable the display to render correct view-dependent effects, such as occlusion, parallax, and specular highlights.}
    \label{tab:data_formats}
\end{table}
\subsection{Visual effects from 3D assets}

The visual experience of a display device, especially for 3D content, depends on the presented content. The epipolar plane image (EPI) in Fig.~\ref{fig:concept} represents a horizontal cross-section image of ray-space ($l(x,u)$), defined with space ($x$) and direction ($u$). Different data formats have limitations: focal slabs (2.5D target) lack spatial information in angles other than the normal angle ($u=0$)~\cite{chang2020towards}. Focal stacks (3D) may overfit to the central view (3D w/ RGB-D) but can be generated with multiple views (3D w/ LF). Light field (4D) offers angle-dependent spatial information but with sparse sampling. The intrinsic nature of the light field enables the reproduction of angle-dependent visual effects such as occlusion and shading. For a comparison of visual effects supported by individual assets, please refer to Table~\ref{tab:data_formats}.

\section{Uniform 3D holographic experience across the eyebox}


Different 3D data formats possess inherent visual effects, and as the dimensions increase, so does the computational complexity. This raises the question of whether 3D near-eye displays necessitate the reconstruction of a 4D light field rather than relying on approximated 3D information since the perceived view-dependent effects highly vary on viewing conditions. Major aspects that affect the experience of view-dependent visuals include the eyebox size of the 3D near-eye display and the pupil status of the human eye. 

In this section, our goal is to determine the most suitable 3D data format for visualizing perceptually realistic 3D scenery and assess it using a holographic near-eye display capable of rendering various 3D assets. To ensure comprehensive results, we conduct simulations and experiments across different scenarios involving variations in eyebox and pupil status.

\subsection{Simulation}
\begin{figure*}[ht!]
  \centering
  \includegraphics[width=\textwidth]{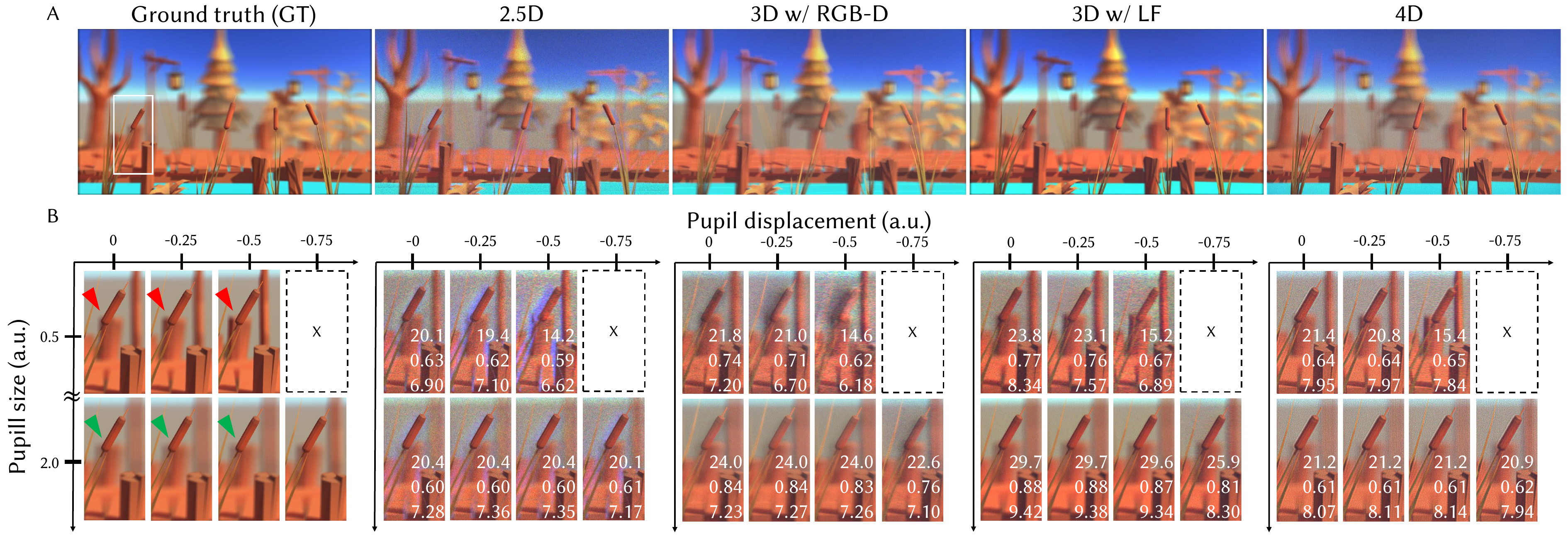}
  \caption{\textbf{Holographic reconstruction with different CGH supervision targets:} (A) Near-depth focused holographic images (2.5D (2nd col, focal slab), 3D w/ RGB-D (3rd col, focal stack with RGB-D), 3D w/ LF (4th col, focal stack with 25$\times$25 LF), 4D (5th col, 9$\times$9 LF)) of the landscape$\_$day scene are reconstructed in the full eyebox condition, respectively with the ground truth (GT) focal stack (1st col). (B) Enlargements of the corresponding holographic scenes reconstructed based on 7 different pupil settings (pupil displacement and size) are presented except the one with the fully vignetted (box with dashed line) condition. Each enlargement is consecutively provided with the quality metrics of PSNR, SSIM (maximum of 1), FovVideoVDP (JOD unit having a maximum of 10) ~\cite{mantiuk2021fovvideovdp} evaluated with the GT focal stack. Here, the pupil displacement (${x}_{p,norm}$) presents the eye pupil's displacement (${x}_{p}$) in the horizontal axis and the pupil size (${D}_{p,norm}$) denotes the diameter of the human eye pupil (${D}_{p}$), and those values are normalized with the width of the eyebox (${w}_{eyebox}$, 2.2 mm). The enlargements with red arrows indicate scenes reconstructed under an overfilled pupil and those with green arrows denote images visualized under an underfilled pupil. (purchased Unity asset: Low Poly Series: Landscape)}
  \label{fig:simulation_pupil}
\end{figure*}


In practice, the eye rotates to gaze at the objects located across the field of view leading to pupil displacement. Moreover, the pupil varies in size depending on the intensity of light entering it, and there is significant variation in pupil size among individuals. The holographic images were reconstructed to examine the impact of different pupil states (displacement and size) as shown in Fig.~\ref{fig:simulation_pupil}. As addressed by previous works~\cite{chakravarthula2022pupil,kim2022accommodative} concerning the eyebox of holographic near-eye displays, the human eye pupil, while not considering its focal state, located in the eyebox domain optically acts as a binary low-pass filter and it samples the display signal.

The holographic images are reconstructed in different pupil states and evaluated with different quality metrics (peak signal-to-noise (PSNR), structural similarity metric (SSIM), FovVideoVDP~\cite{mantiuk2021fovvideovdp}) as provided in Fig.~\ref{fig:simulation_pupil}. Here, we included the FovVideoVDP as it outperformed other quality metrics in terms of evaluation of light field dataset through stereo 3D displays~\cite{kiran2017towards}. The ground truth focal stack is generated with a dense light field (25$\times$25 views) and processed with the identical pupil state. In detail, FovVideoVDP (v1.2.0) is estimated in a non-foveated mode with the condition of 86.2 [pix/deg], Lpeak=100, Lblack=0.1 [${cd}/{{{m}^{2}}}$] and the results are scaled in a unit of Just-Objectionable-Difference (JOD). Note that the difference of 1 JOD refers to the visual difference that 75 percent of subjects choose the option compared to the counterpart and serves as the perceptual difference threshold.

When comparing the supervision of 4D CGH with center-view based CGH supervision (2.5D and 3D w/ RGB-D) using conventional image metrics like PSNR or SSIM, it is observed that 4D-supervised CGH results in relatively similar or sometimes poorer quality across the eyebox. However, the reconstructed results show better FovViodeVDP exceeding around 1 JOD, as described in Fig.~\ref{fig:simulation_pupil}(B). The assessment using FovVideoVDP ensures the improved perceptual quality of 4D across the eyebox.

\subsubsection*{Eyebox-pupil scenarios}
Although determining the field of view and the size of the eyebox is one of the design considerations for holographic near-eye displays with limited \'etendue, the pupil size of the human eye fluctuates based on luminance. This results in various scenarios regarding the ratio of the exit-pupil and ocular-pupil areas~\cite{ratnam2019retinal}. Based on this relation in size, we can categorize into two major eyebox-pupil scenarios: an overfilled pupil when the size of the human eye pupil is smaller than the eyebox and an underfilled pupil when the size of the eye pupil is larger than the eyebox.

In the overfilled pupil scenarios (pupil states indicated by red arrows in Fig.~\ref{fig:simulation_pupil}(B)), the quality of the 3D w/ LF case deteriorates as the pupil is decentered while 4D shows smaller falloffs. If there is a difference of about 1 JOD in the comparison of the cases (3D w/ LF vs. 4D) in the specific pupil state ($({x}_{p,norm},{D}_{p,norm})=$(-0.5, 0.5)), it may affect the quality of the overall viewing experience with eye movements. We additionally provide simulation and experimental results captured with the display system in Supplementary Material.

In contrast to the overfilled pupil scenarios, the 3D w/ LF case outperforms the 4D case in underfilled pupil scenarios (pupil states with green arrows in Fig.~\ref{fig:simulation_pupil}(B)) in terms of the simulated metrics. It is worth noting that evaluation with perceptual metrics may underestimate the impact of human factors and the influence of ocular parallax on perceptual realism since the metrics are built upon 2D image-based assessment. Therefore, in the subsequent subsection, we conduct a user study under the worst-case scenario, where the eyebox is significantly smaller compared to the eye pupil, evaluating various 3D data formats within this context.


\subsection{User evaluation}
\subsubsection*{Hardware and software}
We built a benchtop prototype of a holographic near-eye display with a single SLM having a resolution of 1920 (H) $\times$ 1200 (V), and a full-color laser diode as a perceptual testbed for user validation as Fig.~\ref{fig:userexp1}(A). In addition, we additionally placed an eye tracker to adjust the eye position of participants and simultaneously record the subjects' pupil position and size. The prototype provides an image with a resolution of 1600 $\times$ 900, and the corresponding field of view of 18.6$^{\circ}$ $\times$ 10.5$^{\circ }$. The maximum resolution achieved by the system is 43 cycle per degree (cpd). The eyebox of the near-eye display defined with the blue illumination and a 40-mm focal length eyepiece lens is 2.2 mm $\times$ 1.1 mm as we place a side-band filter to modulate a complex-valued field with amplitude-encoded CGH. We carefully designed the perceptual testbed, ensuring that its maximum resolution surpasses the human visual acuity of 30 cpd~\cite{guenter2012foveated} while maintaining an eyebox size smaller than the average human eye pupil~\cite{de1952pupil}. 

The SLM utilized in the user experiment supports the full-color speckle-reduced image with temporal multiplexing of 24 binary CGHs. The CGH acquisition with various target formats is implemented with Pytorch based on the previous works of the differentiable time-multiplexed CGH optimization frameworks~\cite{choi2022time,lee2022high}. Detailed information on software and hardware can be found in Appendix and  Supplementary Material.

\subsubsection*{Stimuli and Conditions}
\begin{figure}[ht!]
  \centering
  \includegraphics[width=\columnwidth]{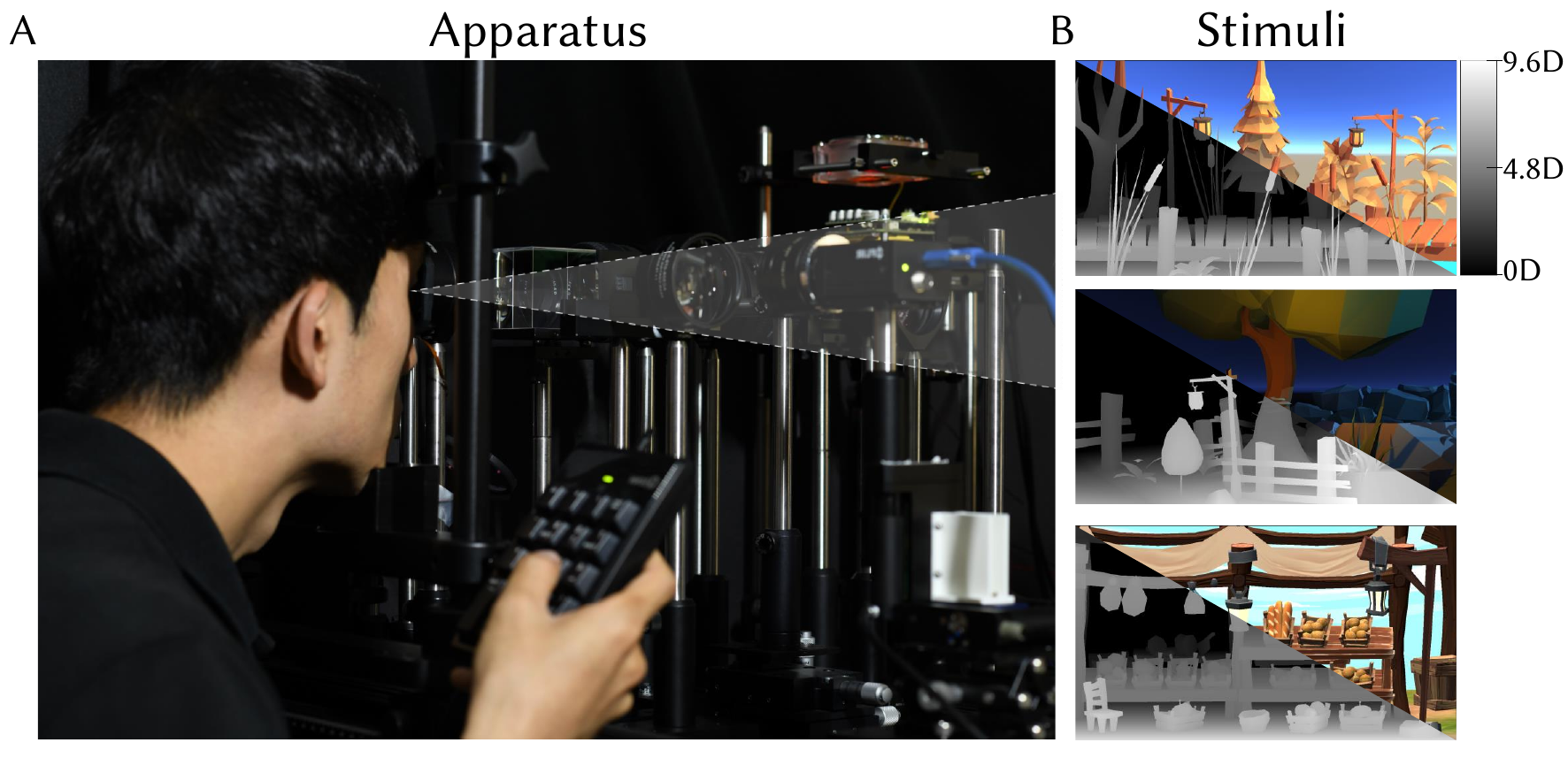} \caption{\textbf{3D holographic perceptual testbed and stimuli.} (A) We conduct the user study using the apparatus shown on the left. Holographic scenes are generated using various CGH methods, using targets rendered with scenes purchased from the Unity Asset Store (Fantastic-Village Pack)} 
  \label{fig:userexp1}
\end{figure}

For the user validation, three volumetric scenes - landscape$\_$day, landscape$\_$night, and village - are used as stimuli as presented in Fig.~\ref{fig:userexp1}(B). The depth range extends from 0 diopter (\emph{D}) to 9.57 \emph{D}, spanning from optical infinity to 11 cm from the eye. This range sufficiently covers the average accommodation range of young adults~\cite{duane1912normal}, and ocular parallax induced by the objects with the given depth range exceeds the minimal angular resolution of the fovea region~\cite{konrad2020gaze}.  The display scheme is explained in Fig. S1. The luminance of each scene is estimated as 2 ${cd}/{{{m}^{2}}}$, and the room was kept dark during the experiment.



\begin{figure*}[ht!]
  \centering
  \includegraphics[width=\textwidth]{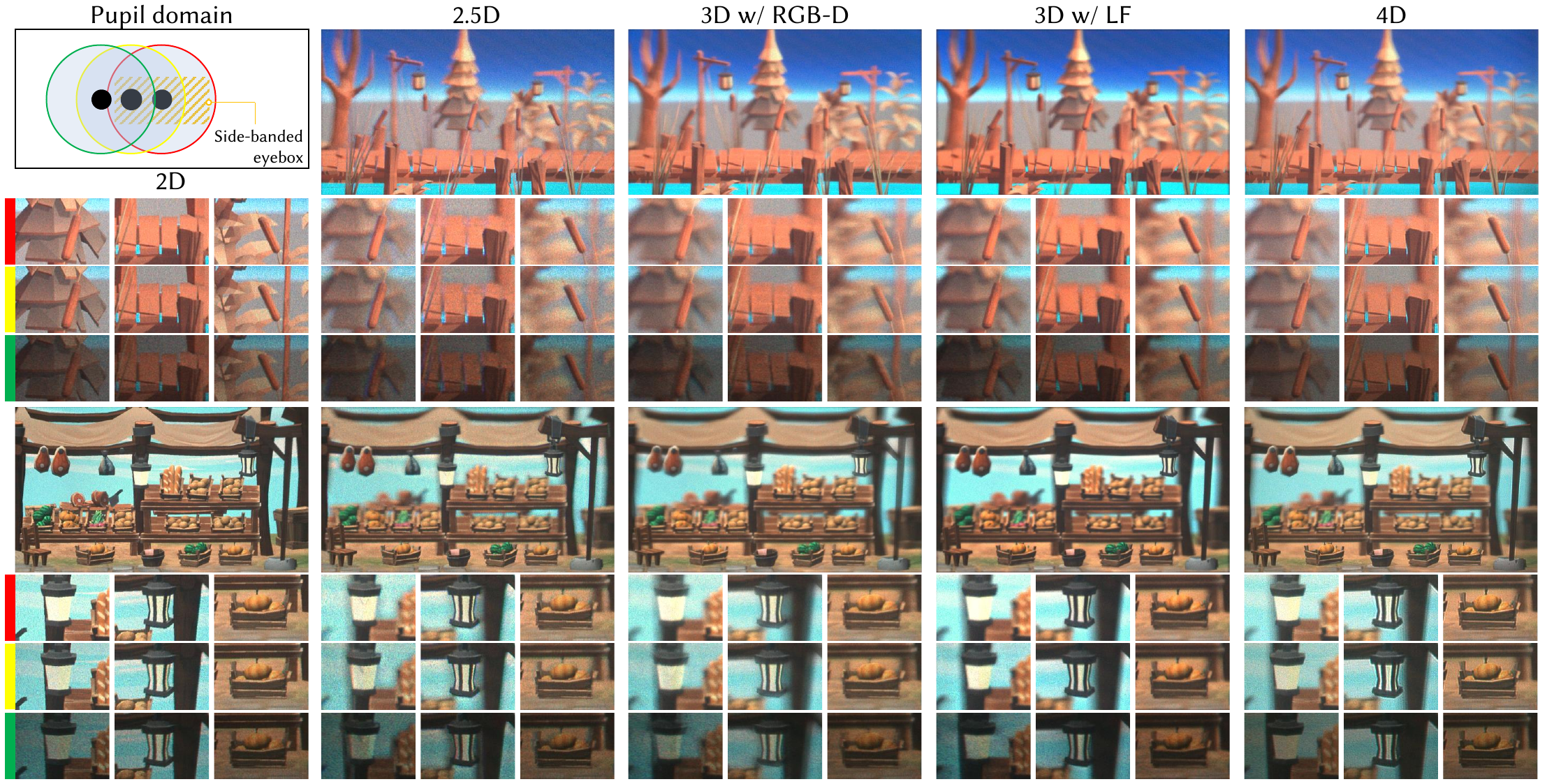}
  \caption{\textbf{Experimental results with different pupil positions.} Holographic scenes supervised with 2D (1st col), 2.5D (2nd), 3D w/ RGB-D (3rd), 3D w/ LF (4th), 4D (5th) targets are captured with different pupil positions (red: $({x}_{p,norm},{D}_{p,norm})=(0,1.1)$, yellow: $(-0.68,1.1)$ and green: $(-1.36,1.1)$). The scenes are photographed with different focal states (landscape$\_$day: 7th, village: 7th) out of 9 distinct focal states equally sampled in diopter. Enlargements are provided with the image focused on the magnified object. The colors of each row for the enlargements indicate the pupil positions (red: center, yellow: decentered, green: vignetted). We intentionally provide the results without modifying the brightness to show the energy across the eyebox. Note that it is hard to discriminate 3D w/ LF case and 4D case with the captured results.}
  \label{fig:exp1}
\end{figure*}

In the case of 2.5D and 3D-supervised CGHs, nine planes equally spaced in a unit of diopter are sampled. For 3D-supervised CGHs, we prepared two scenarios; \emph{3D w/ RGB-D} and \emph{3D w/ LF}. For \emph{3D w/ RGB-D}, the focal stacks are generated with a single RGB-D map blending occlusion boundaries~\cite{lee2022high}. For \emph{3D w/ LF}, we utilized LF with 25$\times$25 orthographic views to generate focal stacks to naturally handle occlusion. Lastly, LF with 9$\times$9 orthographic views is utilized for 4D CGH supervision~\cite{choi2022time}. The captured scenes of each stimulus are provided in Fig.~\ref{fig:exp1} and Fig. S16. 

The experiment is done with four different viewing conditions; \emph{Center} refers to the case when the subjects view 3D contents while placing the pupil at the \textquotesingle sweet spot\textquotesingle\ of the eyebox and this represents the underfilled pupil. \emph{Decentered} and \emph{Vignetted} refer to the condition when the eye is horizontally decentered about 1.25 mm and 2.5 mm from the center, respectively. \emph{w/ head movement} refers to the viewing condition when the subjects perform the task without head movement restriction. 
Note that none of the viewing conditions limited eye movements. We referred to the viewing conditions as \emph{Center}, \emph{Decentered}, \emph{Vignetted} to differentiate the conditions based on the initial placement of the eye.

Before the experiment, six complete pairs with four different CGH supervision cases (\emph{2.5D}, \emph{3D w/ RGB-D}, \emph{3D w/ LF}, \emph{4D}) were prepared, the order was randomly shuffled to eliminate the potential decision bias and each pair was repeatedly provided three times. The complete pairwise comparison was held with three different scenes (landscape$\_$day, landscape$\_$night, and village) in four different viewing conditions (\emph{Center}, \emph{Decentered}, \emph{Vignetted}, \emph{w/ head movement}). The whole number of trials was 216 (6 pairs $\times$ 3 repetitions $\times$ 3 scenes $\times$ 4 viewing conditions).


\subsubsection*{Subjects}
We recruited 28 naïve participants under the age of 40 (ranging from 23 to 36 with an average of 27.6, 12 female) to account for the potential decrease in accommodation range with age. All participants had normal or corrected-to-normal vision and normal color vision. They were rewarded for their participation, and the studies adhered to the Declaration of Helsinki. All subjects provided voluntary written and informed consent, and the experiment was conducted after receiving approval from the Institutional Review Board of the host institution.

\subsubsection*{Procedure}
Before each session, precise head alignment was performed. Subjects were instructed to restrict their head movement in viewing conditions other than \emph{w/  head movement}. The head position of the subjects was controlled by adjusting the components of the chin-and-head rest. During this adjustment procedure, subjects were asked to maintain their gaze over the center object of the sample scene, utilizing eye-tracked data monitored in real-time. For the viewing condition \emph{w/ head movement}, the chin-and-head rest was removed, and subjects were free to move their heads within a range where scenes remained observable. 


In every viewing condition, a two-interval forced choice (2-IFC) \cite{bogacz2006physics} task was conducted, asking subjects to choose the \textquotesingle more realistic 3D\textquotesingle\ 
option after presenting a pair of stimuli in sequence. Subjects were instructed to gaze at different objects and aim for a sharp focus on the gazed object to assess 3D quality, eliminating subjects who maintained focus at a single plane. Each pair of stimuli was displayed for 8 seconds, with a second of a gray noisy image provided in between. Responses were recorded using a keypad, and the next pair was presented after a valid input. After each session, subjects were encouraged to take a break for at least a minute, and the entire experiment took around an hour.


\subsubsection*{Results} 
\begin{figure}[ht!]
  \centering
  \includegraphics[width=\columnwidth]{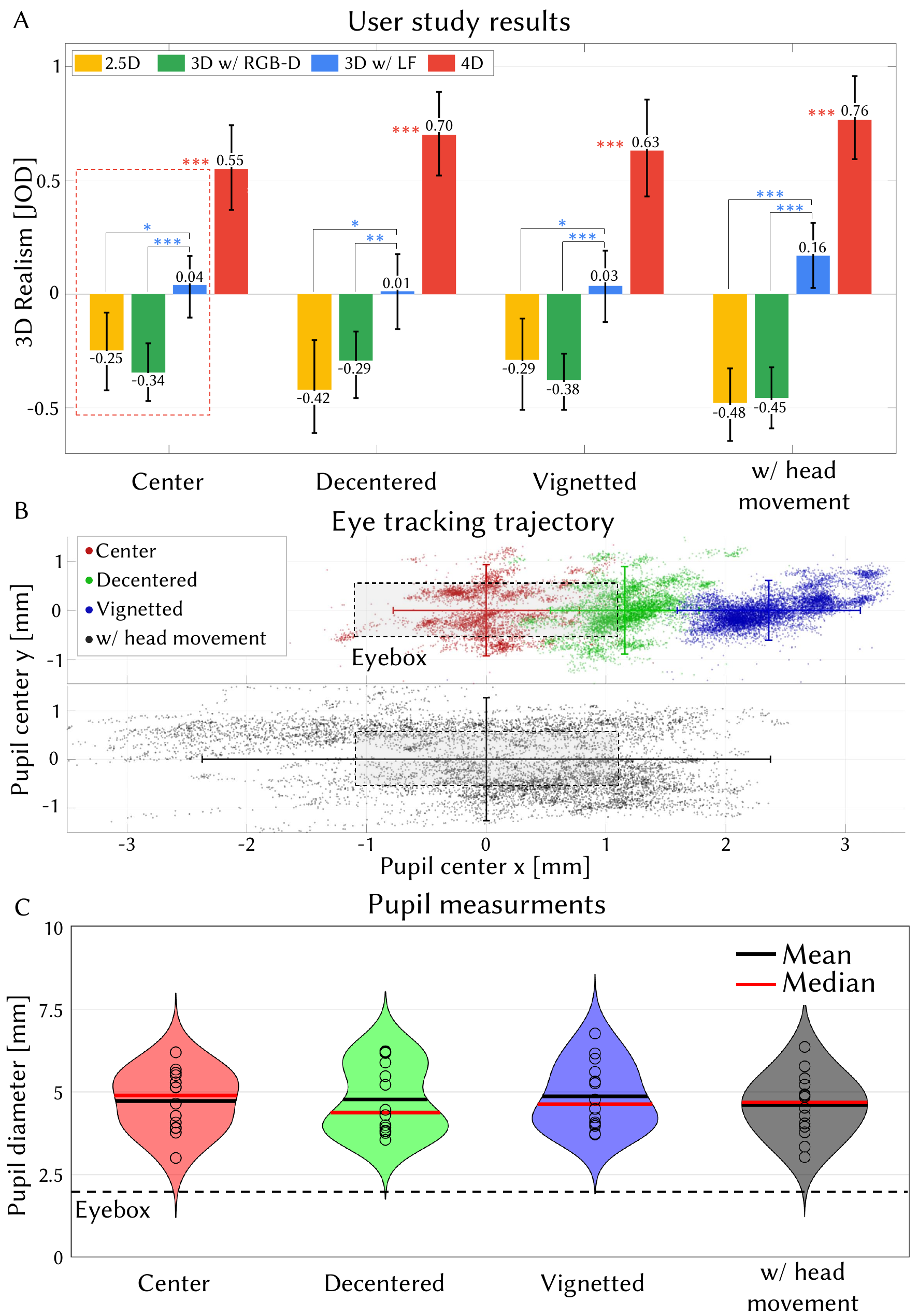}  \caption{\textbf{User experiment results:} (A) 3D realism is assessed using CGHs supervised with four target formats (2.5D in yellow, 3D w/ RGB-D in green, 3D w/ LF in blue, 4D in red) across four viewing conditions (Center, Decentered, Vignetted, and with head movement). The mean JOD is set as zero for each viewing condition. Error bars represent 95$\%$ confidence intervals estimated by bootstrapping 500 samples. Asterisks (blue: 3D w/ LF vs. paired cases, red: 4D vs. other cases) indicate the statistical significance of differences (*: $p$<0.05, **: $p$<0.01, ***: $p$<0.001). (B) The tracked trajectory of the pupil center for one representative subject. Error bars represent the 95$\%$ confidence interval of the pupil displacement. (C) Measured pupil diameters of representative subjects depending on the viewing conditions. The black circle corresponds to the pupil diameter of individual subjects and the dashed line denotes the width of eyebox in our experimental setup.}
  \label{fig:userexp1_result}
\end{figure}

The CGHs supervised with different targets were evaluated in four distinct viewing conditions and compared in terms of perceived 3D realism, as depicted in Fig.~\ref{fig:userexp1_result}(A). The accumulated vote counts from a total of 26 subjects were normalized and scaled using the unit of JOD. The responses of two subjects were excluded after outlier analysis introduced by the work of \citeauthor{perez2017practical} ~\shortcite{perez2017practical}. 


Upon conducting the two-tailed z-test with the scaled JOD scores for each CGH supervision target in each viewing condition, the results indicate that 4D-supervised CGHs exhibit significant improvements in perceived 3D quality across all viewing conditions compared to other forms of CGH supervision. Especially compared with \textit{2.5D} and \textit{3D w/ RGB-D}, the difference exceeds 1 JOD in some conditions. Additionally, \textit{3D w/ LF} is significantly preferred over \textit{2.5D} and \textit{3D w/ RGB-D}, and this preference is even more pronounced in the viewing condition involving head movement. Notably, no significant differences were observed between \textit{2.5D} and \textit{3D w/ RGB-D} in any of the viewing conditions. 

In summary, the results demonstrate significant differences in 3D perceptual realism in every viewing condition when the parallax cues were taken into account in CGH supervision. The 4D-supervised CGH outperformed all other cases by considerable margins, and even the \textit{3D w/ LF} CGH outperformed the other cases with strong evidence of significance.

Throughout the experiment, the position and size of the subjects' pupils were monitored and recorded. Figure~\ref{fig:userexp1_result}(B) presents the measured data of one representative subject in a single session, with different colors indicating different viewing conditions. The measured data demonstrate that the experiments were carried out under various viewing positions. Furthermore, it is noteworthy that even when head movement was restricted, the eye exhibited substantial movement. 

We provide the pupil diameter measured data with the equipped eye-tracker depending on four different viewing conditions as shown in Fig.~\ref{fig:userexp1_result}(C). We excluded the data corrupted by eye blinking by subjecting the measured data achieving the confidence level of 0.85. The average pupil diameters were measured from 4.5 mm to 5 mm. This measured value exceeds pupil diameter of 4.4 mm (${D}_{p,norm}=2.0$) ensuring that most of the pupil positions recorded in the viewing condition of \textit{Center} as the underfilled pupil condition. Due to the imperfect pupil diameter measurement with the eye-tracker, the measured data of 14 subjects are presented. The mean pupil diameter of 5 mm corresponds to the case when the luminance is low around 1 ${cd}/{{{m}^{2}}}$~\cite{napieralski2019modeling}. If the luminance level is as high as the level supported by the conventional VR displays (hundreds of nits)~\cite{mehrfard2019comparative}, the pupil diameter would be smaller and the impact of parallax cues will magnify in the overfilled pupil conditions as observed in Fig.~\ref{fig:simulation_pupil}.

It is intriguing that even within the experiment conducted at the \textit{Center} viewing condition that the pupil is large enough to cover the eyebox, the \emph{4D} approach outperforms the \emph{3D w/ LF} approach in terms of the perceived quality of 3D visuals. Note the retinal image of human eye is based on the focal stack. Notably, the focal stacks in the \emph{3D w/ LF} case are generated with 25$\times$25 views, while \emph{4D} employs 9$\times$9 orthographic views for CGH supervision. Although the VDP estimated in the specific viewing condition is 0.51 JOD which does not exceed 1 JOD value, the discrepancy between the simulated VDP and the actual VDP from user studies could potentially open up a vast research field in 3D quality metrics. This reversal in preference will be further discussed in the discussion section.




\section{How many light field views are required?}
The prior evaluation emphasized the substantial improvement in 3D realism with 4D CGH supervision, especially supporting holographic parallax over other CGH supervision methods in diverse viewing conditions. Then, exploring the ideal number of views for 4D CGH supervision is crucial for efficient rendering, considering the holographic testbed's adaptable specifications—a feature not commonly available in other 3D displays.

\subsection{Experimental results}
\begin{figure*}[ht!]
  \centering

  \includegraphics[width=0.92\textwidth]{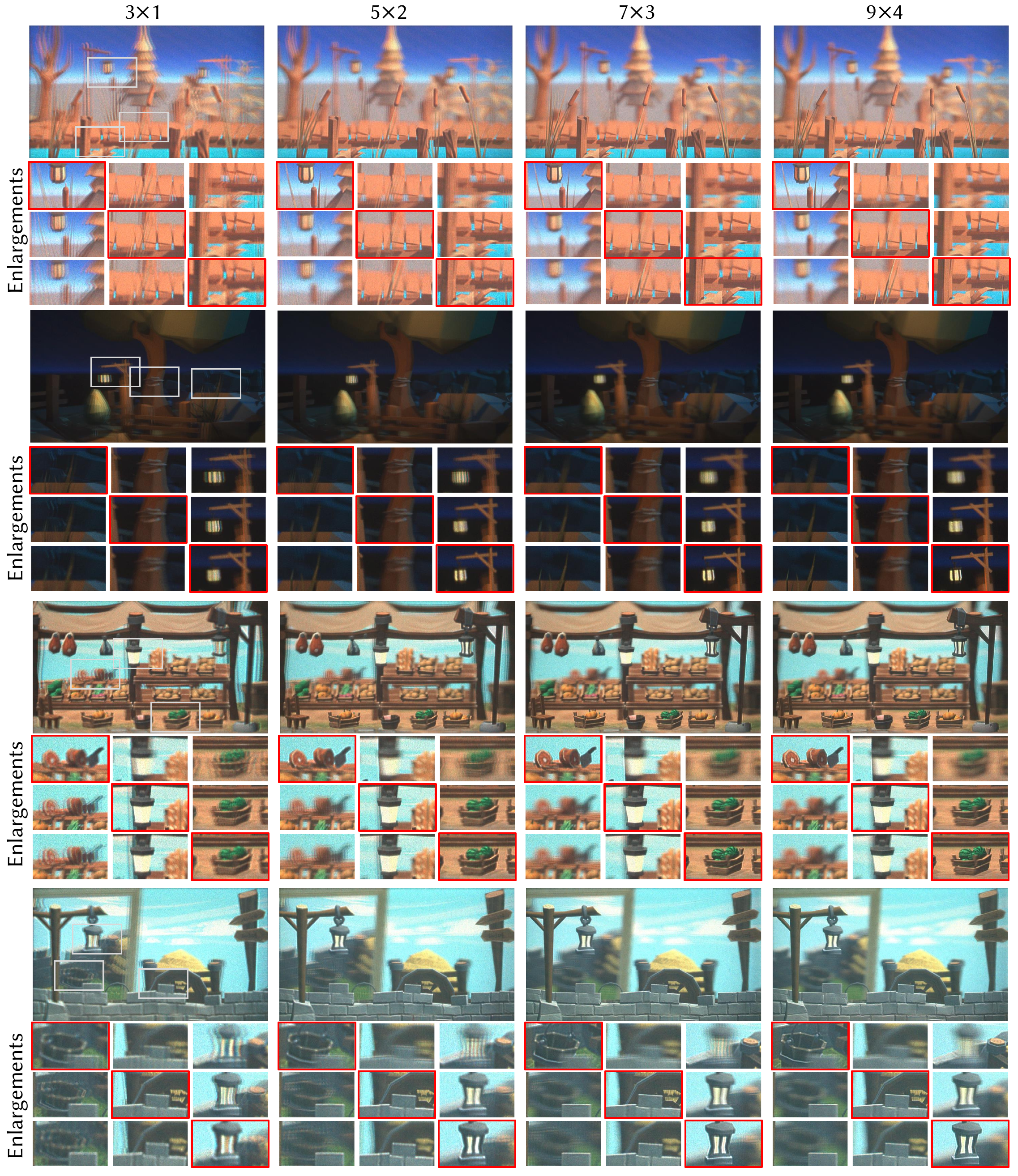}
  \caption{\textbf{Experimental results with the different number of views used for 4D CGH supervision.} The 3D holographic scenes of the sampled focal states are provided. Insets show the images captured with three different focuses randomly chosen for each scene and the images in gray boxes are enlarged. Among the enlarged images, the red boxes indicate that the object is focused.}
  \label{fig:exp2}
\end{figure*}

Fig.~\ref{fig:exp2} presents the experimental results captured with the benchtop prototype of a holographic near-eye display, illustrating the impact of the number of views used in 4D CGH supervision on 3D visualization. Additional reconstructed holographic scenes, varying based on the number of views, can be found in Fig. S12-S13.

\subsection{User study}
\subsubsection*{Procedure}
We obtained CGHs supervised with different view counts: 3$\times$3, 5$\times$5, 7$\times$7, and 9$\times$9. The optical setup, utilizing a side-band filter, resulted in effective view counts of 3$\times$1, 5$\times$2, 7$\times$3, and 9$\times$4, maintaining view gaps. Three scenes were used as stimuli in the initial experiment, and each pairwise comparison was repeated five times, totaling 90 trials lasting approximately thirty minutes. Eyebox centering was ensured before commencing the test, without recording eye-tracking data. All subjects that performed the first user experiment participated in the test and the overall procedure is identical to the first experiment. 

\subsubsection*{Results}
\begin{figure}[ht!]
  \centering
  \includegraphics[width=\columnwidth]{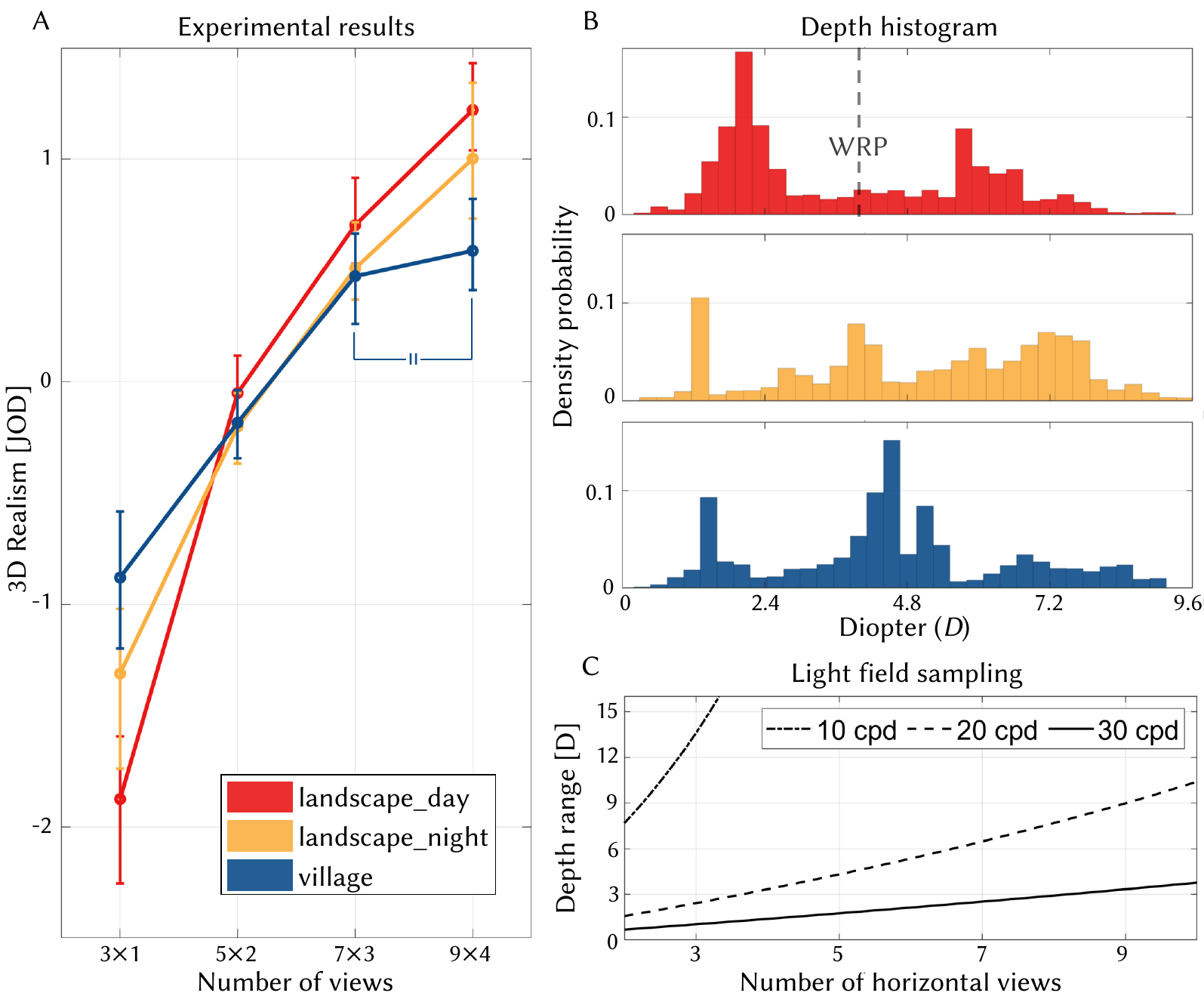}
  \caption{\textbf{The effect of the number of views employed in 4D CGH supervision on the perceived 3D realism:} (A) The JOD-scaled results of the pairwise comparison are provided depending on the scene. The equal symbol indicates that a statistically significant difference is not observed between the paired conditions of the scene. The errorbars indicate the 95$\%$ confidence interval acquired with bootstrapping. (B) Depth histogram profiles of stimuli. The dashed line indicates the WRP's dioptric depth of the system. (C) The relationship between the number of horizontal views used in the 4D CGH supervision and the depth range expressible by the system is plotted with the three distinct spatial bandwidths (dotdash line: 10 cpd, dashed line: 20 cpd, solid line: 30 cpd) of the targets.
  }
  \label{fig:userexp2}
\end{figure}

We conducted an evaluation comparing CGHs supervised using 4D targets with variations in the number of views. The responses of 24 subjects were analyzed, excluding four participants' responses after outlier analysis with JOD scores estimated from accumulated vote counts over scenes. After removing the outliers, we estimated the confidence interval using the bootstrapping method. The statistical test was conducted using a two-tailed z-test on the JOD scores obtained for each viewing condition. The results in Fig.~\ref{fig:userexp2}(A) were dependent on the number of views. Specifically, JOD values in each number of view conditions (3$\times$1, 5$\times$2, 7$\times$3, 9$\times$4) were scaled for three different scenes. Differences in 3D realism between neighboring view number conditions were evaluated by a two-tailed z-test with scaled JOD scores, and significant differences were observed in every paired case except for one (7$\times$3 vs. 9$\times$4 in the village scene, $p$=0.45).

Scene-specific findings regarding the results can be better understood by referring to the depth distribution in Fig.~\ref{fig:userexp2}(B). For the village scene, objects are relatively concentrated near the depth of the wavefront recording plane (WRP) compared to other scenes. The overall depth range can be understood by considering the light field sampling theorem \cite{ng2005light, park2019non}. If the depth range extends from 0 diopters to ${D}_{max}$, the overall depth range supported by the near-eye display, is depicted in a unit of diopter in Fig.~\ref{fig:userexp2}(C). It depends on the scene's spatial bandwidth ($B{x}$), as 
\begin{equation}            
{D}_{max}=\frac{2{{N}_{u}}}{f(2{{N}_{u}}-f\lambda B_{x}^{2})},
\end{equation}
where, $f$ represents the focal length of the eyepiece lens, $\lambda$ is the wavelength of the light source, and $N_{u}$ denotes the angular resolution. With the signal subjected to low-pass filtering, the depth range gets broader. However, if the scene extends to the spatial bandwidth of 30 cpd, more views are required to secure a certain depth range. Interestingly, the analysis of depth representation remains consistent with different optical specifications of the eyepiece lens as provided in Fig. S14.

\section{Discussion}
We conducted user studies using a testbed of modern holographic near-eye displays to determine the ideal 3D formats for providing perceptual reality. Our findings revealed that reconstructing parallax across the eyebox realized with 4D light field as CGH supervision target enhances 3D perceptual realism across various eyebox scenarios in VR near-eye displays.
\subsubsection*{Directional sensitivity of human eye}
The human eye exhibits greater sensitivity to light entering near the center of the pupil than to light near the edge, primarily due to the directional sensitivity of cone photoreceptors. This phenomenon, commonly referred to as the Stiles-Crawford effect~\cite{westheimer2008directional}, can be described by modulating the pupil apodization profile ($A$) as follows:
\begin{equation}
	{{A}}({{x}_{p}},{{y}_{p}})={{A}_{o}({{x}_{p}-{x}_{c}},{{y}_{p}-{y}_{c}})}{{10}^{-p\left( \lambda  \right)(({x}_{p}-{x}_{c})^{2}+({y}_{p}-{y}_{c})^{2})}}.
    \label{eq:2}
\end{equation}
Here, $(x_{p},y_{p})$, $(x_{c},y_{c})$ respectively represents the coordinates of the pupil domain and those of the pupil center in a meter scale. ${A}_{o}$ stands for the original apodization function which is a circular and binary filter, and $p(\lambda)$ is a wavelength-dependent parameter representing the magnitude of the Stiles-Crawford effect. For simulation, we have chosen a constant value of $2.5\cdot {{10}^{4}}$~\cite{westheimer2008directional} for this parameter across all color channels, disregarding wavelength differences.

\begin{figure*}[ht!]
  \centering
  \includegraphics[width=\textwidth]
  {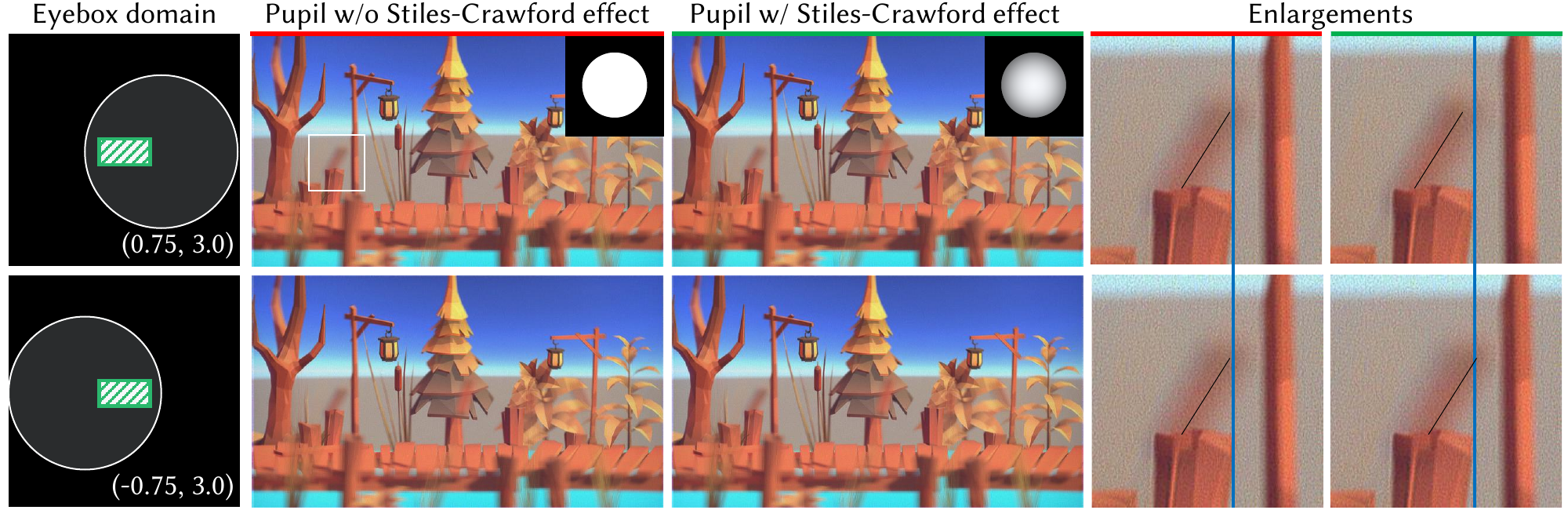}
  \caption{\textbf{Reconstructed images with different pupil apodization profiles} \textbf{of the human eye} (2nd col: diffraction-limited pupil, 3rd col: apodized pupil considering Stiles-Crawford effect).  The images are reconstructed with the 4D supervised CGH of the landscape$\_$day scene when the center of the eye pupil largely deviates from the eyebox and the eye pupil is sufficiently large not to partially sample the eyebox as demonstrated in the (1st col) illustration of the eyebox domain. In the pupil case (1st row), the eye's pupil is decentered to its rightmost extent (a shift of 1.65 mm, ${x}_{p,norm}=0.75$), while in the second scenario (2nd row), the eye is at its leftmost extent (a shift of -1.65 mm, ${x}_{p,norm}=-0.75$). This shift can be converted to the eye rotation of 9.37 degrees, which is almost equivalent to half of the horizontal FoV. In both cases, it is assumed that the eye's pupil is sufficiently large with a diameter of 6.6 mm (${D}_{p,norm}=3.0$) to cover the entire eyebox (2.2 mm $\times$ 1.1 mm). The pupil apodization profile is provided at the top right corner of each column. The identical part of the individual image is cropped and enlarged for better visualization. The blue line is drawn to represent the identical index of the horizontal plane, and the black line is drawn to better visualize the center of the defocused cattail of the scene.}
  \label{fig:SCEpupil}
\end{figure*}

This characteristic of the human eye, unlike the camera,  results in a nontrivial result in the eyebox scenario of an overfilled pupil shown as Fig.~\ref{fig:SCEpupil}. Additional reconstructed results with different CGH algorithms assuming the apodized pupil can be found in Fig. S11. It is worth noting that observing parallax images is also valid even in the extremely large pupil condition as the apodization profile exponentially decreases with the displacement. Although there are individual differences in the optical characteristics of the human visual system, precise optical modeling of downstream optics would help in understanding the perceived image.

\subsubsection*{User study stimuli and apparatus}
The validation could have been performed with low-level psychophysics methodologies~\cite{watson1983quest} to identify the concrete discrimination and detection thresholds for the various viewing parameters such as depth range and number of views. However, these methods require densely sampled evaluation sets with various light field sets and CGHs, which require excessive time and memory. Despite these challenges, our stimuli, comprising complex scenes with depth distribution, effectively elicited natural eye movements provided in Fig. S20. In addition, conducting perceptual studies using binocular holographic near-eye displays can yield more definitive results due to the combined influence of binocular retinal disparity and retinal motion. Finally, conducting VDP simulations with display model parameters (luminance and contrast) matching the actual experimental conditions can potentially improve predictions. However, minor calibration adjustments are unlikely to fully explain the reversal of perceptual realism reported in the experiment, as shown in Fig. S15. Low-level psychophysics experiments conducted on a precisely measured display testbed, which provides \textquotesingle full\textquotesingle\ depth cues, can accelerate exploration into uncharted realms of human 3D perception.

\subsubsection*{Perceptual quality metric of 3D contents}
There have been discrepancies between the realism predicted by the advanced perceptual quality metric~\cite{mantiuk2021fovvideovdp} and the user study results with 3D content. This can be attributed to the fact that the metrics are built upon 2D displays and conventional displays do not typically incorporate 3D visualization. Previous work on gaze-contingent ocular parallax VR rendering~\cite{konrad2020gaze} reported an ocular parallax detection threshold of $\pm $0.36 \emph{D} in eccentricity of 15$^{\circ}$. With the given parameter, ocular parallax detection can be roughly analyzed as discussed in Sec. S5.1.5. This potential integration of perception thresholds in various domains presents an intriguing opportunity for researchers in optics, graphics, and vision science to explore perceptual metrics specifically tailored for evaluating 3D visual stimuli produced by modern displays.

\subsubsection*{Degrees of freedom, the number of constraints, and \'etendue}
Improved perceptual realism achieved through light field optimization stems from the rich spatio-angular information provided by the target, which translates into an increased number of constraints. In our user study, we compared algorithms with the same limited degrees of freedom using our SLM, while there is a tight trade-off between the number of degrees of freedom, the number of constraints, and \'etendue~\cite{monin2022analyzing,monin2022exponentially}. Here, we define the number of degrees of freedom as $\text{(number of optimizable pixels) } \times \text{ (number of frames)}$, and the number of constraints as $\text{ (number of pixels in the target) } \times \text{ (number of views) } \times \text{ (number of planes)}$.

We perform additional simulations on these factors to verify the trend (see Fig. S24); as expected, increased degrees of freedom or a smaller number of constraints lead to low loss values. However, the commonly used mean square error metric should not directly represent the perceptual performance as mentioned in the previous subsection. Moreover, \'etendue expansion is another crucial direction as it would directly increase the possibility of eye displacement, likely magnifying our trend. From this perspective, our results hold significance as the study is conducted in a limited \'etendue setting (underfilled pupil). 

\section{Conclusion}
Our work provides crucial insights on the effectiveness and realism of CGH algorithms that will help guide the community toward passing the visual Turing test of displays using future holographic light field near-eye displays.

\section*{Acknowledgments}
This work was supported by Institute of Information \& communications Technology Planning \& Evaluation (IITP) grant funded by the Korea government(MSIT) (No. 2017-0-00787, Development of vision assistant HMD and contents for legally blind and low visions). Suyeon Choi is supported by a Meta Research Ph.D. Fellowship and a Kwanjeong Scholarship.

\bibliography{main}

\section*{Appendix}
Here, we describe the image formation model and CGH techniques we use in our setup, including 2.5D, 3D, 4D supervisions. For more comprehensive algorithms, we refer to~\cite{park2017recent, choi2022time}. All software is implemented in PyTorch~\cite{paszke2019pytorch}.
\subsubsection*{Image formation model}
In a holographic near-eye display, a coherent light source is incident on an SLM with a source field $\usrc$. The amplitude or phase of the source field is delayed by a spatially-varying input $u_\text{in}$. The manipulated field further propagates, creating a target intensity volume at the desired volume at a distance $z$ off the SLM. We use the angular spectrum method as the free space wave propagation model $\prop$ with the single sideband encoding~\cite{goodman2005introduction, bryngdahl1968single}. The resulting complex-valued field $u_z$ is formulated as follows:

\begin{align}
u_z \left( x, y \right) = \prop \left( u_{\textrm{\tiny SLM}} \left( x, y \right), z \right), \nonumber \\
u_{\textrm{\tiny SLM}} \left( x, y \right) = u_\text{in} \left( x, y \right)  \usrc \left( x, y \right).
\end{align}
\begin{align}
\prop & \left( u,  z \right)  = \iint \Fourier \left( u \right) \cdot \transfer \left(f_x, f_y, z \right) e^{ i 2 \pi (f_x x + f_y y)}d f_x d f_y \, , \nonumber \\
 \transfer &  \left(f_x, f_y, z \right)  =  
 \begin{cases}
    e^{ i \left( \frac{2 \pi}{\lambda} z\sqrt{1 - \left( \lambda f_x \right)^2 - \left( \lambda f_y \right)^2}\right) } & \text{if } f_y \geq 0 \\
    0 & \text{if } f_y < 0,
\end{cases}, 
\end{align}
where $f_x, f_y$ denotes the spatial frequency, $\lambda$ is the wavelength of the light, and $\Fourier$ is the 2D Fourier transform.
\subsubsection*{Optimization for binary amplitude SLMs}
An SLM modulates the complex-valued field with an input amplitude or phase pattern, and the input is usually quantized into a set of levels $\quantset$, (e.g. $\{0, 1\})$. Here, we use a 1-bit SLM in amplitude mode, which only supports output of $q_\text{in} \in \{0, 1\}^{M \times N}$. This SLM can operate at 3600 Hz so the user perceives the time-averaged intensity. In other words, our CGH algorithms aim to obtain the optimal amplitude pattern $q_\text{in} $ for desired target intensity distributions. Since optimizing binary values is a combinatorial optimization problem which is NP-hard, we relax the binary value $q_\text{in}$ as an output of quantization function $\quant$ that takes float value $a_\text{in}$ as input which we optimize for a specific loss function according to the target data:
\begin{align}
u_{\textrm{ in}} \left( x, y \right) = q_{\textrm{ in}}\left( x, y \right) & =  \quant \left( a_\text{in} \left( x, y \right) \right). \nonumber \\
\end{align}

The quantization process is non-differentiable, which does not allow us to use gradient-descent-based methods. To overcome this, we use the Gumbel-Softmax trick~\cite{jang2016categorical} for approximating the gradient of the quantization function. Specifically, we update the amplitude values using the following equation:

\begin{equation}
a_\text{in}^{(k)} \leftarrow {a_\text{in}}^{(k-1)} - \alpha \left( \frac{\partial \loss}{\partial \quant } \cdot \frac{\partial \widehat{\quant} }{\partial a_\text{in}} \right)^T  \loss \left( s \cdot \big| \prop \left( a_\text{in}^{(k-1)} \right) \big|, a_{\text{target}} \right),
\label{eq:phaseupdate_surrogate}
\end{equation}
where $\alpha$ is the step size, $\loss$ is the loss function, $\quant$ is the quantization function, $\widehat{\quant}$ is the relaxed quantization function obtained using the Gumbel-Softmax layer, and $s$ is a scaling factor. We present the implementation results comparing different quantization strategies in Supplementary Material.

%
\subsubsection*{2.5D supervision}
By leveraging the image formation model and utilizing a gradient descent-based update rule, we can optimize the binary amplitude SLM pattern to accommodate various loss functions as described in~\cite{choi2022time}. First, we produce the 2.5D supervision results in our paper employing the multiplane loss function in Eq.~\ref{eq:2.5d}. To implement this approach, we first utilize the closest distance matching technique to create a set of binary masks $M^{(k)}$, corresponding to various distances $z^{(k)}$ from the SLM, using the depth map $D$ obtained from an RGB-D input.
\begin{equation}
	M^{(k)}(x,y)=\begin{cases} 1, \quad \text{if } |z^{(k)}- D(x,y)| <  |z^{(l)}- D(x,y)|, \forall l\neq k, \\ 0, \quad \text{otherwise.}\end{cases}
\end{equation}
Subsequently, we use these binary masks for the multiplane loss, which constrains the wavefront to reconstruct the desired RGB amplitude, denoted as $\target$, at the relevant distances from the SLM, where $\circ$ represents the element-wise product.

\begin{align}
\loss_{\tiny \textrm{2.5D}} = \frac{1}{K} \sum_{k=1}^{K} \loss &\Big( M^{(k)} \circ  s \sqrt{ \frac{1}{T} \sum_{t=1}^T \Big| \prop \left(\quant \left( a_\text{in}^{(t)} \right) , z^{ \left( k \right) } \right)  \Big|^2 }, \nonumber \\
& M^{(k)} \circ \target \Bigg).\label{eq:2.5d}
\end{align}

\subsubsection*{3D supervision}
The 2.5D loss function only restricts the positioning of objects and does not necessarily result in a natural defocus blur for the unconstrained part. To address this, one can assume the amount of defocus occurring at each plane based on the pupil size and penalize all focal slices throughout the volume, ultimately pushing the wavefront toward the desired focal stack using the following loss function:
\begin{equation}
	\loss_{\tiny \textrm{{3D}}} = \loss \left( s \sqrt{ \frac{1}{T} \sum_{t=1}^T \Big| \prop  \left( \quant \left( \amp^{(t)} \right) , z^{ \left\{ j \right\} } \right)  \Big|^2 }, \textrm{fs}_{\tiny \textrm{target}}  \right).
\end{equation}
The target focal stack can be generated using various techniques, such as RGB-D data, off-the-shelf 3D computer graphics software, or light field data. In our paper, we differentiate between 3D supervision techniques based on how the focal stack is produced. Specifically, we generate the focal stack from RGB-D data, which we label as 3D w/ RGB-D supervision. In contrast, when the focal stack target is generated from light field data, which offers more realistic occlusion handling, we refer to it as 3D w/ LF supervision.

\subsubsection*{4D supervision}
It is also possible to obtain an observable light field from the wavefront utilizing the short-time Fourier transform~\cite{zhang2009wigner, padmanaban2019holographic}. The short-time Fourier transform computes the Fourier transform over a small patch surrounding each pixel, providing information about how each pixel appears from different directions. By exploiting this analytical forward relationship between the observable light field and the wavefront, we can directly penalize the wavefront to create the observable light field, incorporating the short-time Fourier transform into the loss function as presented by~\cite{choi2022time}:
\begin{equation}
	\loss_{\tiny \textrm{{4D}}} = \loss \left( s \sqrt{ \frac{1}{T} \sum_{t=1}^T \Big| \textrm{STFT} \left( \prop \left(  \quant \left( \amp^{(t)} \right) , z \right) \right)  \Big|^2 }, \textrm{lf}_{\tiny \textrm{target}}  \right).
\end{equation}

\end{document}


\title{Holographic Parallax Improves 3D Perceptual Realism - Supplementary Material}

\author{Dongyeon Kim}
\authornote{Authors contributed equally to this research.}
\orcid{0000-0003-4141-304X}

\author{Seung-Woo Nam}
\orcid{0000-0001-9031-6049}
\authornotemark[1]
\affiliation{%
  \institution{Seoul National University}
  \country{Republic of Korea}
}

\author{Suyeon Choi}
\orcid{0000-0001-9030-0960}
\authornotemark[1]
\affiliation{%
  \institution{Stanford University}
  \country{USA}
}

\author{Jong-Mo Seo}
\orcid{0000-0002-1889-7405}
\affiliation{%
  \institution{Seoul National University}
  \country{Republic of Korea}
}

\author{Gordon Wetzstein}
\orcid{0000-0002-9243-6885}
\affiliation{%
  \institution{Stanford University}
  \country{USA}
}

\author{Yoonchan Jeong}
\orcid{0000-0001-9554-4438}
\affiliation{%
  \institution{Seoul National University}
  \country{Republic of Korea}
}

\renewcommand{\shortauthors}{Kim, D., Nam, S.-W., and Choi, S. et al.}

\renewcommand{\thesection}{S\arabic{section}} 
\renewcommand{\thefigure}{S\arabic{figure}}
\renewcommand{\theequation}{S\arabic{equation}}
\renewcommand{\thetable}{S\arabic{table}}

\newcommand{\loss}{\mathcal{L}}
\newcommand{\usrc}{u_\text{src}}
\newcommand{\phase}{\phi}
\newcommand{\phasequant}{\psi}
\newcommand{\quant}{q}
\newcommand{\quantset}{\mathcal{Q}}
\newcommand{\prop}{f}
\newcommand{\amp}{a_\text{in}}
\newcommand{\red}[1]{\textcolor{red}{#1}}
\newcommand{\blue}[1]{\textcolor{blue}{#1}}
\newcommand{\propASM}{\mathcal{P}_{\textrm{\tiny ASM}}}
\newcommand{\Fourier}{\mathcal{F}}
\newcommand{\transfer}{\mathcal{H}}
\newcommand{\target}{a_{\text{target}}}

\begin{abstract}
  This is a supplementary material for \textquotesingle Holographic Parallax Improves 3D Perceptual Realism\textquotesingle.
\end{abstract}




\received{20 February 2007}
\received[revised]{12 March 2009}
\received[accepted]{5 June 2009}

\maketitle

\section{System}
In this section, we describe the scheme of the holographic near-eye display system and the setup built for the experiment.
\subsection{Scheme}
\begin{figure}
  \includegraphics[width=\columnwidth]{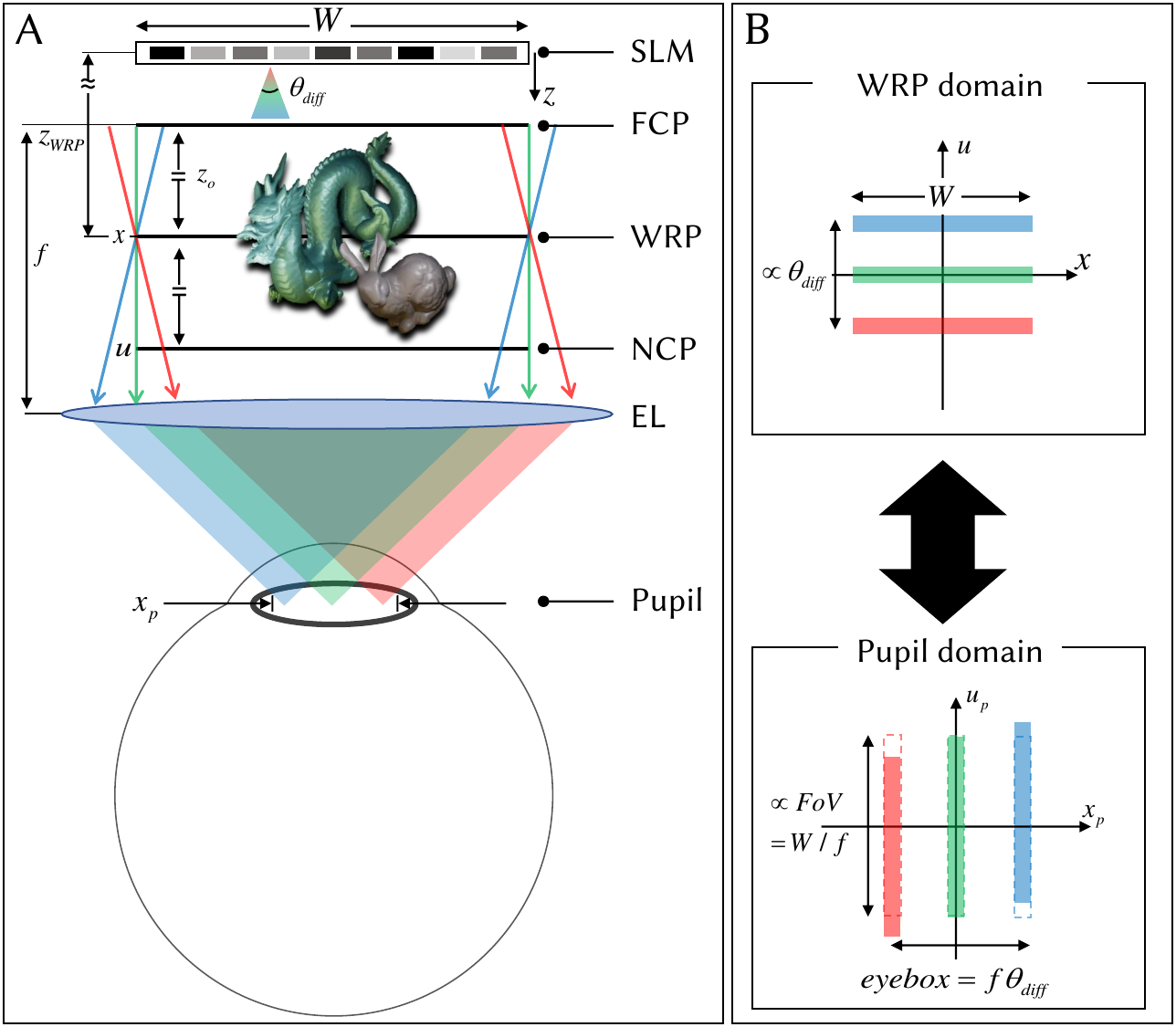}
  \caption{Illustration that describes the (A) schematic of holographic near-eye display. $(x,u)$ represents the spatial dimension of WRP domain and NCP, respectively. Note that $u$ corresponds to the angular dimension of WRP. Likewise, $({x}_{p},{u}_{p})$ is the spatial, angular dimension of the pupil domain, respectively. (B) The light field of the WRP domain and pupil domain shows the relationship between the two domains. The colored light field corresponds to the beam shown in (A). The dashed line shows the relationship of two domains when the WRP is placed at the focal length of EL. We deduced the entire dimension to two for simplicity.}
  \label{fig:subSystemscheme}
\end{figure}

The holographic near-eye display is briefly explained in Figure \ref{fig:subSystemscheme}(A). By using a spatial light modulator (SLM) with a pitch of $p$ and illuminating it with a coherent source of light with a wavelength of $\lambda$, a wave field can be generated within the diffraction angle of ${{\theta }_{diff}}=2{{\sin }^{-1}}({\lambda }/{2{{p}}})$. The SLM field then propagates and reconstructs a wave field at a certain distance, with a width similar to the SLM's width ($W$) and an angle within the diffraction angle. The aim is to reconstruct the intensity profile of $I(x,y,z),\,z\in [{z}_{FCP}\,{z}_{NCP}]$, where $x,y$ represents the horizontal, vertical position, respectively, and $z$ denotes the axial distance from the SLM, within the axial distance between the far clipping plane (FCP, ${z}={z}_{FCP}$) and the near clipping plane (NCP, ${z}={z}_{NCP}$). Additionally, the wavefront recording plane (WRP, ${z}={z}_{WRP}$), which is equivalent to the reference plane of orthographic light fields, is located at the middle of FCP (${z}_{FCP}={z}_{WRP}-{z}_{o}$) and NCP (${z}_{NCP}={z}_{WRP}+{z}_{o}$). We locate the FCP of the rendered volume at the focal length ($f$) of the eyepiece lens (EL). Then, the FCP will be virtually floated at the optical infinity and NCP will be located at the dioptric distance of ${{D}_{NCP}}=1/(f-2{{z}_{o}})-1/f$. 

The beam with the limited diverging angle will form an eyebox, which is an exit pupil of the system. In Fig. \ref{fig:subSystemscheme}(B), the relationship between the WRP domain and the pupil domain is demonstrated. The three beams that propagate in different directions with a small angular bandwidth in the WRP plane are remapped in the pupil domain. It shows the inversion of spatial and angular dimensions as the beams pass the lens. If the WRP domain is filled with the light field having the spatio-angular size of $(W,{{\theta }_{diff}})$, the pupil domain will contain spatio-angular size of $(f{{\theta }_{diff}},{W}/{f})$. The spatial dimension corresponds to the eyebox of the near-eye display, and the field of view (FoV) is proportional to the angular dimension. Note that the product of eyebox and field of view is proportional to the display resolution and wavelength of the beam. 

If the WRP is not placed at the focal length of EL, the projected light field is tilted resulting in the FoV difference depending on the pupil location inside the eyebox. However, placing the WRP plane in the middle is advantageous as the resolution degradation of LF-based hologram is proportional to the distance between the WRP and the depth of an object.

\subsection{Setup}
\begin{figure}
  \includegraphics[width=\columnwidth]{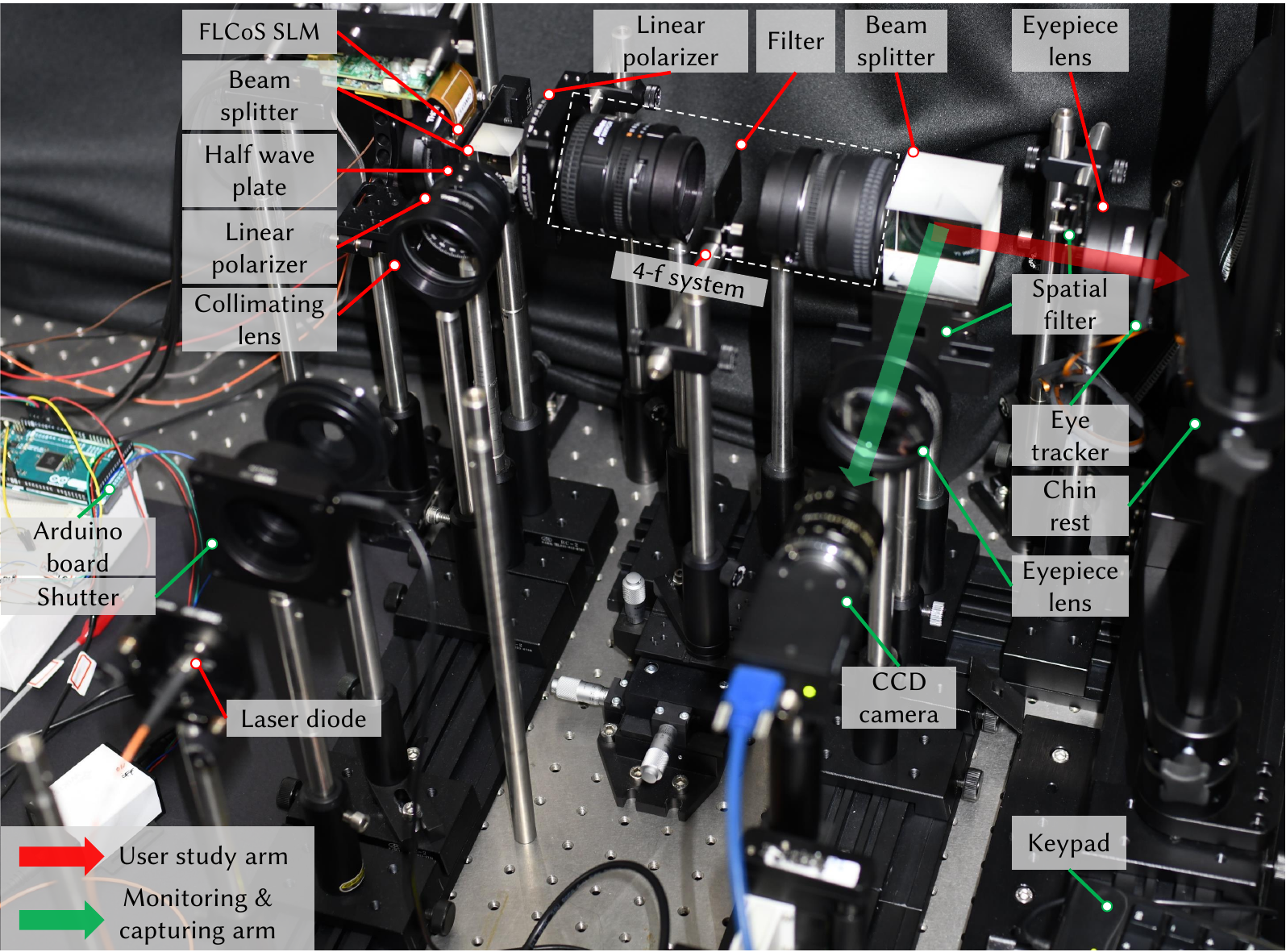}
  \caption{The photograph depicts the testbed of a holographic near-eye display prototype used for user validation. The various components of the testbed are connected to a box. The components highlighted by red lines represent the essential equipment necessary for the holographic near-eye display. Conversely, the components connected by green lines are specifically implemented for the user experiment. The beam path is divided into two paths: the user study arm (indicated by the red arrow) and an additional arm for monitoring and image capturing (indicated by the green arrow).}
  \label{fig:subSystem}
\end{figure}

Figure \ref{fig:subSystem} demonstrates an overview of the holographic near-eye display prototype. A fiber-coupled laser diode of Wikioptics emanates a full-color beam with a central wavelength of 638 nm, 520 nm, and 450 nm. The beam is collimated with a lens (AC-508-200-A, Thorlabs) and the beam is linearly polarized with a series of linear polarizer (LPVISE200-A, Thorlabs) and an achromatic half-wave plate (AHWP10M-600, Thorlabs). We additionally placed a half wave plate to maintain the color balance as there are difference in polarization states by color. The beam is redirected with a 1-inch beam splitter and modulated with a reflective-type spatial light modulator. 

We use a binary ferroelectric liquid crystal on silicon spatial light modulator (FLCoS SLM) to modulate the incident coherent beam. This SLM (QXGA-R10, a product of Forth Dimension Display) operates 1920 $\times$ 1200 pixels with a pitch of 8.2 $\mu m$ at a speed of 3600 Hz to serve 24 full-color binary frames within 1/50 seconds. Placing an analyzer in the beam path allows the operation of SLM in amplitude mode. The field at the SLM plane is relayed with a 4-f system built with two identical camera lenses (AF Nikkor 50mm f/1.4D, Nikkon) facing opposite each other. A filter is placed in the Fourier domain to filter out the high-order signals arising from diffraction and the conjugate noise from complex representation with an amplitude SLM. The filter is fabricated with a rectangle aperture in an aspect ratio of 2:1 with a size determined by the SLM's diffraction angle in blue and the focal length of the 2-f lens. The relayed field is virtually floated by a 2-inch eyepiece lens (AL5040M, Thorlabs) having a focal length of 40 mm to guarantee a wide field of view. Thus, the eyebox size of the near-eye display is 2.2 mm $\times$ 1.1 mm. 

We made an additional beam path by placing a beam splitter after the 4-f system to capture the experimental results and monitor the user experiment. For this arm, a lens (AC508-100-A-ML, Thorlabs) having a focal length of 100 mm is used as an eyepiece lens, and the scenes are captured with a c-mount lens with a 25 mm focal length and charge-coupled device (CCD) camera (BFS-U3-51S5C-C, FLIR) having a resolution of 2448$\times$2048 and a pitch of 3.45$\mu m$. The CCD camera is placed on the two single-axis motorized stages (M-112.1DG1, a product of PI) to capture the image in distinct viewpoints with high accuracy. Note that the eyebox size of this arm is 2.5 times larger than the actual user experiment settings. Thus, we translated the CCD with the converted geometry. Additional spatial filters are placed at the relayed WRPs to eliminate the noise present in the peripheral region. Additional components shown in Fig.~\ref{fig:subSystem} but not described in this section will be explained in Section~\ref{section:userstudy} with the description of user study implementation.

\section{software implementation}
\begin{figure}
  \includegraphics[width=\columnwidth]{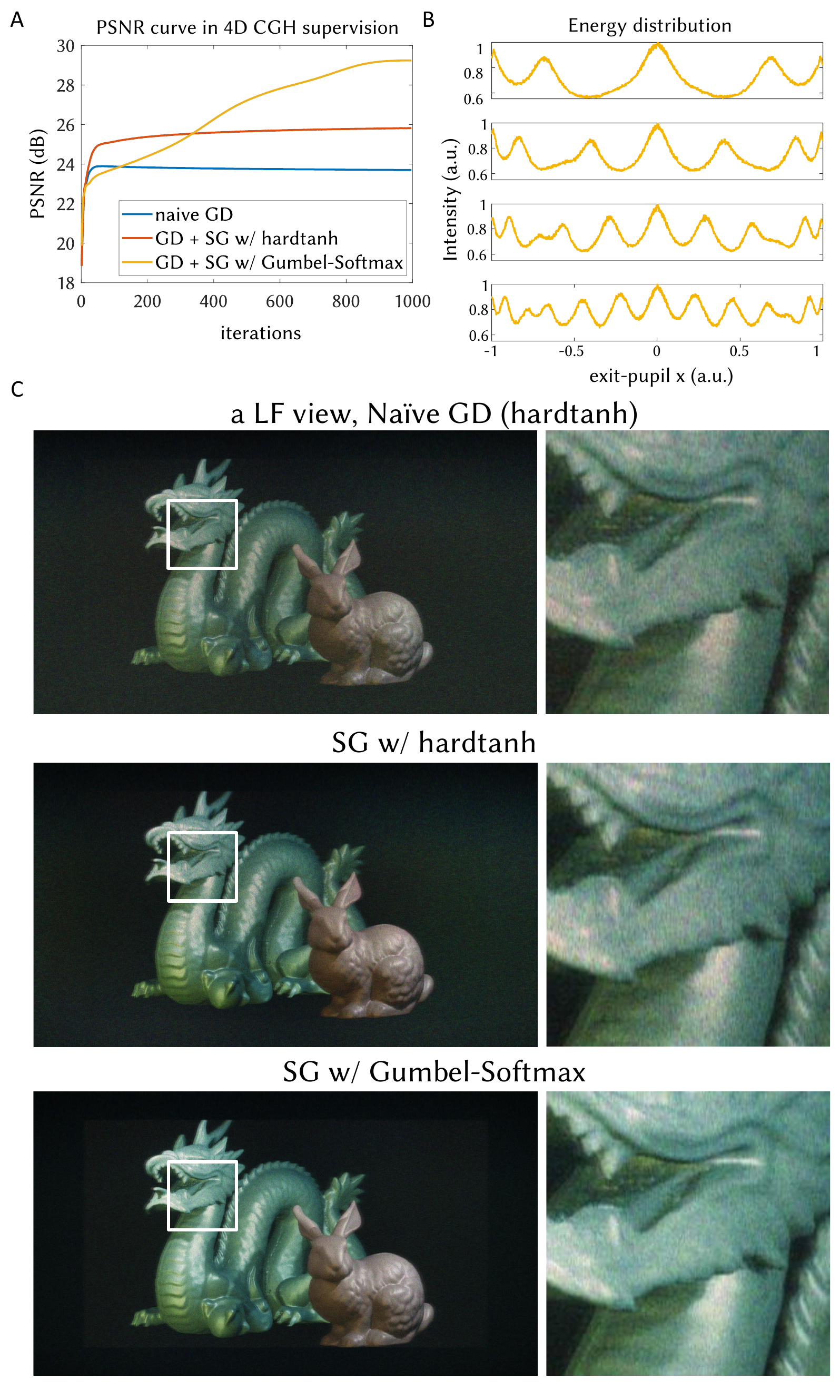}
  \caption{A direct comparison of the surrogate gradient approaches for 4D supervision. We present (A) a convergence graph for binary amplitude SLMs using unit gradient and Gumbel-Softmax gradient methods. On the right, we present the (B) one-dimensional energy distributions across the exit pupil by combining and summing up the intensities at the Fourier plane, with a different number of light field views supervised ($3 \times 3, 5 \times 3, 7 \times 3,\text{ and }9 \times 3$, respectively). We also show (C) a sampled view from reconstructed light field for a qualitative comparison.}
  \label{fig:optimization_comparison}
\end{figure}

While Choi et al.~\shortcite{choi2022time} previously demonstrated the effectiveness of this surrogate gradient method using the Gumbel-Softmax for phase SLMs, our work represents the first application of this technique to binary amplitude SLMs. Figure~\ref{fig:optimization_comparison} demonstrates that this Gumbel-Softmax-based optimization outperforms the previous state-of-the-art binary CGH~\cite{lee2022high}. This approach offers a promising new direction for optimizing binary amplitude SLMs. For the content generation speed and quantitative comparison, please refer to Sec.~\ref{sec:tradeoff}

\section{Light field dataset}
\begin{figure}
  \includegraphics[width=\columnwidth]{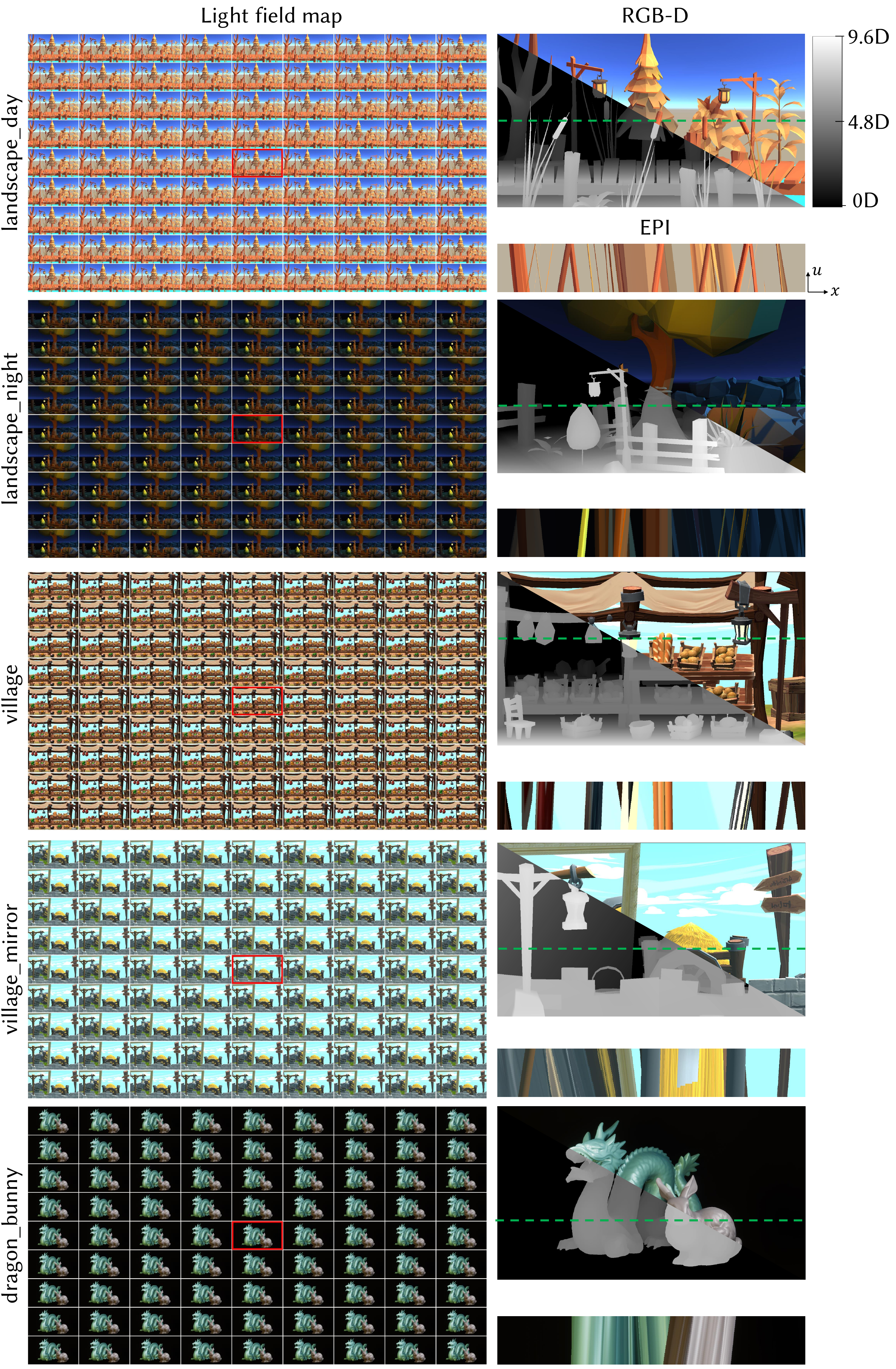}
  \caption{Light field map, RGB-D image, and epipolar plane image (EPI) of the scenes (landscape$\_$day, landscape$\_$night, village, village$\_$mirror, dragon$\_$bunny) used in the paper are demonstrated. 9 $\times$ 9 orthographic images are provided as the light field map. The intensity and depth profile of the corresponding scenes' center view image (red box) is shown. The depth profiles of the orthographic scenes in a metric unit are converted to a unit of diopter considering the optical configuration of the near-eye display system. The EPI of the horizontal section depicted with a green dashed line is provided. The EPIs are drawn based on orthographic light fields and the upright slope implies the object is placed at the WRP. (Low Poly Series: Landscape, Fantastic-Village Pack: purchased unity asset, and Dragon, Bunny: credit to Stanford Computer Graphics Laboratory)}
  \label{fig:subLFDataset}
\end{figure}

We utilized a total of five different scenes in our paper, and these scenes were rendered using Unity. Fig.~\ref{fig:subLFDataset} presents the rendered light field maps and RGB-D images of each scene, along with the corresponding epipolar plane images (EPIs). For the light field maps, we rendered 25$\times$25 orthographic views per color channel. However, in the figure, we have provided a subset of only 9$\times$9 sampled views due to space limitations.

The EPIs provide insights into the angular distribution of the scenes. As depicted in the EPIs shown in Fig.~\ref{fig:subLFDataset}, the individual slices exhibit distinct spatial information along the angular dimension. Notably, certain objects are only visible within specific angular ranges, while they disappear in other angular regions. This disparity in information across angles signifies the presence of "valid parallax," and it emphasizes that parallax containing meaningful information can be obtained when working with data formats that have four or more dimensions.

\section{User study implementation} \label{section:userstudy}
We implemented the overall setup for the user study. We additionally equipped an eye tracker, and a keypad for data acquisition. In addition, a chin-and-head rest, a shutter, an Arduino board, and an optical power meter are equipped to improve the study's accuracy and guarantee the subjects' eye safety. 

We utilized the Add-on eye tracker, developed by Pupil Labs, to measure the displacement of the subject's eye while they viewed the stimulus. Since we recorded the pupil displacement of a single eye, we couldn't utilize the built-in calibration functions designed for tracking both eyes simultaneously. Instead, we calibrated the measured data, which represented the center of the detected pupil, using a scale factor obtained through a pre-calibration procedure. This pre-calibration involved an eye figure with a black pupil that was moved laterally at the eye relief of the near-eye display. The collected data was obtained using the Pupil Labs Network API and saved in comma-separated value (csv) file format. Each trial involved a 2-second recording session, capturing the data at a speed of approximately 120 frames per second. We only included data points with a confidence value higher than 0.85 for further analysis. The response for each pair of options from the participants was received using a keypad and saved in csv file format, with values indicating the options that were compared.

To ensure the absolute position of the subject's head, a chin-and-head rest was employed to restrict head movement. The chin-and-head rest was positioned on a stage that allowed lateral movement. Subjects were instructed to adjust the initial position of their pupil by translating the stage. To address differences in intensity levels between holographic images created with various CGH supervision targets, as well as to ensure safety, we incorporated an Arduino board (Arduino Mega 2560) for two purposes. Firstly, it helped balance the intensity levels by adjusting the pulse width of the light source. The reconstructed images displayed different intensity levels due to variations in the scale factor for CGH supervision. By modulating the width of the rectangular pulse generated by the Arduino board, we could standardize the intensity levels across the images. Secondly, the Arduino board controlled a shutter placed in front of the light source, preventing uncontrolled light emission during the initialization process. 

\subsubsection*{Luminance measurement}
We ensured eye safety by measuring the luminance of the scenes. Directly measuring the luminance of each scene using the luminance meter proved challenging due to the small exit pupil of the holographic near-eye display system, which led to the underfilling of the entrance pupil of the measurement device. Instead, we opted to measure the power of the scene at the eyebox using a power meter (Newport, 818-SL/DB) and converted this data to luminance based on the geometry of the near-eye display system. The luminance (${L}_{v}$) is measured in candela per meter square (nit) and can be calculated using the following equation: 
\begin{equation}
    {{L}_{v}}=\frac{683}{S\cdot \Omega }\int{\Phi (\lambda )V(\lambda )d\lambda }.
\end{equation}
Here, $S$ represents the two-dimensional area where the image is displayed, $\Omega$ denotes the solid angle of the display source, $\Phi (\lambda )$ signifies the measured power at different wavelengths, and $V(\lambda )$ stands for the luminosity function of photopic vision.

The luminosity function varies depending on the light condition as cone cells are nonfunctional in low-light conditions \cite{wandell1995foundations}, but we used the photopic luminosity function as the standard of the scotopic (dark-adapted) vision level ranges below 0.001 nits. The holographic display utilizes a narrow-band source, thus wavelength-dependent power is provided with the power of color primaries. The measured power ranges in hundreds of picowatts, thus the average luminance is approximately converted to 2 nits. This value is significantly lower than the permissible level of laser exposure stated in the cited reference \cite{international1996guidelines}.


\section{Additional simulation results}
In this section, we provide the additional simulation results mostly consisting of the reconstructed results depending on the CGH supervision targets, and the number of views used in 4D CGH supervision with the analysis based on LF sampling theorem.

\subsection{Various 3D CGH supervision target formats}
\subsubsection{Comparison by quality metric}
\begin{figure*}[ht!]
  \centering
  \includegraphics[width=\textwidth]{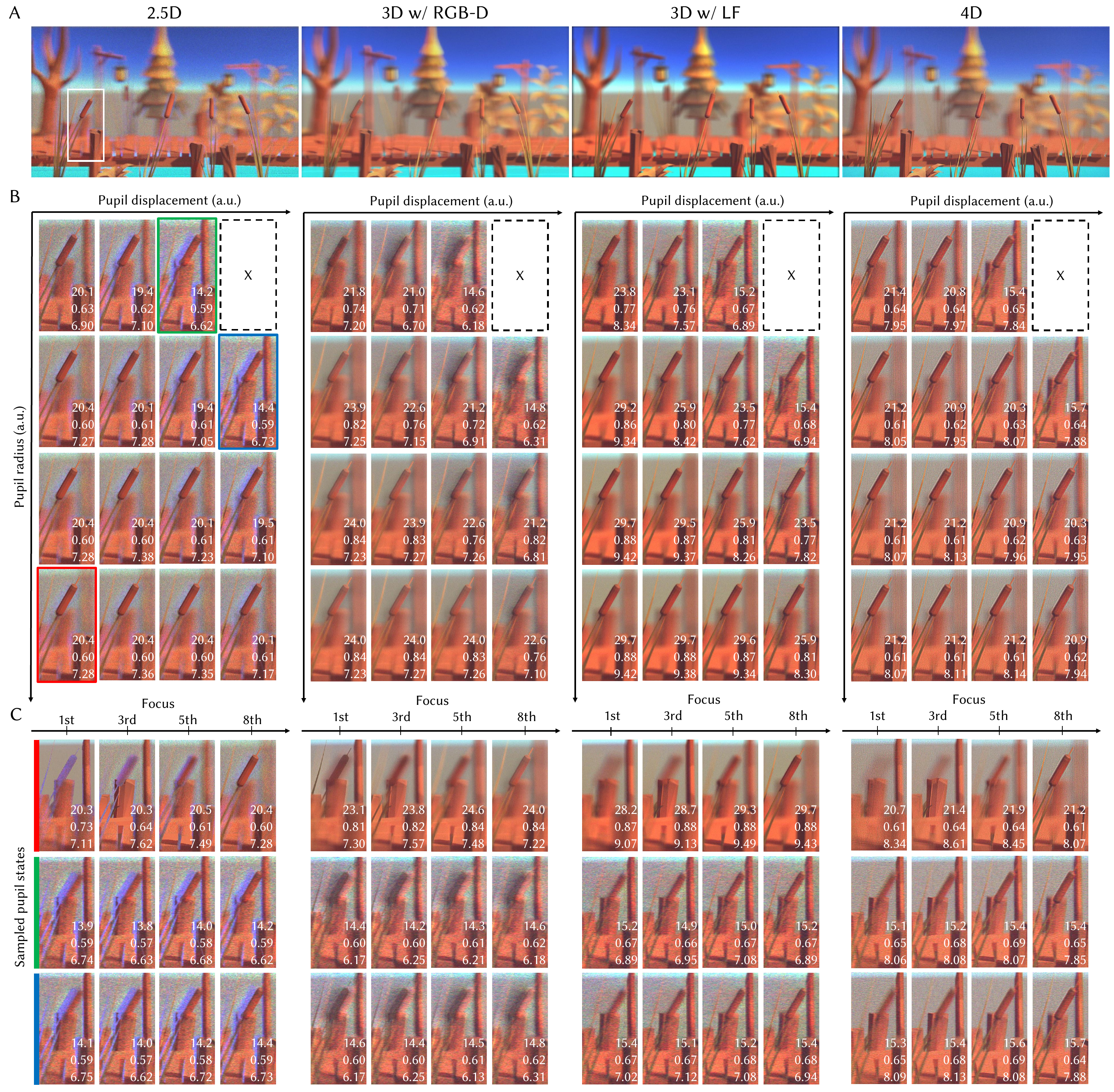}
  \caption{Reconstructed holographic images with various 3D CGH supervision targets: (A) Near-depth reconstructed image of landscape$\_$day (2.5D (1st column), 3D w/ RGB-D (2nd column), 3D w/ LF (3rd column), and 4D (4th column) supervision targets) scene. (B) The white box sections of the images reconstructed with various pupil states (pupil displacement, and pupil radius) are shown. The blank section with a dashed boundary in the figure indicates the that visualization is not capable as the state is fully vignetted. Peak signal-to-noise ratio (PSNR), structural similarity index (SSIM), and FovVideoVDP quality metric in a unit of JOD are consecutively provided on the bottom of every inset. For FovVideoVDP metric, the ground truth image is referenced and demonstrates 10 JOD and the difference of 1 JOD corresponds to a 50 percent preference over the ground truth image. (C) The identical parts of images reconstructed with three different pupil states are provided depending on the focal depths. Four out of the nine depths equally sampled in the diopter are provided for simplicity. The images reconstructed with various pupil shifts, pupil sizes, and focal depths can help understand the impact of those aspects in the realized holographic scenes.}
  \label{fig:subsimulation1}
\end{figure*}

We demonstrate the reconstructed holographic images with various 3D CGH supervision targets as Fig.~\ref{fig:subsimulation1}. In Fig.~\ref{fig:subsimulation1}(A), near-depth (8th out of 9) images of landscape$\_$day scene acquired with different 3D CGH supervision assets are provided. In Fig.~\ref{fig:subsimulation1}, the images are reconstructed with different pupil states and they are evaluated with three image metrics: Peak signal-to-noise ratio (PSNR), structural similarity index (SSIM), and FovVideoVDP quality metric. To calculate the image metrics, we utilize the amplitude of the reconstructed image and the corresponding ground truth images. The ground truth images are cropped to 80$\%$ of the entire FoV to eliminate the effects on the image boundaries from additional propagation. The FovVideoVDP metric is obtained under the identical conditions described in the paper, ensuring consistency in the evaluation process.

In reconstructed holographic images with a 2.5D supervision target, the artifacts get noticeable as the pupil displacement gets larger with defects in color. This is because out-of-focus regions are not penalized for 2.5D supervision. And this effect gets noticeable when a large depth difference between the object and its surroundings is present. Likewise, the reconstructed images with 3D w/ RGB-D target exhibit a similar problem without color artifacts. But, the overall contrast gets dimmer since the occlusion handling in the boundaries affects the contrast of the contents. In the reconstructed images of 3D w/ LF, the parallax is noticeable as the pupil gets decentered, and the overall image quality is far better in terms of metric. This is because the ground truth images are CGH supervision targets in this case. However, the FovVideoVDP metric deteriorates as the pupil size decreases and becomes decentered, deviating from the image formation model in CGH supervision. For cases supervised by 4D content, the image metric worsens as the reconstruction model is based on a plane-to-plane model, while the optimization model relies on a plane-to-perspective model, resulting in differences. Nevertheless, the FovVideoVDP metric remains consistent across various pupil states, and the discrepancy between the simulation results and the user experiment suggests the need for a perceptual quality metric for 3D content as a future research endeavor.

Fig.~\ref{fig:subsimulation1}(C) provides the reconstructed images obtained using sampled pupil states corresponding to different focal depths. When the eyebox is fully sampled by a large pupil positioned at the center, the 3D w/ LF outperforms other 3D CGH supervision approaches, and this trend remains consistent regardless of the reconstructed depth. However, when the pupil becomes smaller and decentered, the cases supervised by 4D assets demonstrate better perceptual metrics.
\subsubsection{Content dependency}
\begin{figure*}[ht!]
  \centering
  \includegraphics[width=\textwidth]{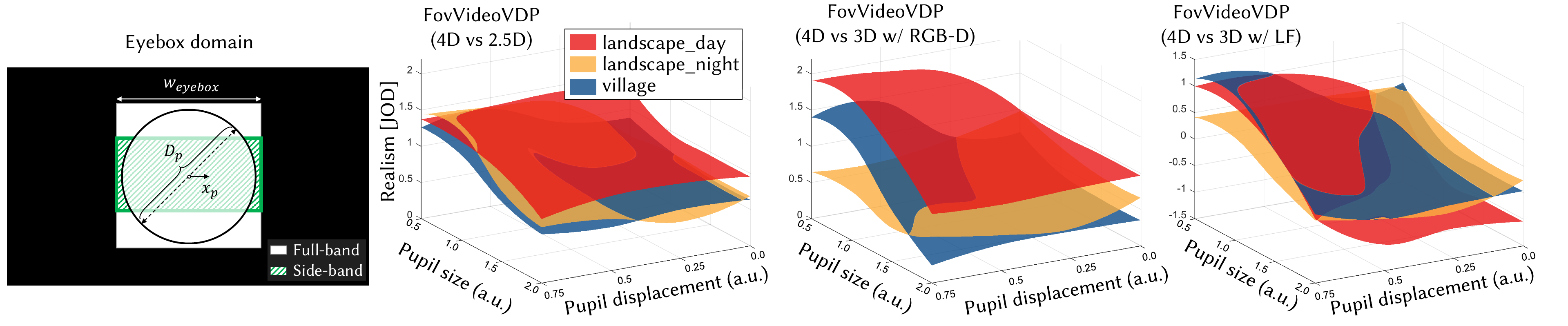}
  \caption{Illustration of the eyebox domain (1st col) of holographic near-eye display with the width of $w_{eyebox}$ and the circular pupil with a diameter of $D_{p}$ and displacement of $({x}_{p},0)$. The side-band eyebox (green) is vertically halved in size relative to the full-band eyebox (white) for complex modulation with a single amplitude SLM. Comparison of 4D CGH supervision with 2.5D (2nd col), 3D w/ RGB-D (3rd col), 3D w/ LF (4th col) CGH supervision is conducted with the image reconstructed with various normalized pupil displacement (${{x}_{p,norm}}= {{x}_{p}}/{{{w}_{eyebox}}}$) and normalized pupil size (${{D}_{p,norm}}= {{D}_{p}}/{{{w}_{eyebox}}}$). Note that the scale of the grid differs for the last figure.}
  \label{fig:supeyeboxJOD}
\end{figure*}

The FovVideoVDP metric extracts data from the image at the center depth across 15 distinct pupil states within the eyebox domain. Comparing the JOD value of 4D CGH supervision with 2.5D, 3D w/ RGB-D, and 3D w/ LF involves subtracting values, depicted in Fig.~\ref{fig:supeyeboxJOD}, alongside an illustration of the eyebox domain. These plots showcase three scenes (landscape$\_$day in red, landscape$\_$night in orange, and village in blue).

While some variations occur based on the content, the comparison consistently highlights the superior performance of 4D supervision over other approaches in overfilled pupil conditions. However, the increment diminishes as the pupil becomes underfilled, even showing JOD differences below zero in the case of overfilled pupils when comparing 4D vs. 3D w/ LF.

\subsubsection{Eyebox evaluation}
The accurate provision of a 4D light field across the eyebox requires attention to two critical aspects. First, it is necessary to ensure that the entire energy is distributed across the eyebox. Unlike other types of near-eye displays, holographic near-eye displays can adjust the effective size of the eyebox by controlling the phase randomness of the reconstructed field. It is important to note that manipulating the randomness of the phase profile of the reconstructed field impacts the effective size of the eyebox, the depth of field of the display, and the gain of dynamic accommodation response~\cite{kim2022accommodative}. Therefore, the size of the eyebox must be maximized, as determined by the product of the maximum angle and the focal length of the eyepiece lens. The eyeboxes created by the different CGH supervision approaches are measured and the captured results are provided in Fig.~\ref{fig:subeyeboxexp}. Additionally, the perspective images with designated carrier frequency should provide scenes with desired energy distribution.



\begin{figure}[ht!]
  \centering
  \includegraphics[width=\columnwidth]
  {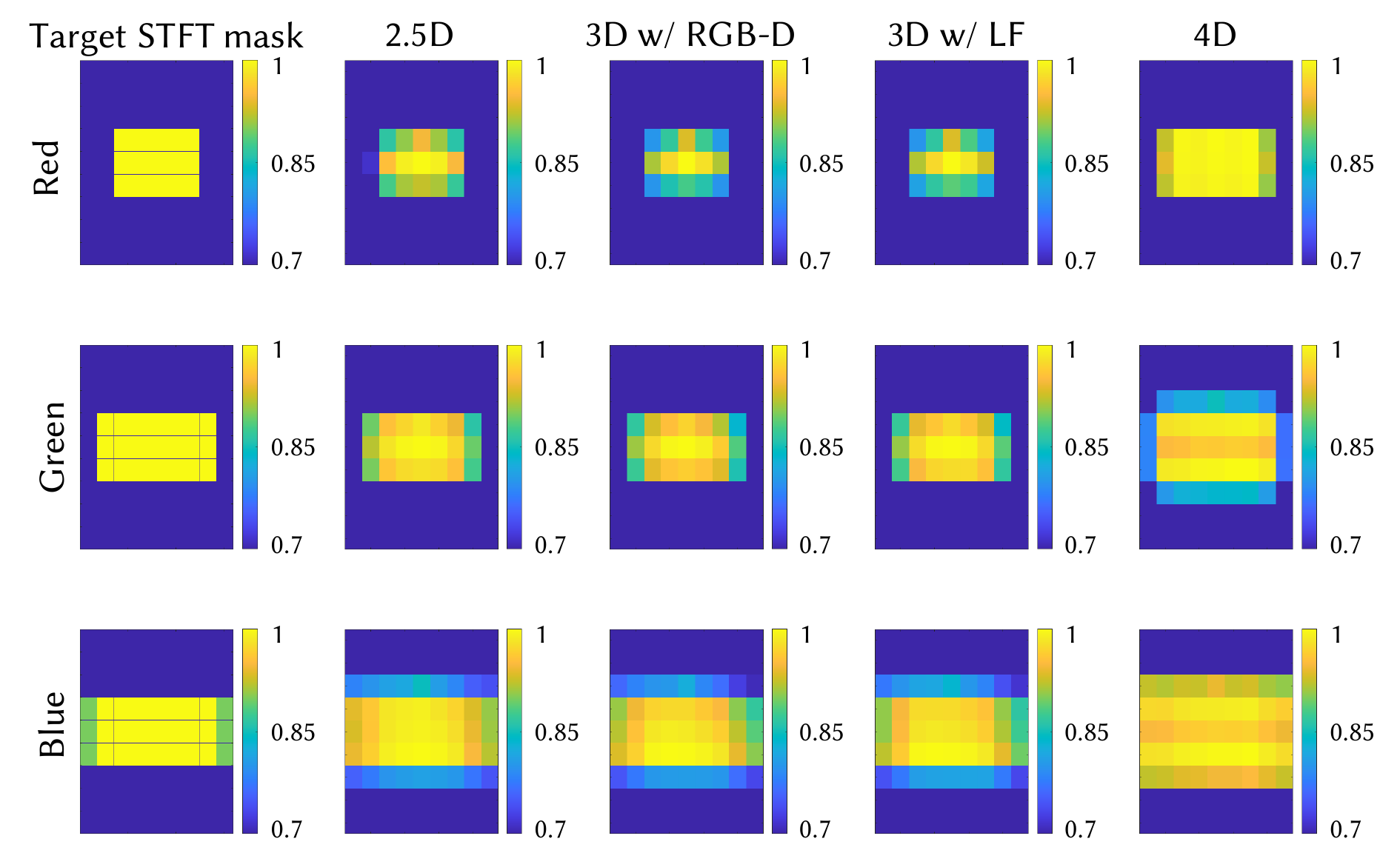}
  \caption{Energy distribution of the 9$\times$9-STFT-reconstructed images of village scene. The 9$\times$9 masks are tiled based on the index of the carrier frequency and provided depending on the various 3D CGH supervision approaches (2nd col: 2.5D, 3rd col: 3D w/ RGB-D, 4th col: 3D w/ LF, 5th col: 4D) with the target STFT weight mask (1st col). These energy distributions are provided based on each color channel (1st row: red, 2nd row: green, 3rd row: blue). The overall intensity is normalized and the scale bar is clipped with a minimum value of 0.7.}
  \label{fig:supstftenergy}
\end{figure}

For an accurate analysis of the energy carried by each individual localized beam, each of which carries the signal of the discrete light field, it is essential to examine the energy of each orthogonal view. The energy distribution of the reconstructed view images is illustrated in Fig.~\ref{fig:supstftenergy}. This figure demonstrates the energy distribution of 9$\times$9 STFT-reconstructed images of the village scene using different CGH supervision approaches. Each 9$\times$9 tile represents the averaged intensity of the STFT-reconstructed image with a specific direction.

From the figure, it becomes evident that both 2.5D supervision and 3D supervision struggle to accurately reconstruct the light field with a discrete carrier frequency, as the energy of the views near the boundary decreases. In contrast, 4D supervision uniformly generates the light field with a discrete carrier frequency. The target STFT mask varies in color to match the physical eyebox of the system. It's worth noting that the weight of the target STFT mask is adjusted at the boundary to mitigate the impacts of undesirable diffraction at the edge. 

\subsubsection{Eye movement}
\begin{figure*}[ht!]
  \centering
  \includegraphics[width=\textwidth]
  {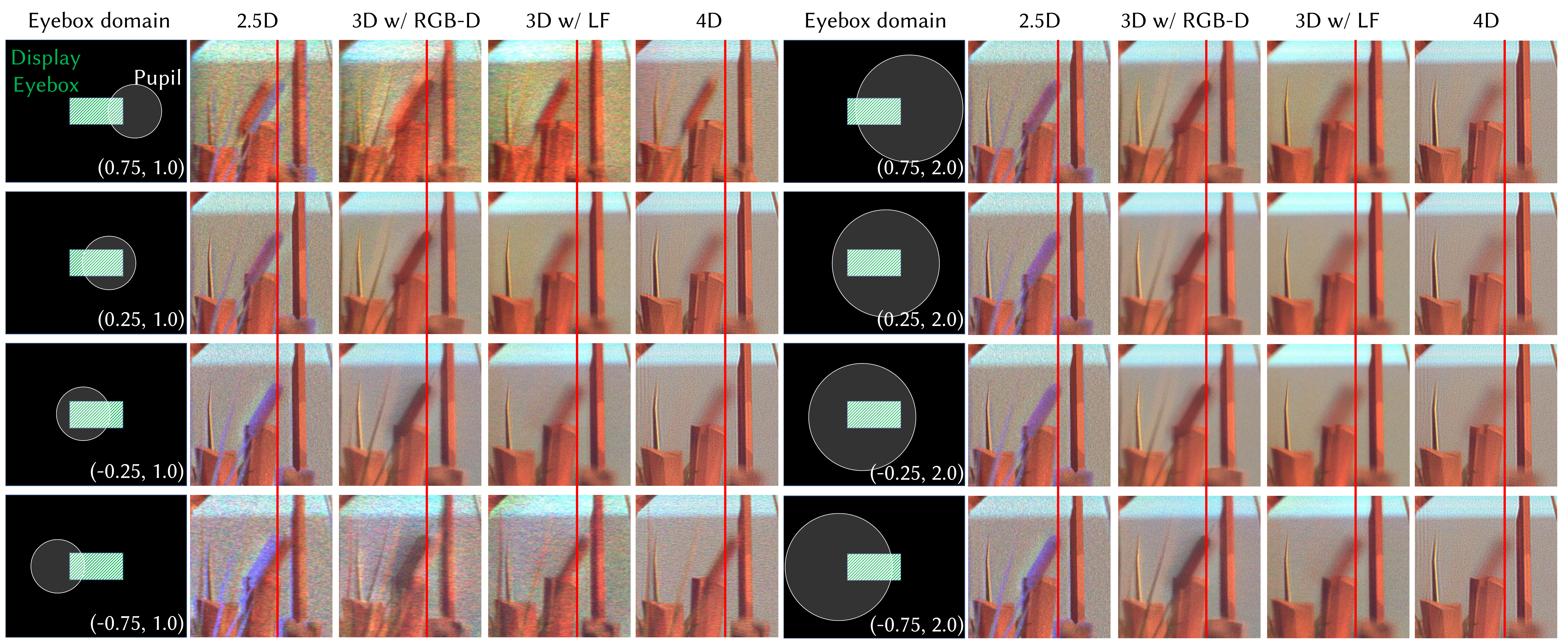}
  \caption{Reconstructed scenes depending on the state of the pupil inside the eyebox domain when the human eye pupil is simulated as diffraction-limited. The green section represents the eyebox of the near-eye display system and the pupil is represented with a white circle. For each state of the eye position and diameter $({x}_{p,norm},{D}_{p,norm})$, parts of landscape$\_$day scene with focal depth of 2nd depth are provided. The scenes assume the pupil functions as a diffraction-limited aperture. Red lines are placed on purpose to emphasize the disparity of the objects.}\label{fig:subParallax}
\end{figure*}

To simulate how scenes are perceived as the eye moves within the eyebox, we present simulated results using different 3D CGH supervision approaches based on various eye positions and diameters, as depicted in Fig.~\ref{fig:subParallax}. When the eye's pupil is sufficiently small, it samples only a portion of the eyebox signal, making 4D supervision highly tolerant, regardless of the eye's position. In detail, the object near to the eye is placed relatively leftward compared to the object placed far when the pupil is placed in the left area of the eyebox. Furthermore, parallax information is well-preserved in cases of 4D supervision, as discussed in Fig.~\ref{fig:subsimulation1}.

If the eye's pupil enlarges due to practical conditions such as changes in light conditions, the view-dependent parallax effect naturally diminishes. Moreover, when the eye's pupil becomes large enough to sample the entire eyebox, no differences are observed as the eye moves. This is because we assume that the eye pupil is a diffraction-limited pupil without any fluctuation in transmittance, in contrast to real-world conditions. This assumption downsizes the effect of displacement-dependent parallax, which has been discussed in the main paper.

\subsubsection{Ocular parallax detection analysis}
\begin{figure}[ht!]
  \centering
  \includegraphics[width=\columnwidth]
  {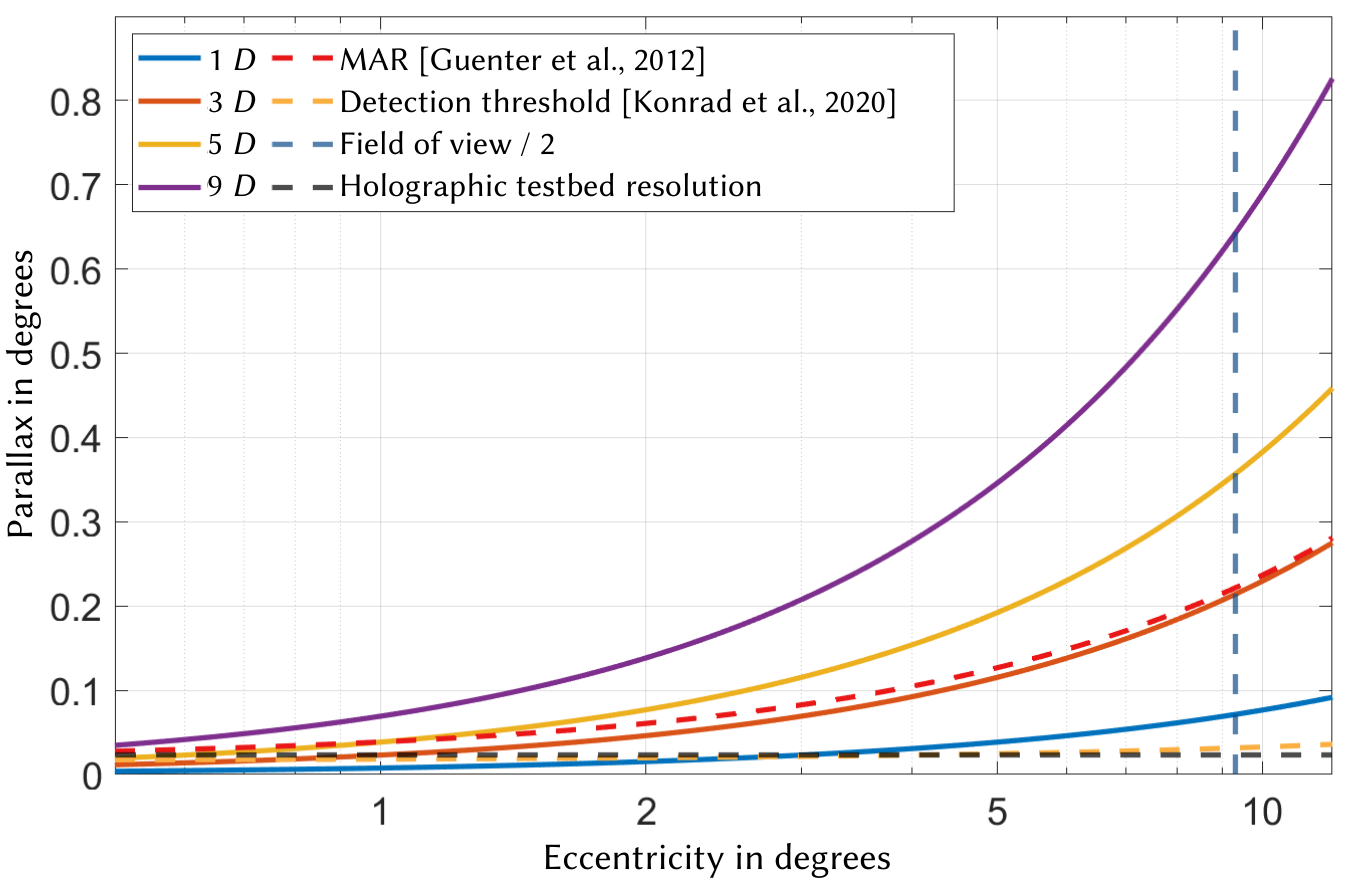}
  \caption{The amount of ocular parallax, estimated in a unit of degrees of visual angles, depending on the retinal eccentricity and the dioptric disparity of two objects placed at different depths (solid lines). Minimum angle of resolution (MAR) in the work of Guenter et al.~\shortcite{guenter2012foveated} (red, dashed line) and the ocular parallax detection threshold investigated in the work of Konrad et al.~\shortcite{konrad2020gaze} (orange, dashed line) are present along with the half of the field of view (blue, dashed line), and minimum resolution supported by holographic near-eye display testbed (black, dashed line). Note that the ocular parallax detection threshold is far smaller than the MAR. }\label{fig:supocularparallax}
\end{figure}
Figure~\ref{fig:supocularparallax} provides the amount of the ocular parallax realized by two objects placed at different depths. The depth of the reference object is assumed as 0.5 \emph{D} for simplicity. We refer to the work of Konrad et al.~\shortcite{konrad2020gaze} for the details on the ocular parallax simulation. 

Along with the parallax of two objects, we demonstrate the minimum angle of resolution (MAR, $\omega$)~\cite{guenter2012foveated} depending on the eccentricity ($e$) as $\omega =0.022e+{{\omega }_{0}}$, where, ${\omega}_{0}$ denotes the minimum angular resolution as 1/60 (20/20 vision) in a unit of degrees. In addition, the detection threshold of ocular parallax was measured as 0.36 \emph{D} in the eccentricity of 15 degrees in the user evaluation of the work~\cite{konrad2020gaze}. Here we assume the threshold also follows the linear model of the eccentricity-dependent resolution falloff and the slope is fitted as 0.0016. Other than these thresholds, we additionally provide the halved horizontal FoV and the minimum angular resolution supported by the testbed.

\begin{figure*}[ht!]
  \centering
  \includegraphics[width=\textwidth]{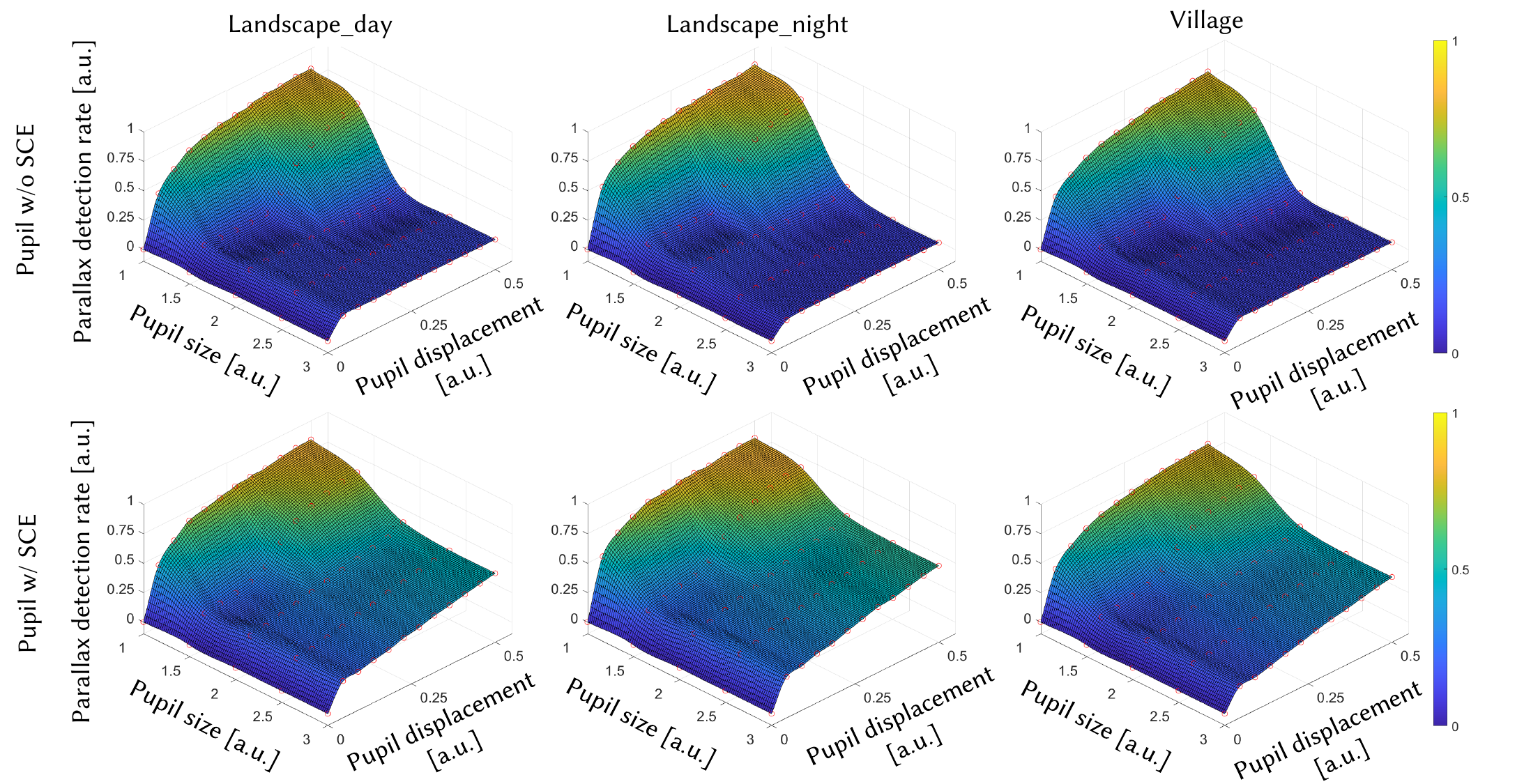}
  \caption{The ocular parallax detection rate is estimated using three different scenes (landscape$\_$day, landscape$\_$night, and village) processed with 60 pupil states (depicted by red circles) and interpolated across the pupil state domain, considering pupil size and pupil displacement. The reconstructions are performed under two human eye pupil apodization profiles: (\textit{top}) pupil w/o Stiles-Crawford effect (SCE) and (\textit{bottom}) pupil w/ SCE.}
  \label{fig:supSCEPupil_parallax}
\end{figure*}

As observed in Fig.~\ref{fig:supocularparallax}, the detection threshold is far below the MAR measured with the static stimuli as the detection threshold involves motion-based perception, which may result in detection in the periphery. Although the depth range of the stimuli used for the user study spans 9.6 \emph{D}, the ocular parallax can still be detected even with the objects having a narrower depth range. This ocular parallax estimation leaves a question; Does the images perceived with the eye movement present parallax larger than the detection threshold?    



To answer the raised concern, we roughly quantify ocular parallax detection using two frames processed with specified pupil displacement and pupil diameter. We employ a technique commonly used in finding 3D stereo pairs - feature point matching~\cite{liu2010sift} to calculate disparities in these frames.  

In detail, if a set of feature points ($\mathbf{X}$) is extracted from a single image ($I$), it can be formulated as $\mathbf{X}\left( I \right)=\left\{ \mathbf{x}|\mathbf{x}\in F\left( I \right) \right\}$, where,  $\mathbf{x}\in {{\mathbb{R}}^{2\times 1}}$ is a vector presented the two-dimensional angular displacement from the center axis and $F\left( \cdot  \right)$ is the feature point extraction operator. Then, a set ($\mathbf{P}$) of pairs $({\mathbf{X}}_{1},{\mathbf{X}}_{2})$ extracted from two different frames $({I}_{1},{I}_{2})$ can be presented depending on the threshold value (${\theta }_{th}$) as follows:
\begin{equation}
    {\mathbf{P}}_{\theta }({{\mathbf{X}}_{1}},{{\mathbf{X}}_{2}})=\left\{ \mathbf{x}|{{\mathbf{x}}_{1}}\in {{\mathbf{X}}_{1}}\,,{{\mathbf{x}}_{2}}\in {{\mathbf{X}}_{2}},\,\left\| {{\mathbf{x}}_{1}}-{{\mathbf{x}}_{2}} \right\|\ge {{\theta }_{th}} \right\}.
\end{equation}
Here, the subscript in the single-frame image denotes the pupil state ($p$) comprised of pupil displacement and diameter and the focal state ($j$) of the eye. Here, each pair of feature points is evaluated whether the $l$-2 norm ($\left\| \cdot  \right\|$) of the difference in the angle space exceeds the detection threshold.  

We normalized the count of feature point pairs meeting the specified condition by dividing it by the total number of extracted pairs. This normalization process entailed traversing through the sampled focal states and dividing by the total extracted feature pairs. The resulting value represents the parallax detection rate and can be presented as
\begin{equation}
    {{R}_{\theta }\left({p}_{1},{p}_{2} \right)}=\frac{\bigcup\limits_{j=1}^{J}{{{\mathbf{P}}_{\theta}}\left( {{\mathbf{X}}_{{{p}_{1}},j}},{{\mathbf{X}}_{{{p}_{2}},j}} \right)}}{\bigcup\limits_{j=1}^{J}{\mathbf{P}\left( {{\mathbf{X}}_{{{p}_{1}},j}},{{\mathbf{X}}_{{{p}_{2}},j}} \right)}}.
    \label{eq:ParallaxDetectionRate}
\end{equation} 

We simulate the ocular parallax detection rate under the conditions emulating an ideal 3D display capable of supporting 25$\times$25 perspective images, matching the viewing conditions of our experimental setup. This step helps eliminate potential errors in the feature point extraction process, particularly in holographic images containing speckle noise. Furthermore, we extend this evaluation to encompass an ideal 3D display scenario that does not experience angular resolution degradation, achieved through a dense distribution of view images. This quantitative analysis broadens the scope of validity of our findings.

In our analysis, we utilized the SIFT flow~\cite{liu2010sift} method for feature point extraction. We gathered paired images representing nine distinct focal states for our investigation. To simplify our computations, we reduced the dimensions of the parallax detection rate function described in Eq.~\ref{eq:ParallaxDetectionRate} from six to two. We assumed that both pupils were horizontally aligned. Furthermore, we maintained consistent pupil diameters in both states (${p}_{1}=({x}_{p,norm},{D}_{p,norm}), {p}_{2}=(0,{D}_{p,norm})$), given that the pupil size is primarily influenced by the scene's luminance.

The parallax detection rate is calculated with 5 (pupil diameter) $\times$ 12 (pupil displacement) discrete pupil states and interpolated across the pupil state domain as provided in Fig.~\ref{fig:supSCEPupil_parallax}. In the figure, the pupil diameter and pupil displacement are normalized with the width of the eyebox, respectively.

The parallax detection rate is simulated based on two different pupil apodization modes (pupil w/o Stiles-Crawford effect (diffraction-limited pupil) and pupil w/ Stiles-Crawford effect). The figures demonstrate how the parallax detection rate would be affected by the human eye apodization profile. For the detection threshold value, we chose the eccentricity-dependent model, which is approximated as a linear function of eccentricity with the given parameter of the work of Konrad et al.~\shortcite{konrad2020gaze}. 


In Fig.~\ref{fig:supSCEPupil_parallax}, it is clear that the parallax is hardly detected when the pupil displacement is minimal regardless of the pupil diameter. However, there are large differences when the eye is in motion. Especially, if the pupil is assumed to a diffraction-limited, the parallax detection rate does not vary depending on the pupil displacement when the normalized pupil size exceeds 2.0. On the other hand, the parallax detection rate begins to saturate at a certain value when the pupil is apodized. It can be observed in the bottom figures of Fig.~\ref{fig:supSCEPupil_parallax} plotting the projected line showing the relationship between pupil size and parallax detection rate. Note that the average of the measured subjects' pupil diameters exceeds 4.4 mm (${D}_{p,norm}=2.0$) in the actual experiment.

Here, we do not claim that the given ocular parallax detection rate is built upon an accurate model with measurements nor it represents the absolute detection probability, but we evaluate with the given simulation to show the difference in terms of ocular parallax. Interestingly, there is limited exploration into the detection and discrimination threshold of ocular parallax. Accurate modeling of the ocular parallax and quantifying the impact on 3D realism presents an intriguing avenue for future research. 

\subsubsection{Stiles-Crawford effect depending on CGH supervisions}
\begin{figure}[ht!]
  \centering
  \includegraphics[width=\columnwidth]{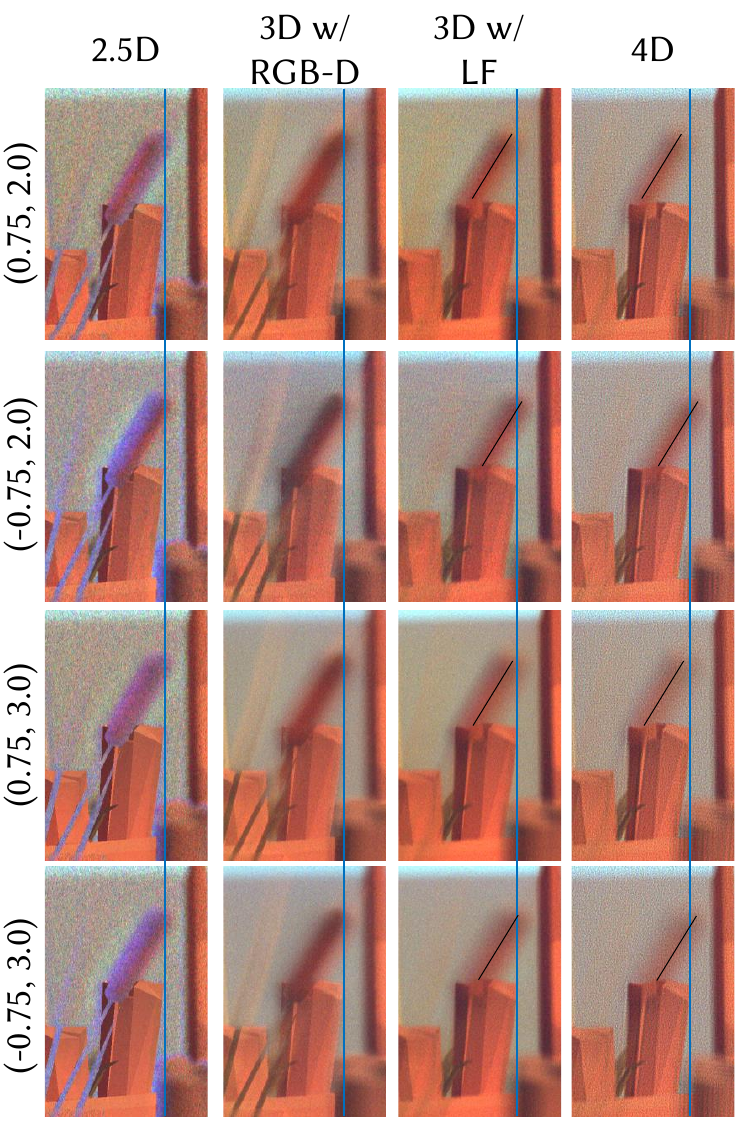}
  \caption{Enlargements of reconstructed results with apodized pupil having Stiles-Crawford effect depending on CGH algorithms (1st col: 2.5D, 2nd: 3D w/ RGB-D, 3rd: 3D w/ LF, 4th: 4D). They are reconstructed in different pupil states (${x}_{p,norm}$,${D}_{p,norm}$). For visibility, the blue, and black lines are additionally drawn to emphasize the disparity of the objects.}
  \label{fig:supSCE_CGH}
\end{figure}

We present the reconstructed results of different CGH algorithms assuming the pupil with Stiles-Crawford Effect in Fig.~\ref{fig:supSCE_CGH}. They are provided with different pupil states, where two of the pupil states ((${x}_{p,norm}$,${D}_{p,norm}$) = (0.75, 3.0), (-0.75, 3.0)) show fully underfilled pupil cases. The disparity of the objects that corresponds to the depth difference can be seen in the results, while LF-based supervision targets (3D w/ LF and 4D) presented images with robust defocus blur. 
In addition, it can be observed that the additional adoption of SC pupil reconstructs different perspectives even in the case when the pupil is completely underfilled. The 3D w/ LF and 4D cases show similar results while the parallax is slightly more noticeable in the results of 4D case. Note that the difference among the algorithms can be found in Fig.~\ref{fig:subParallax}.

\subsection{Number of views for 4D CGH supervision}
\begin{figure*}[ht!]
  \centering
  \includegraphics[width=0.82\textwidth]{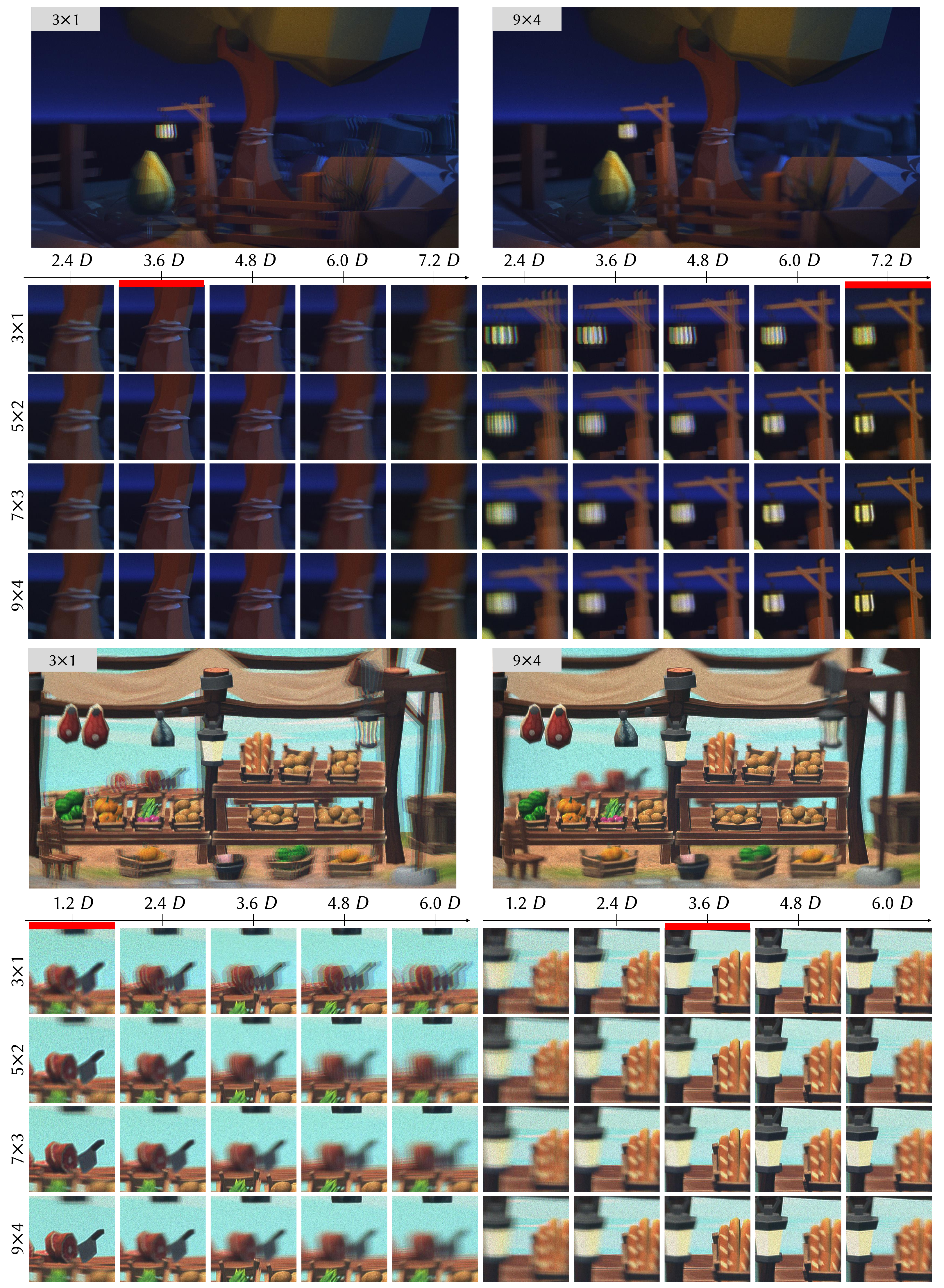}
  \caption{Reconstructed holographic images of landscape$\_$night (\textit{top}) and village (\textit{bottom}) scene supervised with sparse light field (\textit{left}) and dense light field (\textit{left}). Two different sections of the holographic images supervised with different view numbers (3$\times$1, 5$\times$2, 7$\times$3, 9$\times$4) are provided with five different focal states. The red bar placed at the top of the column indicates that the enlarged object is best focused at the depth.}
  \label{fig:supexp2_1}
\end{figure*}

\begin{figure*}[ht!]
  \centering
  \includegraphics[width=0.82\textwidth]{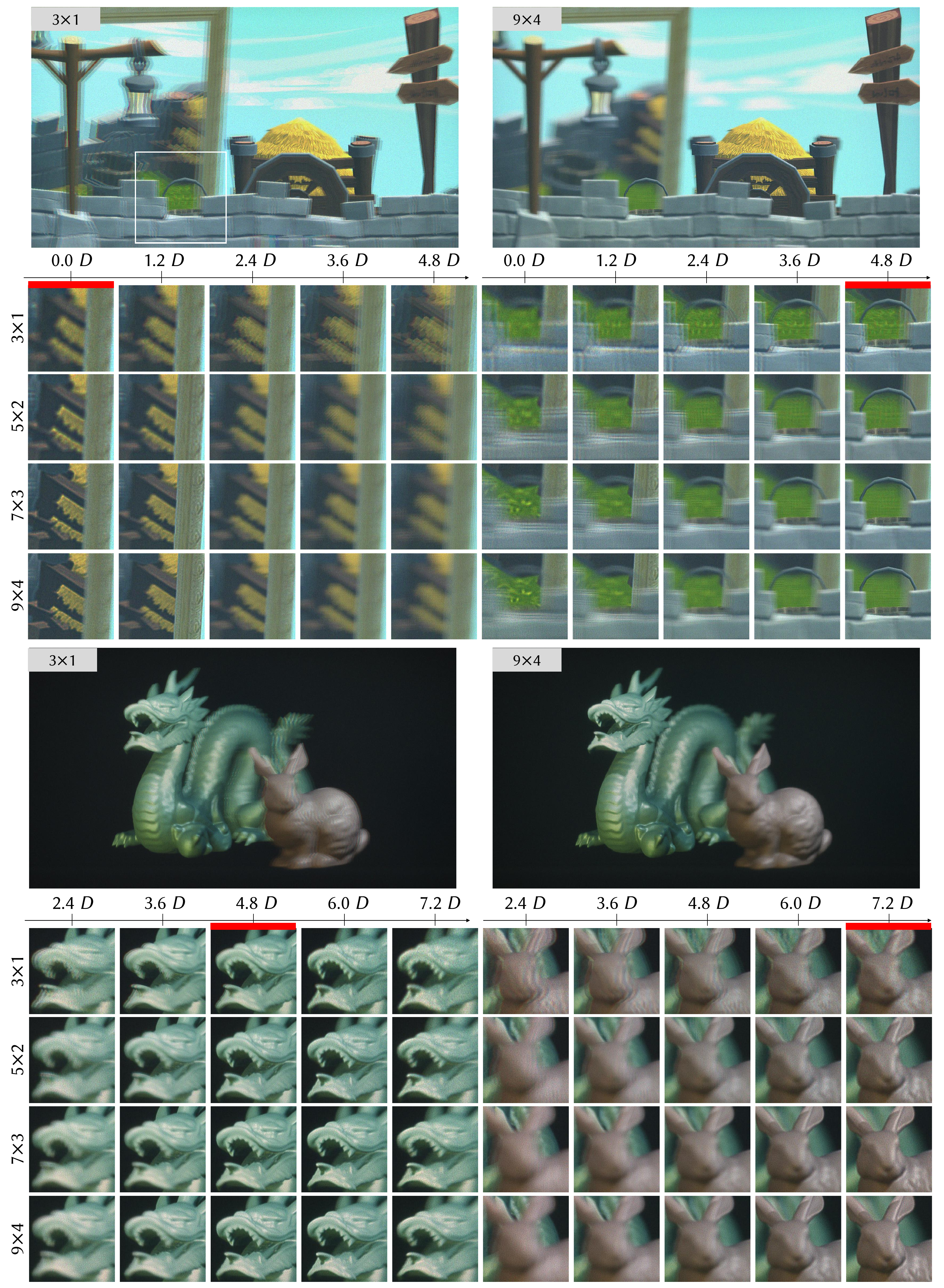}
  \caption{Reconstructed holographic images of village$\_$mirror (\textit{top}) and dragon$\_$bunny (\textit{bottom}) scene supervised with sparse light field (\textit{left}) and dense light field (\textit{left}). Two different sections of the 4D-supervised holographic images with different numbers of views (3$\times$1, 5$\times$2, 7$\times$3, 9$\times$4) are provided with five different focal states. The red bar placed at the top of the column indicates that the enlarged object is best focused at the depth.}
  \label{fig:supexp2_2}
\end{figure*}

In section 5, we investigate the number of views required for 4D CGH supervision through both camera-incorporated experiments and a user study. Each condition is easily understood with the simulated images of various scenes provided in Fig.~\ref{fig:supexp2_1}-\ref{fig:supexp2_2}. 

By examining these images, it becomes apparent that the overall quality of the reconstructed 3D scene improves as more views are incorporated into the 4D CGH supervision. However, the objects placed near the WRP are reconstructed with a high resolution (see the enlarged images of tree, baguette in Fig.~\ref{fig:supexp2_1}, and handle of basket, head of dragon in Fig.~\ref{fig:supexp2_2}) and do not demonstrate the noticeable difference when the scenes are rendered with denser views. On the other hand, the image quality suffers for the objects lying at planes that deviated from the WRP (see the enlarged images of street light, fruit in Fig.~\ref{fig:supexp2_1} and wagon reflected by a mirror, head of bunny in Fig.~\ref{fig:supexp2_2}) especially when the views are sparsely sampled. Therefore, the reconstructed results visually demonstrate that the placement of objects relative to the WRP and the number of views directly affect the resolution of the reconstructed scene.

The resolution of CGHs exhibits a depth-dependent characteristic due to the reconstruction of individual view images in 4D-supervised CGHs. However, when CGHs are generated using a plane-to-plane model, the highest resolution remains consistent regardless of changes in depth. Surprisingly, the results obtained from the user study present contrasting findings, even though the experiments are conducted with scenes featuring a large distribution of depth. These outcomes indirectly suggest that the assessment of 3D perceptual realism cannot be solely determined by the resolution of objects located at varying distances from the user.

\subsection{Number of views in LF sampling analysis}

\begin{figure}[ht!]
  \centering
  \includegraphics[width=\columnwidth]{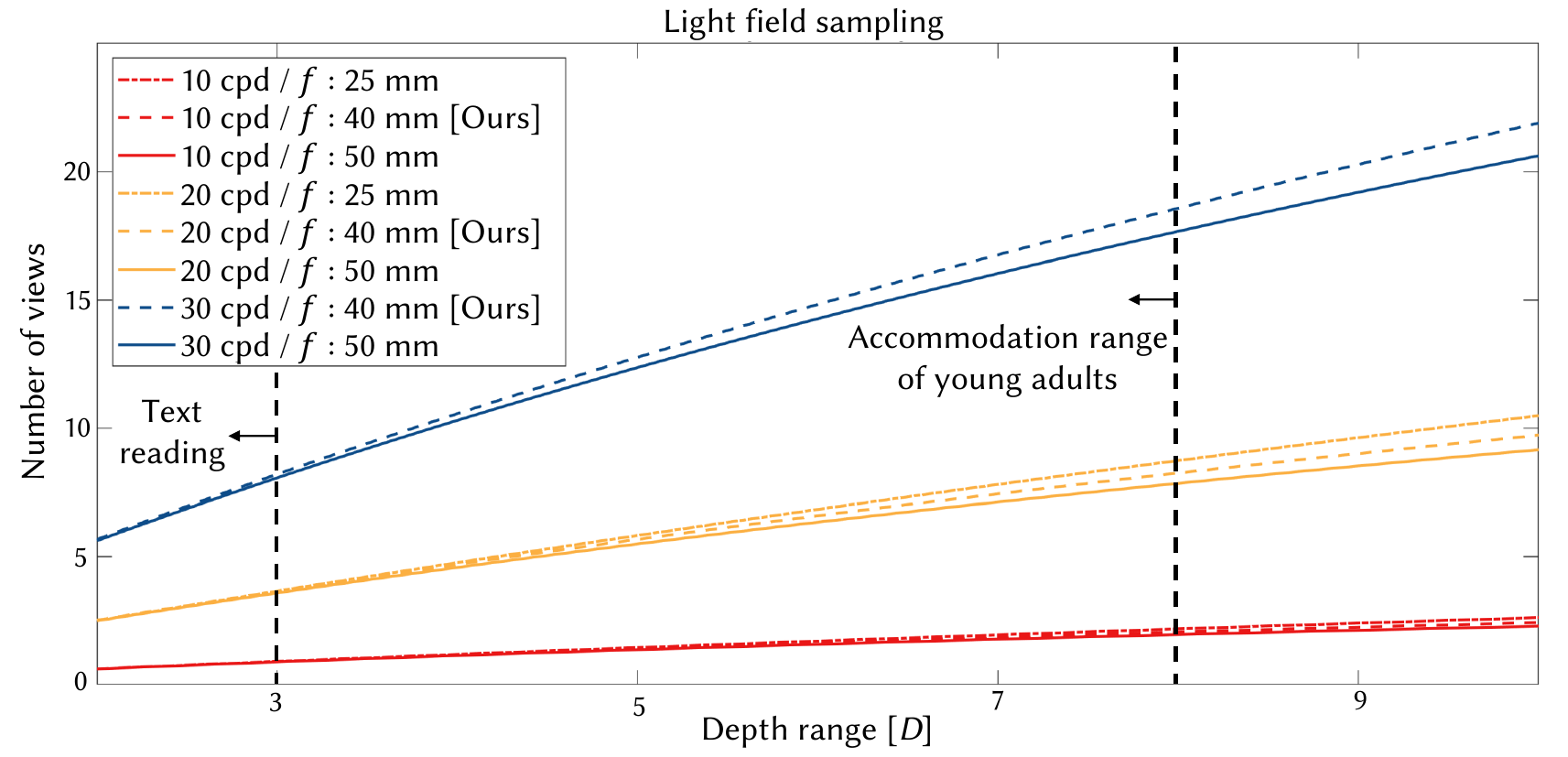}
  \caption{Required number of views in horizontal for 4D CGH supervision based on light field sampling theorem. It differs by the spatial bandwidth of the scene, and the depth range of the 3D scene. The graph is plotted with three different spatial bandwidths (red: 10 cpd, orange: 20 cpd, blue: 30 cpd) and the optical configurations having eyepiece lenses with different focal lengths (dotdash line: $f:$ 25 mm, dashed line: $f:$ 40 mm, and solid line: $f:$ 50 mm). The wavelength is assumed as 532 nm for the simulation. The simulation is conducted with an SLM having a pixel pitch of 8.2 $\mu m$ and a horizontal resolution of 1920, which is identical to the experimental setup. We additionally placed black dashed lines in 3 \emph{D} and 8 \emph{D},  each of which indicates the dioptric range of text reading and accommodation range~\cite{duane1912normal}, respectively. Note that the case of 30 cpd / $f:$ 50 mm is not plotted as the cut-off frequency of the condition is below 30 cpd.}
  \label{fig:sublfsampling}
\end{figure}

We analyze the number of views required for 4D CGH supervision based on light field sampling theorem~\cite{park2017recent,zhang2009wigner}. In the theorem, the depth range covered by the light field is proportional to the angular resolution. We deviate the analysis with the optical configuration of near-eye displays.
Let's assume the situation when the WRP is placed at a certain distance, and the FCP is located at the focal length of the eyepiece lens as Fig.~\ref{fig:subSystemscheme}. The metric distance between FCP and NCP is $2z_{o}$ which makes the depth coverage from 0 \emph{D} to $D_{max}$ and the relationship between the two variables is as follows:

\begin{equation}
	{{z}_{o}}=\frac{f}{2}-\frac{1}{2\left( {{D}_{\max }}+\frac{1}{f} \right)}=\frac{{{f}^{2}}{{D}_{max}}}{2\left( f{{D}_{\max }}+1 \right)}.
    \label{eq:s10}
\end{equation}

Here, the angular resolution ($N_{u}$) of the light field required to reconstruct the image with spatial bandwidth of ($B_{x}$) in the distance of $z_{o}$ can be obtained as:
\begin{equation}
    {{N}_{u}}\ge \lambda {{z}_{o}}B_{x}^{2}
    \label{eq:s11}
\end{equation}
Integration of the equations (Eq.~\ref{eq:s10} and Eq.~\ref{eq:s11}) allows us to calculate the maximum depth range supported by the given angular resolution of the light field and the spatial bandwidth of the scene. If the spatial bandwidth is bounded to a certain range, the low-pass filtered spatial bandwidth (${B}_{x,v}$) can be simply acquired with the ratio of spatial frequency (${v}$) and cut-off frequency (${v}_{cutoff}$) as ${{B}_{x,v}}={{B}_{x}}\frac{v}{{{v}_{cutoff}}}$. The cut-off spatial frequency can be acquired with the optical configuration of the near-eye display.

Therefore, the graph labeled as Fig.~\ref{fig:sublfsampling} illustrates the number of horizontal perspectives required for 4D-supervised CGH based on LF sampling analysis, depending on the depth range. We sampled three different spatial bandwidth regions, and the results are depicted using three different eyepiece lenses. We sampled three different spatial bandwidth regions, and the results are drawn with three different eyepiece lenses. Although the eye relief of the near-eye display, which we assume that the focal length of the eyepiece lens is identical to the eye relief, is down to the conventional range, the estimated value for the required number of views is similar to the parameter obtained with our system's configuration. 

Furthermore, if the entire depth range is reduced to 3 \emph{D}, which corresponds to the depth at which we often position books for reading, the required number of horizontal views is approximately 8, supporting a resolution of 30 cpd. However, if the 3D content aims to cover the full depth range of 8 \emph{D} with high resolution, more than 15 views are necessary. It is important to note that this analysis does not consider other potential factors such as diffraction from the eye pupil or aberration in an individual's eye, which can affect the point spread functions and ultimately impact the results.

\subsection{VDP simulation with matched display model}
The comparisons of VDP in Fig. 3, Fig. S4-S5 are performed based on the luminance and contrast conditions of conventional VR headsets, while the experimental conditions differ from the simulated conditions. First, the experiment was conducted under low-luminance lighting conditions due to safety concerns regarding eye safety. Additionally, the contrast may be lower for holographic displays due to imperfect black level expression. These mismatches in the display model between the simulation and experimental conditions may result in different outcomes.

\begin{figure*}[ht!]
  \centering
  \includegraphics[width=\textwidth]{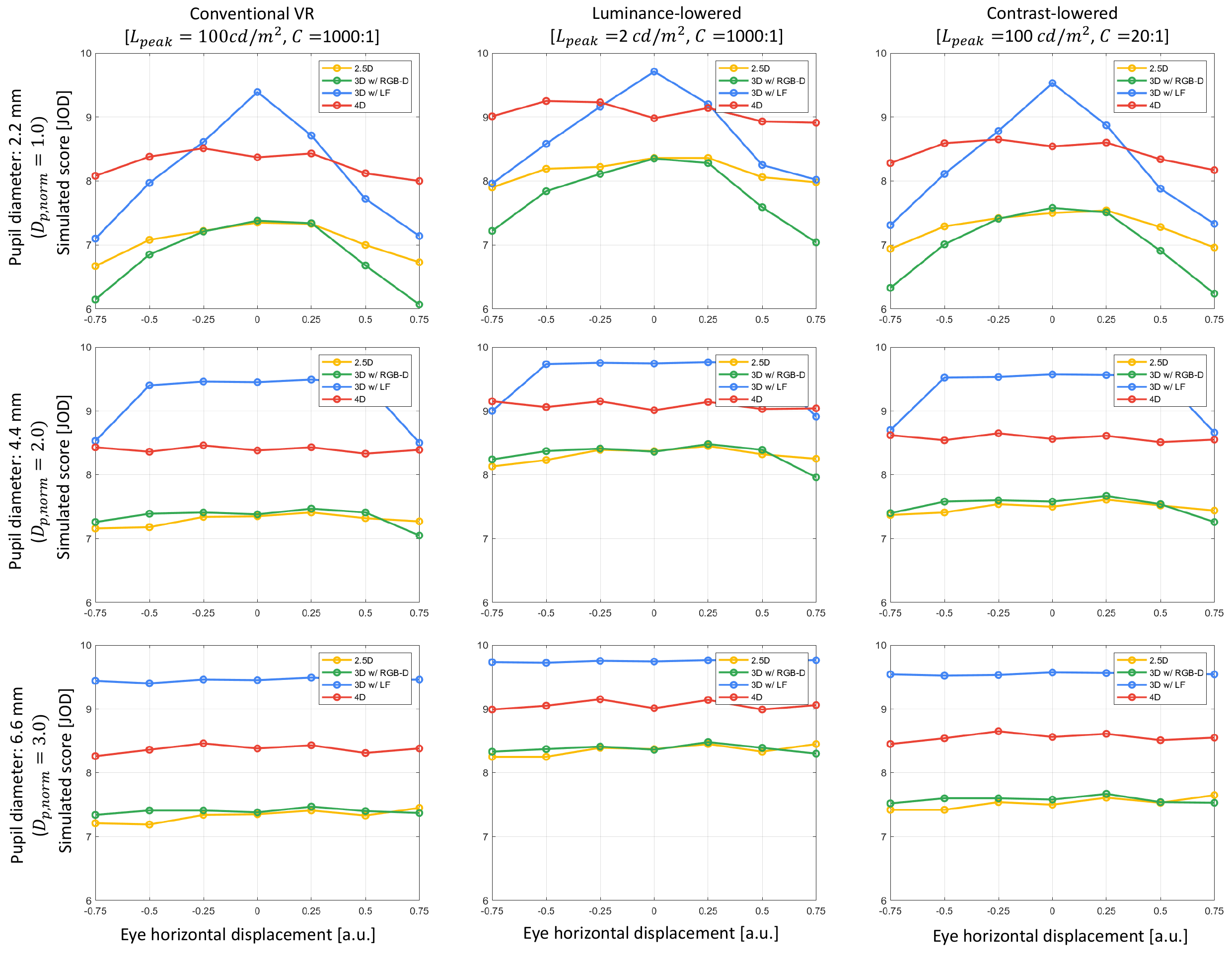}
  \caption{FovVideoVDP simulation of different CGH algorithms (yellow: 2.5D, green: 3D w/ RGB-D, blue: 3D w/ LF, and red: 4D) under different luminance (${L}_{peak}$) and contrast (${C}={L}_{peak}:{L}_{min}$) conditions. The first column denotes the conventional VR model having luminance of 100 ${cd}/{{{m}^{2}}}$ and contrast level of 1000. The second column corresponds to the low luminance condition matched with the experimental condition. The third column demonstrates the condition when the contrast level is worse. They are demonstrated with three different pupil diameter conditions. The values are extracted with the reconstructed image of fifth focal stack of the landscape$\_$day scene assuming diffraction-limited pupil condition. The range of -0.5 to 0.5 in the eye horizontal displacement corresponds to the width of the eyebox.}
  \label{fig:supJOD}
\end{figure*}

Fig.~\ref{fig:supJOD} demonstrates the simulated FovVideoVDP results of various CGH algorithms conducted under three different conditions: the conventional VR condition, with a luminance of 100 ${cd}/{{{m}^{2}}}$ and a contrast level of 1000:1; the luminance-lowered condition to match the peak luminance with the experiment, corresponding to 2 ${cd}/{{{m}^{2}}}$ and a contrast level of 1000:1; and the contrast-lowered condition, with a luminance level of 100 ${cd}/{{{m}^{2}}}$ and a contrast level of 20:1. These are simulated with seven different pupil displacements and the results are plotted based on pupil sizes (${D}_{p,norm}=1.0, 2.0, 3.0$).

It can be observed that as the luminance level decreases, the absolute values of JOD tend to increase. This is due to suppressed contrast sensitivity at low luminance levels, which allows the simulation to neglect grainy noise that is prevalent in holographic images, especially when the contrast level is lower. However, the overall trend does not change, nor does it reverse. This implies that the minute mismatch of luminance or contrast specifications is not the major reason for the superiority of the 4D case over the others.

\section{Additional experimental results}
In this section, we additionally demonstrate the experimental results that are not provided in the paper. 
\subsection{Additional captured results}
\begin{figure*}[ht!]
  \centering
  \includegraphics[width=\textwidth]{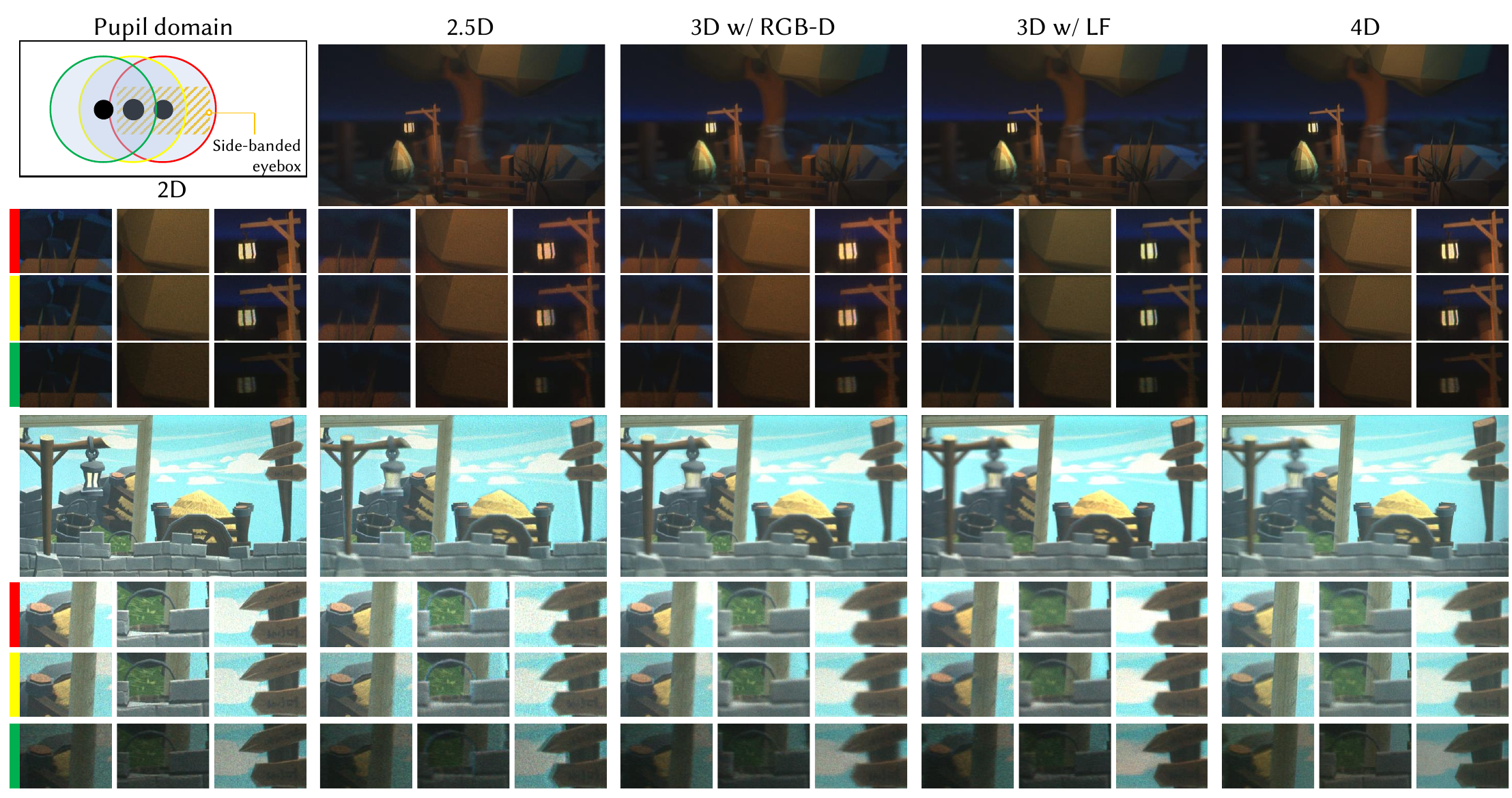}
  \caption{Additional experimental results with different pupil positions. Holographic scenes supervised with 2D (1st col), 2.5D (2nd), 3D w/ RGB-D (3rd), 3D w/ LF (4th), 4D (5th) targets are captured with different pupil positions (red: $({x}_{p,norm},{D}_{p,norm})=(0,1.1)$, yellow: $(-0.68,1.1)$ and green: $(-1.36,1.1)$). The scenes are photographed with different focal states (landscape$\_$night: 7th, village$\_$mirror: 3rd) out of 9 distinct focal states equally sampled in diopter. Enlargements are provided with the image focused on the magnified object. We intentionally provide the results without modifying the brightness to show the energy across the eyebox. Note that it is hard to discriminate 3D w/ LF case and 4D case with the captured results.}
  \label{fig:supexp1}
\end{figure*}

Figure~\ref{fig:supexp1} additionally presents the images captured while changing the camera positions. For the village$\_$mirror scene, the depth difference of the objects reflected by the mirror is observed when comparing RGB-D-based approaches and LF-based approaches. This also presents the limitation of the RGB-D-based presentation of volumetric scene. 

\subsection{Captured results with pupil movement contour}

\begin{figure*}[ht!]
  \centering
  \includegraphics[width=\textwidth]{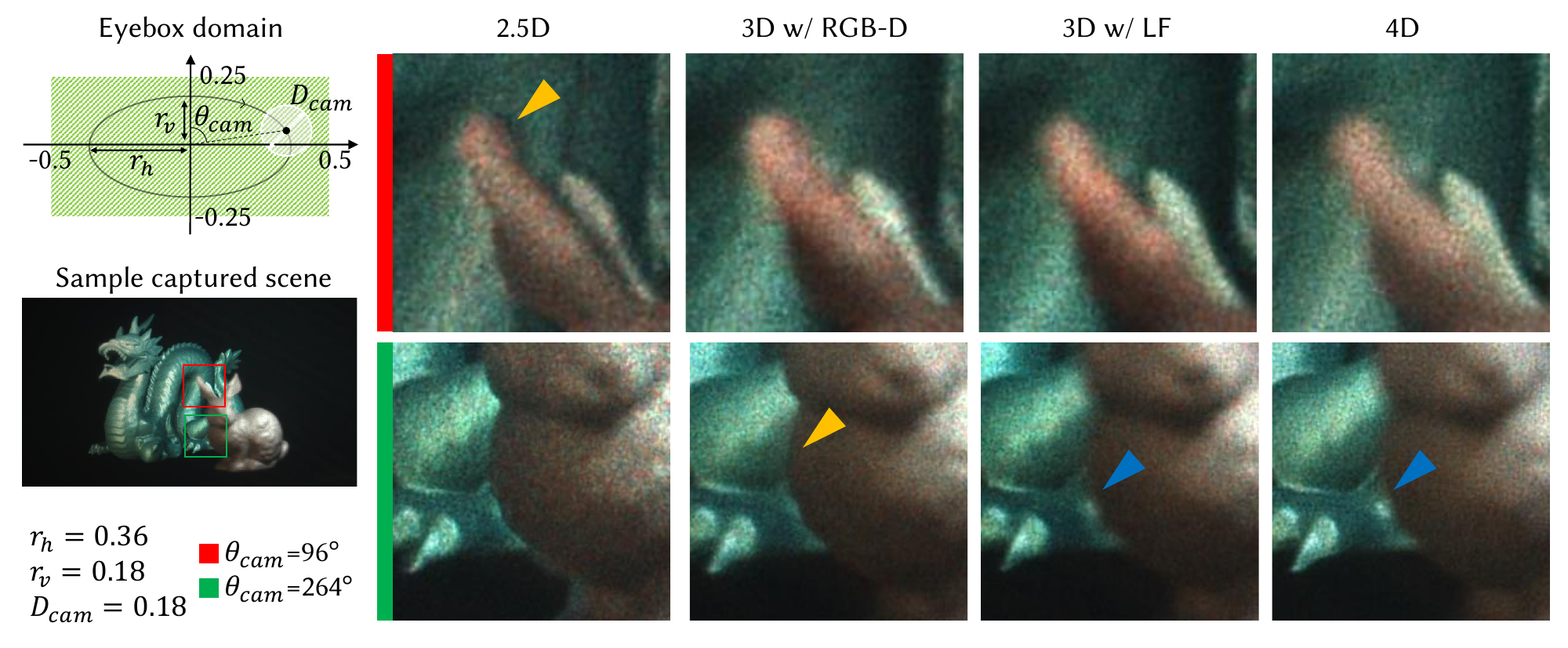}
  \caption{Additional captured results showcasing different 3D CGH supervision approaches under distinct acquisition conditions (red and green) are presented below the sample captured scene. To maintain consistency in the specifications regarding pupil states, we have provided them with the normalized coordinates. Enlargements near the bunny's ear are provided for images captured under the red acquisition condition, while those near the bunny's body are provided for images captured under the green acquisition condition. These captured results offer a clearer insight into the issues within each CGH supervision method. Yellow arrows highlight the occlusion boundary problem, with the 2.5D case displaying discontinuity in the occlusion boundary, while the 3D w/ RGB-D still reconstructs a sharp occlusion boundary. Blue arrows in the 3D w/ LF and 4D cases indicate the issue with focal-stack-based targets even if they are processed with dense LF. This averaging in the focal stack generation procedure limits the reconstruction of view-dependent visual effects in the case of 3D w/ LF, particularly in sections with view-dependent visual effects.}
  \label{fig:supexpparallax}
\end{figure*}


We present the captured frames of two different pupil states of the camera at the eyebox domain as shown in Fig.~\ref{fig:supexpparallax}. In detail, the camera is placed at the rightmost (${{\theta }_{k}}={96}^\circ $) and leftmost (${{\theta }_{k}}={264}^\circ $). For better visualization, we provide enlarged images of various 3D CGH supervision approaches. Refer to Fig. 2 in the paper, 2.5D and 3D have occlusion boundary problems, and 3D w/ LF demonstrates images with averaged intensity across the view images, which inherently suppresses the visualization of view-dependent effects.

\subsection{Captured eyebox}
\begin{figure}[ht!]
  \centering
  \includegraphics[width=\columnwidth]{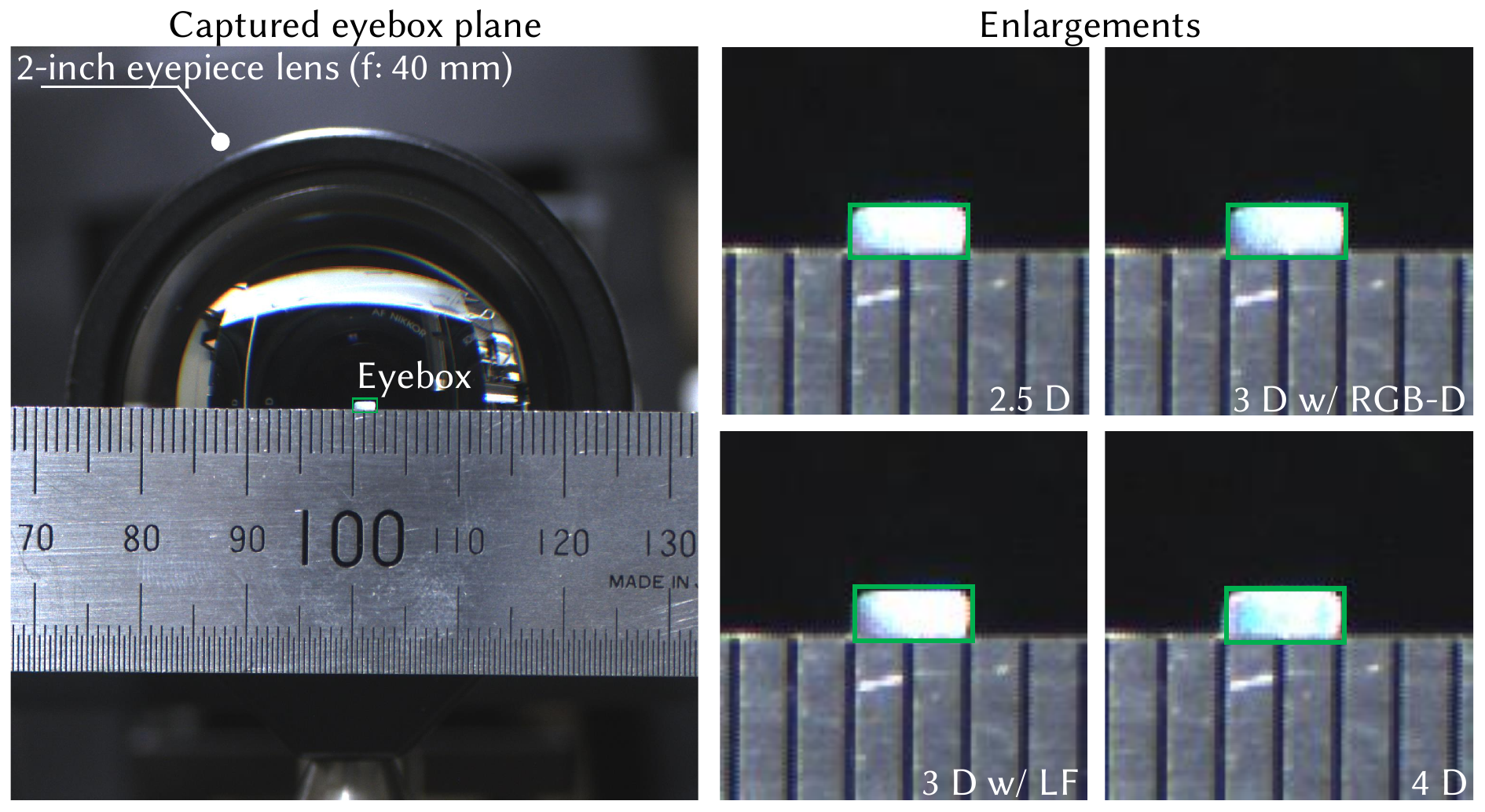}
  \caption{Captured eyebox plane with camera is shown. The enlargements of captured eyebox (green rectangle) are provided depending on the CGH supervision approaches (2.5D, 3D w/ RGB-D, 3D w/ LF, and 4D). Note that the physical size of the entire eyebox is 2.2 mm $\times$ 1.1 mm for the illumination of 450 nm (blue channel) as the work exploits side-band filtering for complex modulation with an amplitude-only SLM. Slight misalignment between the near-eye display system and the camera that captured eyebox presents asymmetrical energy distribution, which is not the main focus of the figure.}
  \label{fig:subeyeboxexp}
\end{figure}

We additionally provide the captured eyebox of the holographic near-eye display system with various CGH supervision targets as Fig.~\ref{fig:subeyeboxexp}. At the same time, the ruler with the millimeter scale is placed at the eyebox plane to roughly measure the physical size of the eyebox. 

Some previous works on holographic displays showed some results limiting the eyebox size while improving the 2D quality of holographic scene at the sampled depth. However, we employed binary SLM that results in complex-valued field with random phase distribution that eventually supports full eyebox irrespective of the CGH supervision targets.  

\subsection{3D realism depending on the CGH supervision targets (user experiment 1)}
Following the pairwise comparison, the responses from two subjects out of twenty-eight were identified as outliers and subsequently excluded from the analysis. To guide the outlier analysis, we referred to the work of Perez-Ortiz and Mantiuk~\shortcite{perez2017practical}. After removing the outliers, we estimated the confidence interval using the bootstrapping method. The statistical test was conducted using a two-tailed z-test on the JOD scores obtained for each viewing condition.

In detail, in the viewing condition of \textit{Center}, statistically significant differences in JOD scores were found in most of the paired conditions ($p$<0.001: \textit{4D} vs the other cases, \textit{3D w/ LF} vs \textit{3D w/ RGB-D}, \textit{3D w/ LF} vs \textit{2.5D}, $p$<0.05: \textit{3D w/ LF} vs \textit{2.5D}) except the \textit{2.5D} vs \textit{3D w/ RGB-D} ($p$= 0.37). In the viewing condition of \textit{Decentered}, the significant results were found in the paired conditions ($p$ <0.001: \textit{4D} vs the other cases, $p$<0.01: \textit{3D w/ LF} vs \textit{3D w/ RGB-D}, $p$<0.05: \textit{3D w/ LF} vs \textit{2.5D}) except the \textit{2.5D} vs \textit{3D w/ RGB-D} ($p$=0.36). In case of \textit{Vignetted}, significant differences were present in the paired conditions ($p$<0.001: \textit{4D} vs the other cases, \textit{3D w/ LF} vs \textit{3D w/ RGB-D}, $p$<0.05: \textit{3D w/ LF} vs \textit{2.5D}) except the paired condition of \textit{2.5D} vs \textit{3D w/ RGB-D} ($p$=0.38). Lastly, in the case of \textit{w/ head movement}, the JOD scores of paired conditions were significantly different ($p$<0.001: \textit{4D} vs the other cases, \textit{3D w/ LF} vs \textit{3D w/ RGB-D}, \textit{3D w/ LF} vs \textit{2.5D}) except the paired condition of \textit{2.5D} vs \textit{3D w/ RGB-D} ($p$=0.85). 


\subsubsection*{Additional raw data of pairwise comparison}
\begin{figure*}[ht!]
  \centering
  \includegraphics[width=\textwidth]{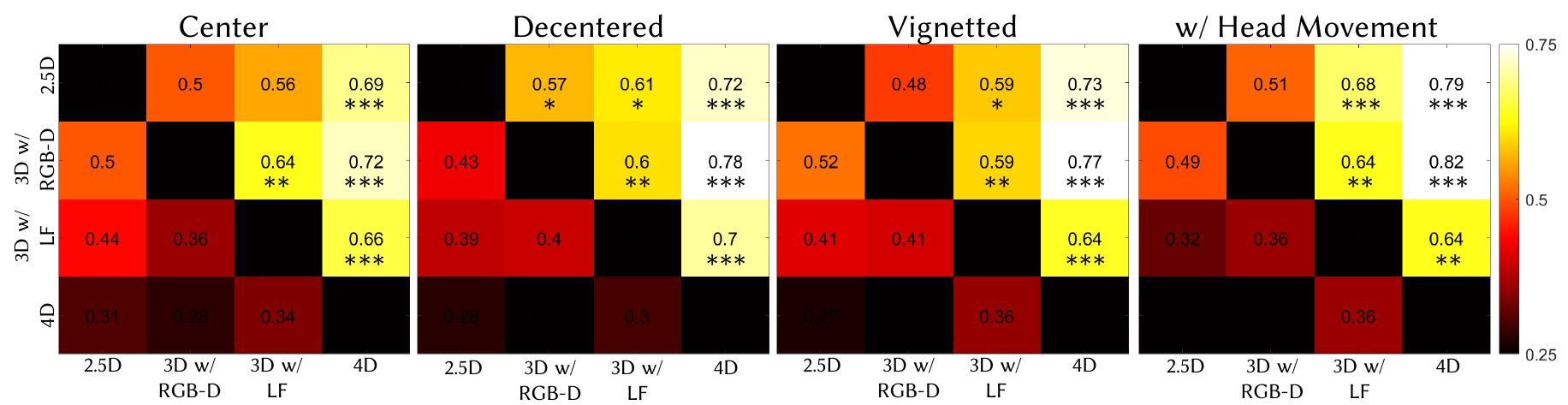}
  \caption{Normalized matrices of comparisons based on varying viewing conditions are provided, reflecting preferences estimated by the vote counts favoring the column option over the row option. Additionally, the figure presents statistical significance determined via the one-tailed Wilcoxon signed rank test using the preferences of twenty-six subjects (*: $p$<0.05, **: $p$<0.01, and ***: $p$<0.001). The colorbar spans from 0.25 to 0.75, representing the range between -1 JOD and +1 JOD in the converted scale.}
  \label{fig:subfigurevotecount}
\end{figure*}

In the paper, we initially presented scaled JOD results in Fig. 4. Here, we provide the raw data of vote counts obtained from pairwise comparisons, illustrated in Fig.~\ref{fig:subfigurevotecount}. These counts depict the preference for the column option over the row option across four distinct viewing conditions accumulated with three evaluation scenes. The diagonal elements of the matrices are zero since the comparisons are only performed between different CGH supervision targets.

Additionally, we performed one-tailed Wilcoxon signed rank tests using the vote counts obtained from twenty-six subjects. It is important to note that these statistical results may slightly differ from those in Sec. 3, which were based on scaled JOD rather than raw vote counts. Even preferences below 0.75, equivalent to 1 JOD in the scaled unit, exhibit strong statistical significance in the non-parametric significance test. These findings further highlight the superior performance of 4D CGH supervision, surpassing even the performance of 3D w/ LF case under the \textit{Center} viewing condition.

\subsection{Eye tracking trajectory} 
\begin{figure}[ht!]
  \centering
  \includegraphics[width=\columnwidth]{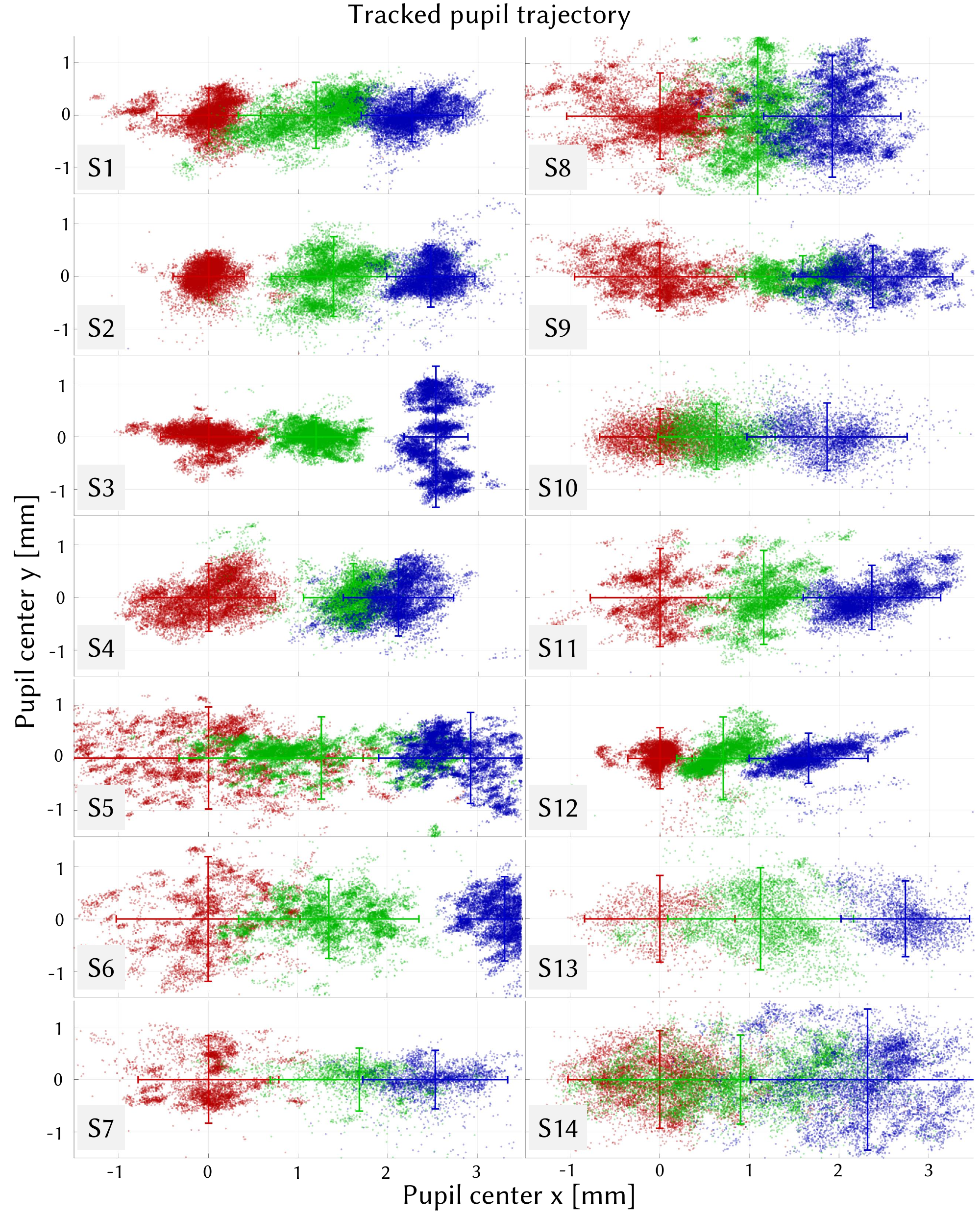}
  \caption{Recorded eye trajectory of fourteen subjects during a single session when experiencing the holographic image of landscape$\_$day scene. Each color dot indicates the recorded pupil position with different viewing conditions (red: \textit{Center}, green: \textit{Decentered}, blue: \textit{Vignetted}). The horizontal displacement of the \textit{Center} condition is regarded as zero, whereas the vertical displacements have been adjusted to have an average of zero for improved visualization. The error bars represent the 95$\%$ confidence interval of the pupil displacement recorded in each viewing condition.}
  \label{fig:subexp1_eyetrack}
\end{figure}

To validate our assumptions regarding the continuous eye movements during the experience of holographic near-eye displays, we utilized an eye tracker to record the participants' pupil movements while they performed visual tasks. Fig.~\ref{fig:subexp1_eyetrack} illustrates the recorded eye movement trajectories of fourteen subjects during a single session, corresponding to each viewing condition.

The presented figure clearly demonstrates that the experiments were carried out in distinct sections of the eyebox as intended. Prior to the experiment, the center of the eyebox was determined in the global coordinate system and served as the reference point. The initial position of the pupil for each session was adjusted to achieve the desired deviations of 1.25 mm for the \textit{Decentered} condition and 2.5 mm for the \textit{w/ head movement} condition from the center position. However, there were slight variations in the average measurements among individuals. It is important to note that the average displacement for each viewing position is less than 1 mm away from the desired placement. Due to potential factors such as occlusion by the eyelid and blinking, we have only provided the measured pupil displacement of a few subjects for better visualization.

\begin{figure}[ht!]
  \centering
  \includegraphics[width=\columnwidth]{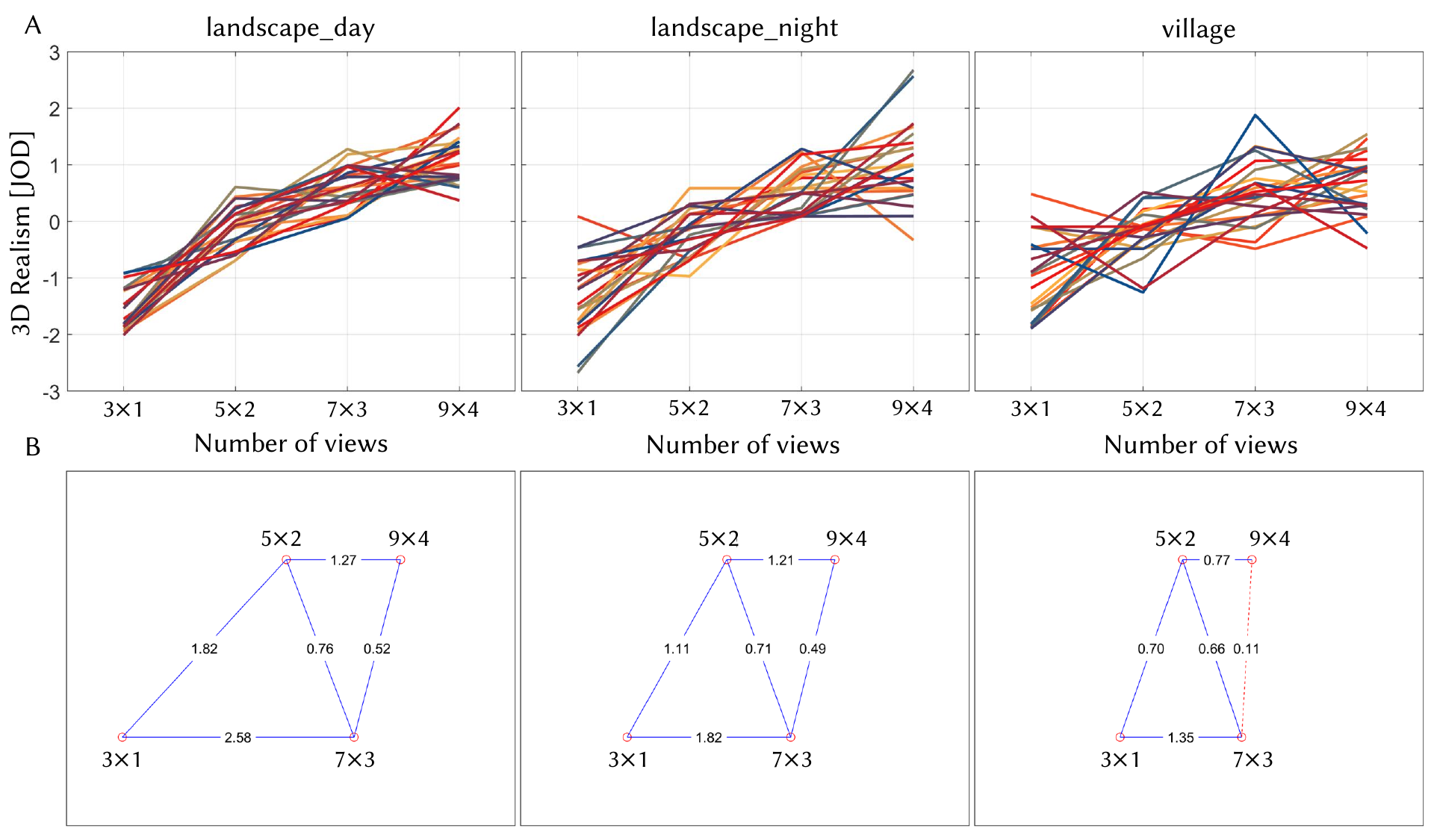}
  \caption{Additional experimental results of user experiment $\#$2 (1st column: landscape$\_$day, 2nd column: landscape$\_$night, 3rd column: village): (1st row) JOD scores based on the responses from individual subjects. Each line indicates the JOD scores of each view number case with an average of zero. (2nd row) The graphical demonstrations of the JOD scaling. The red circle indicates the number of view conditions, and the lines are interconnected with neighboring options. The blue line represents the statistical significance of the JOD score difference ($p$<0.05) between the paired conditions evaluated with a two-tailed z-test, as opposed to the red dashed line. The value on the individual line indicates the mean JOD difference of the paired conditions.}
  \label{fig:subexp2}
\end{figure}

\subsection{3D realism depending on the number of views for 4D CGH supervision (user experiment 2)}
In Sec. 4, the 3D realism of the holographic scene is evaluated depending on the number of views used in CGH supervision, and the JOD value is provided depending on the target stimuli. The landscape$\_$day scene shows -1.88, -0.05, 0.70, and 1.22 JOD. The landscape$\_$night scene demonstrates -1.31, -0.20, 0.51, and 1.00 JOD. Lastly, the village scene resulted in -0.88, -0.18, 0.47, and 0.59 JOD as shown in Fig. 5(A).

We provide additional experimental results as Fig.~\ref{fig:subexp2} acquired with the pairwise comparisons with conditions that differ in the number of views used in 4D CGH supervision. Fig.~\ref{fig:subexp2}(A) demonstrates the scaled JOD scores in individual subjects depending on the stimuli. Notably, the experiment conducted with the landscape$\_$day scene exhibited consistent and prominently positive slopes in individual preference results. In contrast, the experiment featuring the landscape$\_$night scene displayed relatively smaller slopes, accompanied by greater variability in responses among individuals. Furthermore, the disparity in 3D realism based on the number of views proved to be notably minimal, and a significant number of subjects exhibited an inverted JOD in the paired option (7$\times$3 vs 9$\times$4). 

The effect of perceived 3D realism depending on the number of views used for 4D CGH supervision is evaluated with a two-tailed z-test on the scaled JOD scores. Fig.~\ref{fig:subexp2}(B) shows the graphical representation of the scaling and simultaneously demonstrates the statistical results. Most paired options elicited very strong statistical significance on the difference ($p$<0.001). The paired option of 7$\times$3 vs 9$\times$4 in the landscape$\_$night scene showed strong evidence ($p$<0.01), while the paired option of 7$\times$3 vs 9$\times$4 in the village scene showed no evidence of the significance ($p$=0.45).

\section{Discussion}
In this section, we provide additional discussions not held in the main paper. 

\subsection{Display types for 3D perceptual testbeds}

\definecolor{mg}{rgb}{0.639,0.984,0.722}
\definecolor{my}{rgb}{0.996,0.875,0.643}
\definecolor{mr}{rgb}{0.941,0.561,0.620}
\begin{table*}[ht!]
    \centering
    \begin{tabular}{c|c|c|c|c|c|c|}
         \multirow{2}{*}{} & \multirow{2}{*}{resolution} & eye-tracking &  retinal blur &  monocular & image  & \multirow{2}{*}{eyebox} \\
                           &                             & required     &  class        &  occlusion/parallax              & quality & \\
         \hline
        fixed focus ~\cite{cakmakci2006head} & \cellcolor{mg} high & \cellcolor{mg} no & \cellcolor{mr} incorrect & \cellcolor{mr} not supported & \cellcolor{mg} high & \cellcolor{mg} wide \\
        \hline
        varifocal ~\cite{mercier2017fast} & \cellcolor{mg} high & \cellcolor{mr} yes & \cellcolor{mr} rendered & \cellcolor{mr} not supported & \cellcolor{mg} high & \cellcolor{mg} wide \\
        \hline
        fixed multifocal ~\cite{zhong2021reproducing} & \cellcolor{my} moderate & \cellcolor{mr} yes & \cellcolor{my} near-correct & \cellcolor{my} optimized & \cellcolor{my} moderate & \cellcolor{mr} narrow \\
        \hline
        attenuated layers ~\cite{huang2015light} & \cellcolor{mr} low & \cellcolor{my} optional & \cellcolor{my} near-correct & \cellcolor{mg} correct & \cellcolor{my} moderate & \cellcolor{my} moderate \\
        \hline
        integral imaging ~\cite{lanman2013near} & \cellcolor{my} moderate & \cellcolor{mg} no & \cellcolor{my} near-correct & \cellcolor{mg} correct & \cellcolor{mg} high & \cellcolor{my} moderate \\
        \hline
        holographic ~\cite{kim2022accommodative} & \cellcolor{mg} high & \cellcolor{mg} no & \cellcolor{mg} correct & \cellcolor{mg} correct & \cellcolor{mg} mid-high & \cellcolor{mr} narrow \\
        \hline
    \end{tabular}
    \caption{Assessment of various displays that support accommodation for a perceptual testbed, based on optical and perceptual criteria. A large portion of the criteria and evaluations are adapted from Matsuda et al.~\shortcite{matsuda2017focal}. It is noteworthy that recent advances in holographic displays have significantly improved image quality, which we classify as mid-high~\cite{peng2020neural,shi2021towards}. Eyebox defined as `moderate' ranges from 5–10 mm. We exclude computational complexity and form factor from the comparison, as perceptual studies can be conducted on prototypes with pre-rendered content.}
    \label{tab:display_types}
\end{table*}

As summarized in Table 1, data formats and CGH supervision techniques can be evaluated based on standard visual cues and scene representation capacities like defocus blur and view-dependent effects. Light field and its supervision for holographic displays uniquely support accurate view-dependent effects, including occlusion, parallax, and specular highlights. Table~\ref{tab:display_types} compares various accommodation-supporting displays for 3D scene perceptual testbeds~\cite{shibata2011zone, hoffman2008vergence, zhong2021reproducing, mercier2017fast}. Holographic displays, might now offer the best testbed for perceptual studies, with their flexibility to represent arbitrary data formats, and their improved image quality with recent advances~\cite{peng2020neural,shi2021towards}. Despite the limitations of a small eyebox, our study finds that parallax content (4D light field) greatly enhances perceptual realism, even with a narrow eyebox. We expect the perceptual impact to grow as the \'etendue of holographic displays increases.

\subsection{Trade-off between computation efficiency and quality}\label{sec:tradeoff}
\begin{figure}[ht!]
  \centering
  \includegraphics[width=\columnwidth]
  {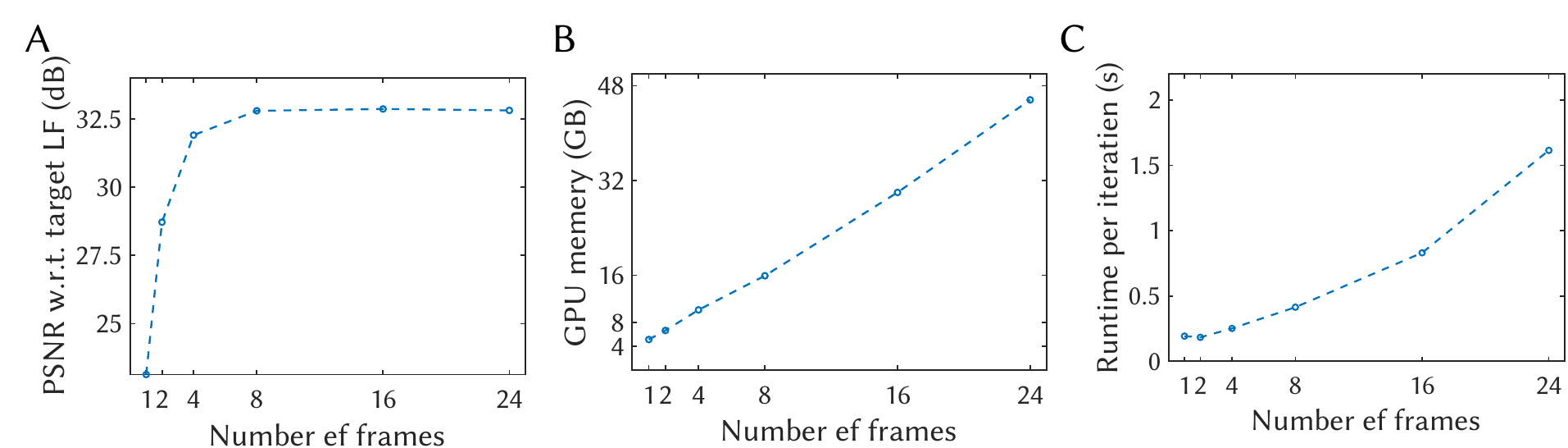}
  \caption{Trade-off in computation efficiency and performance in time-multiplexed holographic displays. Here, we show (A) light field fidelity measured by PSNR, (B) reserved GPU memory, and (C) runtime per iteration using various numbers of frames.}
  \label{fig:tradeoff}
\end{figure}

\begin{figure}[ht!]
  \centering
  \includegraphics[width=\columnwidth]
  {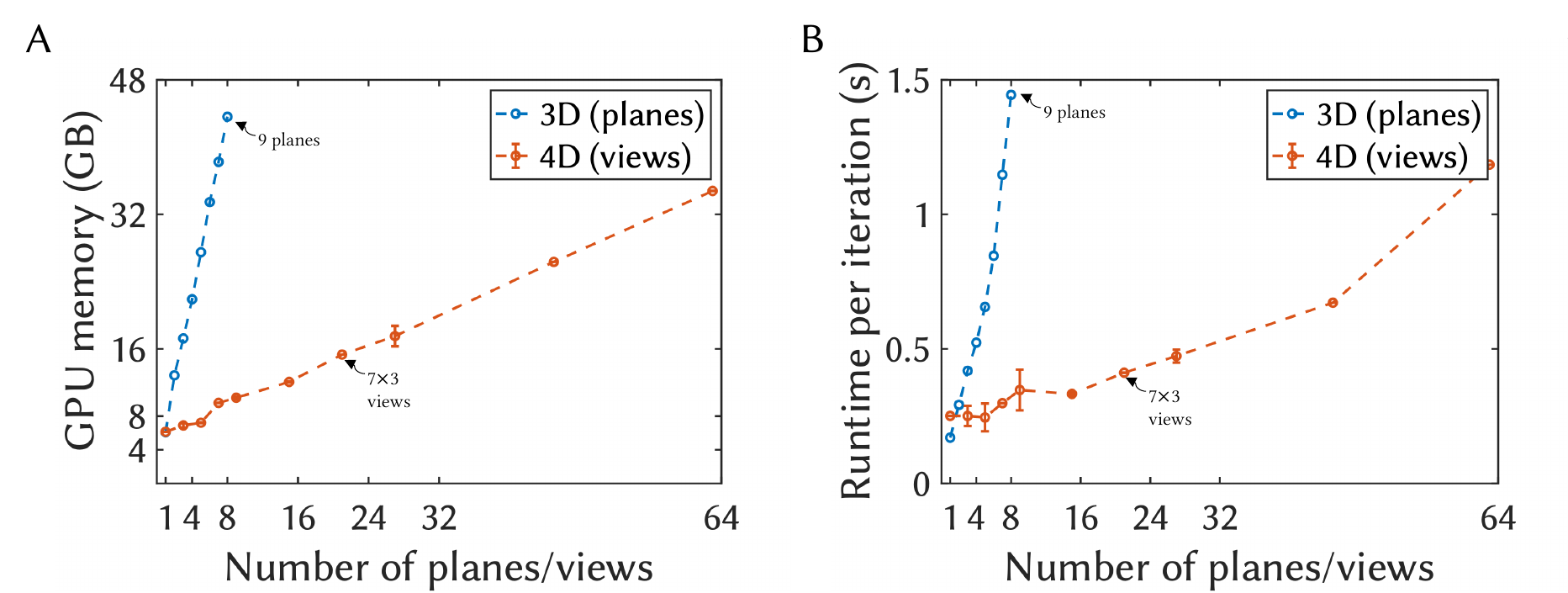}
  \caption{(\textit{left}) GPU memory and (\textit{right}) computation time as a function of the number of planes for 3D (blue, focal stack) and views for 4D (red, light field) CGH optimizations. The memory and runtime for each optimization type are represented on the same graph, with the x-axis indicating the number of planes or views, respectively. The errorbar in 4D CGH supervision represents the standard deviation of each measured value with combinatorial candidates (e.g. 9: 9$\times$1, 3$\times$3).}
  \label{fig:supp_compute_3d_4d}
\end{figure}

In this work, we have demonstrated that light field optimization using gradient-descent-based methods, which directly optimize for the raster light fields, is the most effective approach for achieving perceptual realism in 3D holographic displays. Notably, this achievement is enabled by time-multiplexed holographic display engines utilizing fast SLMs~\cite{kim2022accommodative, choi2022time}. Consequently, the computational resources required for calculating multiple frames of phase/amplitude patterns increase. Here, we present data on the time, performance, and memory usage in relation to the number of frames. In Fig.~\ref{fig:tradeoff}, we report our results from simulating our holographic display system with varying numbers of frames across four different scenes. We calculate the average performance (A), reserved GPU memory (B), and runtime per iteration (C). We run 2000 iterations for $7\times3$ views on an NVIDIA RTX A6000 GPU for this comparison, and the optimization almost converged after 1000 iterations (See Fig. S3). We note that the fidelity plateaus after 8 frames for our binary case, which implies that we could effectively allocate frames for different figures of merit, such as \textquotesingleétendue\textquotesingle. This would be an interesting avenue for future work. It is important to note that when using 24 frames, the simulation may take hours to complete thousands of iterations. Faster generation of light field holograms is probably one of the most interesting problems to tackle in future work, which we expect to solve using the deep learning-based method~\cite{peng2020neural, shi2021towards} that already has shown exciting progress.

In Fig.~\ref{fig:supp_compute_3d_4d}, we also compare the computation speed (per iteration) and memory usage for focal stack and light field optimizations on the same GPU. Using 8 frames, we were only able to optimize up to 9 planes using a 48 GB GPU, whereas for light field optimizations, we could optimize up to 63 views $(=9\times7)$. We could optimize a larger number of views than planes because the STFT operations used in light field optimizations share correlations between the graphs.

Similar to CGH supervision using light field data, 3D CGH supervision encounters limitations in memory and runtime per iteration. These approaches necessitate substantial memory allocation to retain the focal stack targets for the sampled depths. Additionally, the wave propagation relies on the fast Fourier transform-based angular spectrum method~\cite{goodman2005introduction}, causing resource demands to increase proportionally with the number of planes.

\subsection{Trade-off between degrees of freedom, the number of constraints, and \'etendue}\label{sec:tradeoff_etendue}
\begin{figure}[ht!]
  \centering
  \includegraphics[width=\columnwidth]
  {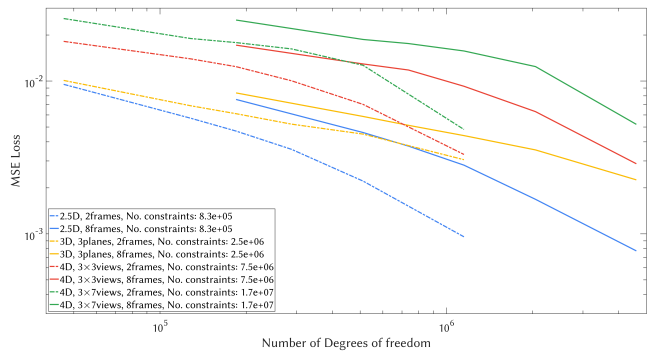}
  \caption{Loss values vs the number of degrees of freedom. We perform additional simulations on these factors to verify the trend. As expected, increased degrees of freedom or a smaller number of constraints lead to low loss values.}
  \label{fig:supp_fig_dof}
\end{figure}
In Fig.~\ref{fig:supp_fig_dof}, we present the loss values for various numbers of degrees of freedom, constraints, and CGH methods. Specifically, we run the gradient-descent-based optimization for CGH methods as described in the Appendix of the manuscript with different degrees of freedom in SLMs and number of frames. The number of degrees of freedom is calculated as $\text{(Number of optimizable pixels)} \times \text{(Number of frames)}$. To tune the number of optimizable pixels, we assume the SLMs with the same size but with larger pixel pitches. In practice, we set superpixels, so it has optimizable pixels number of $1920 \times 1200, 960 \times 600, \cdots, 192 \times 120$, and upsample them so that they match the original SLM pixel resolution in the optimization pipeline. This implementation allows us to match the feature size in the simulation pipeline while varying the number of optimizable pixels. Thus, having the same number of degrees of freedom with different frame counts implies larger superpixels, leading to lower performance within the same color category. The number of constraints is calculated by $\text{ (Number of pixels in the target (ROI)) } \times \text{ (number of views) } \times \text{ (number of planes) }$. We ran 1,000 iterations, set the ROI as central $1190 \times 700$ pixels, and the learning rate to $0.1$ for 2 frames and $0.4$ for 8 frames. As expected, increased degrees of freedom or a smaller number of constraints lead to low loss values. However, the commonly used mean square error metric should not directly represent the perceptual performance and exploring metric functions for constraining the spatial-angular information of light or perceptual realism would be an interesting direction for future study~\cite{kiran2017towards}.

\subsection{Eye rotation requirements}
\begin{figure}[ht!]
  \centering
  \includegraphics[width=\columnwidth]
  {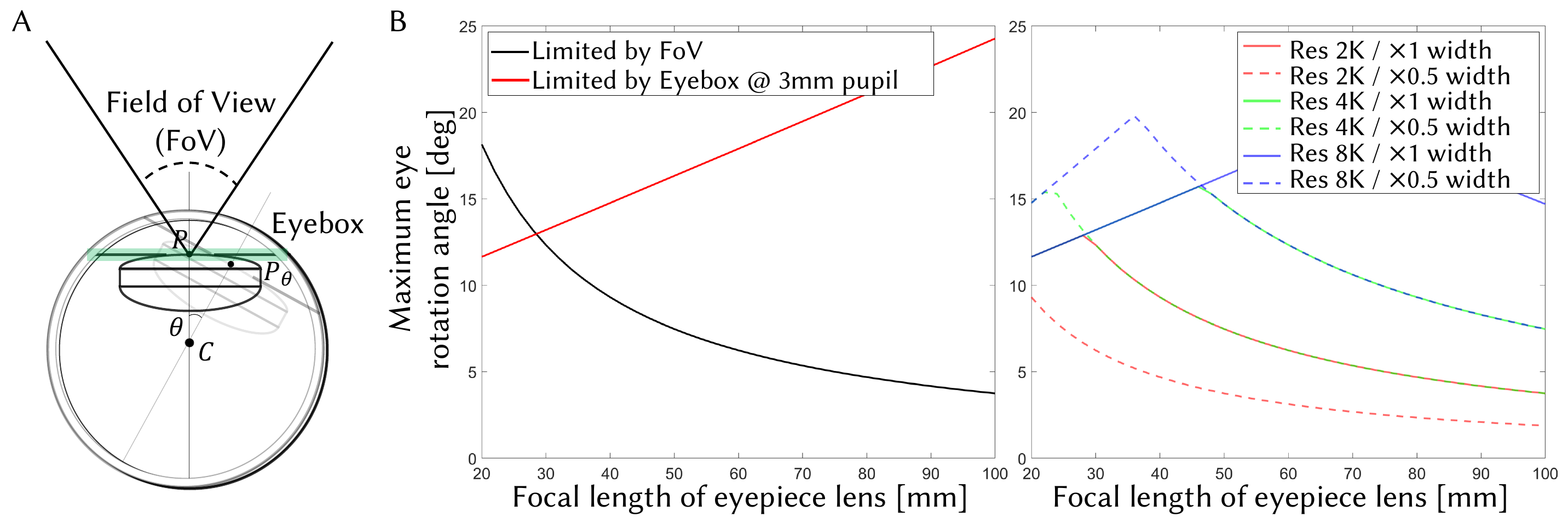}
  \caption{Required eye rotation angle in the given holographic near-eye displays that presents a trade-off relationship between FoV and size of eyebox. (A) Schematic of eye when the near-eye displays present an image with a specific field of view (FoV) and eyebox. The eye rotates based on the center of rotation ($C$). Here, the FoV and the eyebox are defined by the focal length of the eyepiece lens and the SLM's resolution and pixel pitch. (B) (left) The FoV and the eyebox limit the overall maximum rotation angle and (right) the maximum rotation angle can be plotted in different SLM specifications.}
  \label{fig:supeyerotation}
\end{figure}

Throughout the user study, our primary assumption revolved around the near-eye display's eyebox size being smaller than the pupil size, which we refer to underfilled pupil. However, we highlight the effectiveness of 4D CGH supervision when the eyebox surpasses the eye pupil, allowing partial sampling and clearer image disparity. It is important to note that not all near-eye display systems facilitate proper eye rotation movement. To elucidate the necessity of eye rotation in near-eye display configurations, we provide Figure~\ref{fig:supeyerotation}. In Figure~\ref{fig:supeyerotation}(A), a schematic eye interacting with a near-eye display, featuring a specific field of view and eyebox, is illustrated. Here, our assumption involves the eye rotating around the center of rotation ($C$) without additional translational movement, and aligns the visual axis and optical axis. The center of rotation is approximated as 10 mm behind the center of the iris ($P$). 

In this context, the eye's rotation range is constrained by two key factors: the FoV and the eyebox. The maximum eye rotation within the FoV refers to the highest angle the visual axis can pivot to reach the edge of the FoV. Simultaneously, the eye's rotation within the eyebox is computed by multiplying the distance between the eye pupil and the rotation center ($\overline{PC}$) by the rotation angle ($\theta$), ensuring it remains within the limits of the eyebox. We also account for an additional rotational angle allowance of half of the pupil diameter. The determination of the near-eye display system's maximum required rotation angle involves selecting the smaller value between these two calculations.

Analyzing Fig.~\ref{fig:supeyerotation}(B), as the focal length of the eyepiece lens increases, the rotation angle limited by the FoV decreases, while the restriction imposed by the eyebox expands. However, our primary focus remains on the minimum value between these limitations. For this simulation, a pupil diameter of 3 mm was assumed. Consequently, a shorter focal length for the eyepiece, which has a large chance of being an underfilled pupil, induces maximum eye rotation in the near-eye display setup. Conversely, opting for a longer focal length eyepiece, resulting in an overfilled pupil scenario, does not significantly prompt eye rotation due to its highly restricted FoV. This trend relaxes in an étendue-expanded scenario with high-resolution SLMs, as depicted in Fig.~\ref{fig:supeyerotation}. However, a decrease in the pixel pitch of the SLM to suit the near-eye display setup necessitates an eyepiece with a shorter focal length. Note that recent near-eye displays develop in shortening the eye relief along with the focal length to minimize the overall size.

\subsection{Debate in multi-focal vs. multi-view}
In conventional autostereoscopic 3D displays, specific data formats are mandated by systematic constraints. For instance, a multi-layer scheme supports focal-stack-based imagery exclusively, while the multi-view scheme displays multiple view images but with reduced spatial resolution. However, holographic displays can reconstruct both 3D data formats (multi-focal and multi-view) and the spatial-angular resolution trade-off of the multi-view scheme is relatively relaxed compared to the displays with integral imaging. Please refer to the Table~\ref{tab:display_types} for the comparisons. This intriguing capability of holographic display sparks a debate that warrants thorough discussion.

The target of $i$-th focal stack (${fs}_{i}$) can be acquired with the given light field map as follows:
\begin{equation}
    {{fs}_{i}}=\sum\limits_{v}{{{w}_{v}}\cdot {{l}_{v}}\left( x-{{x}_{i,v}},y-{{y}_{i,v}} \right)},
\end{equation}
where, ${l}_{v}$ is a $v$-th 2D slice of the light field, ${w}_{v}$ is a constant stating the view-dependent weight, and $({x}_{i,v},{y}_{i,v})$ is the coordinate translation depending on the view and depth. This indicates that the focal stack is a linear combination of the translated 2D slice of the light field and the view-dependent weight is usually unitary.

The perceived retinal image depending on the pupil state ($p$) and the focal state ($j$) of the eye can be simplified as follows:
\begin{align}
    {{I}_{p,j}} & =\sum\limits_{i}{{{psf}_{p,i,j}}*{{fs}_{i}}} \nonumber \\
    & = \sum\limits_{i}{{psf}_{p,i,j}}*{\sum\limits_{v}{{{w}_{v}}\cdot {{l}_{v}}}} \nonumber \\
    & \ne \sum\limits_{v}{{{psf}_{p,j,v}}*{{l}_{v}}}
\end{align}
Here, $psf_{p,i,j}$ is the displacement-dependent point spread function depending on the depth of the focal stack and the focal state of the eye. As stated in Sec. S5.1.4, the point spread function is also a function of pupil displacement, and it should not be pre-defined in the rendering stage for precise visualization. Thus, the approximation of a volumetric scene into a set of focal stacks can suppress the reconstruction of view-dependent information and ultimately deteriorate the 3D visual experience.



\section{Appendix}
In our study, we opted to scale the perceptual difference based on vote counts obtained through pairwise comparisons, as opposed to employing a direct rating system (mean-opinion score, as referenced by Hossfeld et al.\cite{hossfeld2016qoe}). Direct rating requires pre-trained subjects to assign a unified score to results with multi-dimensional differences, and individual rating scores tend to vary. In contrast, pairwise comparison is a simpler method that is well-suited for non-experts. Furthermore, this approach offers results with low measurement errors~\cite{shah2016estimation} and can utilize sparse sampling with adaptive experimental procedures~\cite{mantiuk2012comparison}.

\subsection{Statistics in pairwise comparison}
For the analysis of the results obtained from the pairwise comparison, we referred to the work of Perez-Ortiz and Mantiuk~\shortcite{perez2017practical} and utilized their released code from the GitHub repository \\(https://github.com/mantiuk/pwcmp). Here, we briefly summarize the analysis for a better understanding.

Analyzing the statistical difference using JOD-scaled data is more complex compared to the statistical test with direct rating experiments or accumulated vote counts. This is because the JOD scores are interconnected and not independent from one another. In detail, the correlation between the scaled JOD scores among the options can be determined by examining the covariance matrix ${C}$ obtained during JOD scaling. In that case, if we consider the pairwise comparison of $n$ conditions, which are scaled in JOD scores as ${q}=({{q}_{1}},\ldots ,{{q}_{n}})$, the score difference between two conditions, say $i$ and $j$, can be calculated as ${q}_{ij}={q}_{i}-{q}_{j}$. The variance for this score difference is given by ${{v}_{ij}}={{c}_{ii}}+{{c}_{jj}}-2{{c}_{ij}}$, where ${{c}_{ii}}$ and ${{c}_{jj}}$ represent the diagonal elements and ${{c}_{ij}}$ represents the off-diagonal element in the covariance matrix ${C}$. Based on this, we can assume that the z-value of the score difference follows a normal distribution, represented as ${{z}_{ij}}={{{q}_{ij}}}/{\sqrt{{{v}_{ij}}}}\sim N(0,1)$. A two-tailed z-test is employed to determine if we can reject the null hypothesis that there is no difference in JOD scores between the two conditions, with a specified level of confidence.



\bibliography{supplement}

\appendix